\documentclass[longauth]{aa} % for the long lists of affiliations 
\usepackage{graphicx}
\usepackage{txfonts}
\usepackage{pdflscape}

\usepackage{natbib,twoopt}
\usepackage{amsmath} % For multiple line equations
\usepackage[breaklinks=true]{hyperref} %% to avoid \citeads line fills
\bibpunct{(}{)}{;}{a}{}{,} %% natbib format for A&A and ApJ
\makeatletter
\newcommandtwoopt{\citeads}[3][][]{\href{http://adsabs.harvard.edu/abs/#3}%
{\def\hyper@linkstart##1##2{}%
\let\hyper@linkend\@empty\citealp[#1][#2]{#3}}}
\newcommandtwoopt{\citepads}[3][][]{\href{http://adsabs.harvard.edu/abs/#3}%
{\def\hyper@linkstart##1##2{}%
\let\hyper@linkend\@empty\citep[#1][#2]{#3}}}
\newcommandtwoopt{\citetads}[3][][]{\href{http://adsabs.harvard.edu/abs/#3}%
{\def\hyper@linkstart##1##2{}%
\let\hyper@linkend\@empty\citet[#1][#2]{#3}}}
\newcommandtwoopt{\citeyearads}[3][][]%
{\href{http://adsabs.harvard.edu/abs/#3}
{\def\hyper@linkstart##1##2{}%
\let\hyper@linkend\@empty\citeyear[#1][#2]{#3}}}
\makeatother

\usepackage{color}
\definecolor{mygreen}{RGB}{0,128,0}

\hypersetup{colorlinks=true,linkcolor=blue,citecolor=blue,urlcolor=blue}

 \makeatother
\newcommand{\gacs}[1]{{\footnotesize\texttt{#1}}}
\usepackage{wasysym}

\newcommand{\phat}[1]{{#1}}

\begin{document}

   \title{\textit{Gaia} Early Data Release 3}
   \subtitle{The astrometric solution}

\author{
     L.~Lindegren                     \inst{\ref{inst:0001}}
\and S.A.~Klioner                       \inst{\ref{inst:0002}}
\and J.~Hern\'{a}ndez                 \inst{\ref{inst:0003}}
\and A.~Bombrun                       \inst{\ref{inst:0004}}
\and M.~Ramos-Lerate                  \inst{\ref{inst:0005}}
\and H.~Steidelm\"{ u}ller            \inst{\ref{inst:0002}}
\and U.~Bastian                       \inst{\ref{inst:0006}}
\and M.~Biermann                      \inst{\ref{inst:0006}}
\and A.~de Torres                     \inst{\ref{inst:0004}}
\and E.~Gerlach                       \inst{\ref{inst:0002}}
\and R.~Geyer                         \inst{\ref{inst:0002}}
\and T.~Hilger                        \inst{\ref{inst:0002}}
\and D.~Hobbs                         \inst{\ref{inst:0001}}
\and U.~Lammers                       \inst{\ref{inst:0003}}
\and P.J.~McMillan                      \inst{\ref{inst:0001}}
\and C.A.~Stephenson                    \inst{\ref{inst:0007}}
\and J.~Casta\~{n}eda                 \inst{\ref{inst:0008}}
\and M.~Davidson                      \inst{\ref{inst:0009}}
\and C.~Fabricius                     \inst{\ref{inst:0010}}
\and G.~Gracia-Abril                  \inst{\ref{inst:0011},\ref{inst:0006}}
\and J.~Portell                       \inst{\ref{inst:0010}}
\and N.~Rowell                        \inst{\ref{inst:0009}}
\and D.~Teyssier                      \inst{\ref{inst:0007}}
\and F.~Torra                         \inst{\ref{inst:0008}}
\and S.~Bartolom{\'e}                 \inst{\ref{inst:0010}}
\and M.~Clotet                        \inst{\ref{inst:0010}}  
\and N.~Garralda                      \inst{\ref{inst:0010}}
\and J.J.~Gonz\'{a}lez-Vidal            \inst{\ref{inst:0010}}
\and J.~Torra$^\dagger$               \inst{\ref{inst:0010}}
\and U.~Abbas                         \inst{\ref{inst:0012}}
\and M.~Altmann                       \inst{\ref{inst:0006},\ref{inst:0013}}
\and E.~Anglada Varela                \inst{\ref{inst:0014}}
\and L.~Balaguer-N\'{u}\~{n}ez        \inst{\ref{inst:0010}}
\and Z.~Balog                         \inst{\ref{inst:0006},\ref{inst:0015}}
\and C.~Barache                       \inst{\ref{inst:0013}}
\and U.~Becciani                      \inst{\ref{inst:0016}}
\and M.~Bernet                        \inst{\ref{inst:0010}}
\and S.~Bertone                       \inst{\ref{inst:0017},\ref{inst:0018},\ref{inst:0012}}
\and L.~Bianchi                       \inst{\ref{inst:0019}}
\and S.~Bouquillon                    \inst{\ref{inst:0013}}
\and A.G.A.~Brown                         \inst{\ref{inst:0020}}
\and B.~Bucciarelli                   \inst{\ref{inst:0012}}
\and D.~Busonero                      \inst{\ref{inst:0012}}
\and A.G.~Butkevich                     \inst{\ref{inst:0012}}
\and R.~Buzzi                         \inst{\ref{inst:0012}}
\and R.~Cancelliere                   \inst{\ref{inst:0021}}
\and T.~Carlucci                      \inst{\ref{inst:0013}}
\and P.~Charlot                       \inst{\ref{inst:0022}}
\and M.-R. L.~Cioni                         \inst{\ref{inst:0023}}
\and M.~Crosta                        \inst{\ref{inst:0012}}
\and C.~Crowley                       \inst{\ref{inst:0004}}
\and E.F.~del Peloso                    \inst{\ref{inst:0006}}
\and E.~del Pozo                      \inst{\ref{inst:0024}}
\and R.~Drimmel                       \inst{\ref{inst:0012}}
\and P.~Esquej                        \inst{\ref{inst:0025}}
\and A.~Fienga                        \inst{\ref{inst:0026},\ref{inst:0027}}
\and E.~Fraile                        \inst{\ref{inst:0025}}
\and M.~Gai                           \inst{\ref{inst:0012}}
\and M.~Garcia-Reinaldos              \inst{\ref{inst:0003}}
\and R.~Guerra                        \inst{\ref{inst:0003}}
\and N.C.~Hambly                        \inst{\ref{inst:0009}}
\and M.~Hauser                        \inst{\ref{inst:0015},\ref{inst:0028}}
\and K.~Jan{\ss}en                    \inst{\ref{inst:0023}}
\and S.~Jordan                        \inst{\ref{inst:0006}}
\and Z.~Kostrzewa-Rutkowska           \inst{\ref{inst:0020},\ref{inst:0029}}
\and M.G.~Lattanzi                      \inst{\ref{inst:0012},\ref{inst:0030}}
\and S.~Liao                          \inst{\ref{inst:0012}}
\and E.~Licata                        \inst{\ref{inst:0012}}
\and T.A.~Lister                        \inst{\ref{inst:0031}}
\and W.~L\"{ o}ffler                  \inst{\ref{inst:0006}}
\and J.M.~Marchant                      \inst{\ref{inst:0032}}
\and A.~Masip                         \inst{\ref{inst:0010}}
\and F.~Mignard                       \inst{\ref{inst:0033}}
\and A.~Mints                         \inst{\ref{inst:0023}}
\and D.~Molina                        \inst{\ref{inst:0010}}
\and A.~Mora                          \inst{\ref{inst:0024}}
\and R.~Morbidelli                    \inst{\ref{inst:0012}}
\and C.P.~Murphy                        \inst{\ref{inst:0003}}
\and C.~Pagani                        \inst{\ref{inst:0034}}
\and P.~Panuzzo                       \inst{\ref{inst:0035}}
\and X.~Pe\~{n}alosa Esteller         \inst{\ref{inst:0010}}
\and E.~Poggio                        \inst{\ref{inst:0012}}
\and P.~Re Fiorentin                  \inst{\ref{inst:0012}}
\and A.~Riva                          \inst{\ref{inst:0012}}
\and A.~Sagrist\`{a} Sell\'{e}s       \inst{\ref{inst:0006}}
\and V.~Sanchez Gimenez               \inst{\ref{inst:0010}}
\and M.~Sarasso                       \inst{\ref{inst:0012}}
\and E.~Sciacca                       \inst{\ref{inst:0016}}
\and H.I.~Siddiqui                      \inst{\ref{inst:0036}}
\and R.L.~Smart                         \inst{\ref{inst:0012}}
\and D.~Souami                        \inst{\ref{inst:0037},\ref{inst:0038}}
\and A.~Spagna                        \inst{\ref{inst:0012}}
\and I.A.~Steele                        \inst{\ref{inst:0032}}
\and F.~Taris                         \inst{\ref{inst:0013}}
\and E.~Utrilla                       \inst{\ref{inst:0024}}
\and W.~van Reeven                    \inst{\ref{inst:0024}}
\and A.~Vecchiato                     \inst{\ref{inst:0012}}
}
\institute{
     Lund Observatory, Department of Astronomy and Theoretical Physics, Lund University, Box 43, 22100 Lund, Sweden\relax                                                                                                                                                                                        \label{inst:0001}
\and Lohrmann Observatory, Technische Universit\"{ a}t Dresden, Mommsenstra{\ss}e 13, 01062 Dresden, Germany\relax                                                                                                                                                                                               \label{inst:0002}
\and European Space Agency (ESA), European Space Astronomy Centre (ESAC), Camino bajo del Castillo, s/n, Urbanizacion Villafranca del Castillo, Villanueva de la Ca\~{n}ada, 28692 Madrid, Spain\relax                                                                                                           \label{inst:0003}
\and HE Space Operations BV for European Space Agency (ESA), Camino bajo del Castillo, s/n, Urbanizacion Villafranca del Castillo, Villanueva de la Ca\~{n}ada, 28692 Madrid, Spain\relax                                                                                                                        \label{inst:0004}
\and Vitrociset Belgium for European Space Agency (ESA), Camino bajo del Castillo, s/n, Urbanizacion Villafranca del Castillo, Villanueva de la Ca\~{n}ada, 28692 Madrid, Spain\relax                                                                                                                            \label{inst:0005}
\and Astronomisches Rechen-Institut, Zentrum f\"{ u}r Astronomie der Universit\"{ a}t Heidelberg, M\"{ o}nchhofstr. 12-14, 69120 Heidelberg, Germany\relax                                                                                                                                                       \label{inst:0006}
\and Telespazio Vega UK Ltd for European Space Agency (ESA), Camino bajo del Castillo, s/n, Urbanizacion Villafranca del Castillo, Villanueva de la Ca\~{n}ada, 28692 Madrid, Spain\relax                                                                                                                        \label{inst:0007}
\and DAPCOM for Institut de Ci\`{e}ncies del Cosmos (ICCUB), Universitat  de  Barcelona  (IEEC-UB), Mart\'{i} i  Franqu\`{e}s  1, 08028 Barcelona, Spain\relax                                                                                                                                                   \label{inst:0008}
\and Institute for Astronomy, University of Edinburgh, Royal Observatory, Blackford Hill, Edinburgh EH9 3HJ, United Kingdom\relax                                                                                                                                                                                \label{inst:0009}
\and Institut de Ci\`{e}ncies del Cosmos (ICCUB), Universitat  de  Barcelona  (IEEC-UB), Mart\'{i} i  Franqu\`{e}s  1, 08028 Barcelona, Spain\relax                                                                                                                                                              \label{inst:0010}
\and Gaia DPAC Project Office, ESAC, Camino bajo del Castillo, s/n, Urbanizacion Villafranca del Castillo, Villanueva de la Ca\~{n}ada, 28692 Madrid, Spain\relax                                                                                                                                                \label{inst:0011}
\and INAF - Osservatorio Astrofisico di Torino, via Osservatorio 20, 10025 Pino Torinese (TO), Italy\relax                                                                                                                                                                                                       \label{inst:0012}
\and SYRTE, Observatoire de Paris, Universit\'{e} PSL, CNRS,  Sorbonne Universit\'{e}, LNE, 61 avenue de l’Observatoire 75014 Paris, France\relax                                                                                                                                                              \label{inst:0013}
\and ATG Europe for European Space Agency (ESA), Camino bajo del Castillo, s/n, Urbanizacion Villafranca del Castillo, Villanueva de la Ca\~{n}ada, 28692 Madrid, Spain\relax                                                                                                                                    \label{inst:0014}
\and Max Planck Institute for Astronomy, K\"{ o}nigstuhl 17, 69117 Heidelberg, Germany\relax                                                                                                                                                                                                                     \label{inst:0015}
\and INAF - Osservatorio Astrofisico di Catania, via S. Sofia 78, 95123 Catania, Italy\relax                                                                                                                                                                                                                     \label{inst:0016}
\and Center for Research and Exploration in Space Science and Technology, University of Maryland Baltimore County, 1000 Hilltop Circle, Baltimore MD, USA\relax                                                                                                                                                  \label{inst:0017}
\and GSFC - Goddard Space Flight Center, Code 698, 8800 Greenbelt Rd, 20771 MD Greenbelt, United States\relax                                                                                                                                                                                                    \label{inst:0018}
\and EURIX S.r.l., Corso Vittorio Emanuele II 61, 10128, Torino, Italy\relax                                                                                                                                                                                                                                     \label{inst:0019}
\and Leiden Observatory, Leiden University, Niels Bohrweg 2, 2333 CA Leiden, The Netherlands\relax                                                                                                                                                                                                               \label{inst:0020}
\and University of Turin, Department of Computer Sciences, Corso Svizzera 185, 10149 Torino, Italy\relax                                                                                                                                                                                                         \label{inst:0021}
\and Laboratoire d'astrophysique de Bordeaux, Univ. Bordeaux, CNRS, B18N, all{\'e}e Geoffroy Saint-Hilaire, 33615 Pessac, France\relax                                                                                                                                                                           \label{inst:0022}
\and Leibniz Institute for Astrophysics Potsdam (AIP), An der Sternwarte 16, 14482 Potsdam, Germany\relax                                                                                                                                                                                                        \label{inst:0023}
\and Aurora Technology for European Space Agency (ESA), Camino bajo del Castillo, s/n, Urbanizacion Villafranca del Castillo, Villanueva de la Ca\~{n}ada, 28692 Madrid, Spain\relax                                                                                                                             \label{inst:0024}
\and RHEA for European Space Agency (ESA), Camino bajo del Castillo, s/n, Urbanizacion Villafranca del Castillo, Villanueva de la Ca\~{n}ada, 28692 Madrid, Spain\relax                                                                                                                                          \label{inst:0025}
\and Universit\'{e} C\^{o}te d'Azur, Observatoire de la C\^{o}te d'Azur, CNRS, Laboratoire G\'{e}oazur, Bd de l'Observatoire, CS 34229, 06304 Nice Cedex 4, France\relax                                                                                                                                         \label{inst:0026}
\and IMCCE, Observatoire de Paris, Universit\'{e} PSL, CNRS, Sorbonne Universit{\'e}, Univ. Lille, 77 av. Denfert-Rochereau, 75014 Paris, France\relax                                                                                                                                                           \label{inst:0027}
\and TRUMPF Photonic Components GmbH, Lise-Meitner-Stra{\ss}e 13,  89081 Ulm, Germany\relax                                                                                                                                                                                                                      \label{inst:0028}
\clearpage
\and SRON, Netherlands Institute for Space Research, Sorbonnelaan 2, 3584CA, Utrecht, The Netherlands\relax                                                                                                                                                                                                      \label{inst:0029}
\and University of Turin, Department of Physics, Via Pietro Giuria 1, 10125 Torino, Italy\relax                                                                                                                                                                                                                  \label{inst:0030}
\and Las Cumbres Observatory, 6740 Cortona Drive Suite 102, Goleta, CA 93117, USA\relax                                                                                                                                                                                                                          \label{inst:0031}
\and Astrophysics Research Institute, Liverpool John Moores University, 146 Brownlow Hill, Liverpool L3 5RF, United Kingdom\relax                                                                                                                                                                                \label{inst:0032}
\and Universit\'{e} C\^{o}te d'Azur, Observatoire de la C\^{o}te d'Azur, CNRS, Laboratoire Lagrange, Bd de l'Observatoire, CS 34229, 06304 Nice Cedex 4, France\relax                                                                                                                                            \label{inst:0033}
\and School of Physics and Astronomy, University of Leicester, University Road, Leicester LE1 7RH, United Kingdom\relax                                                                                                                                                                                          \label{inst:0034}
\and GEPI, Observatoire de Paris, Universit\'{e} PSL, CNRS, 5 Place Jules Janssen, 92190 Meudon, France\relax                                                                                                                                                                                                    \label{inst:0035}
\and Department of Astrophysical Sciences, 4 Ivy Lane, Princeton University, Princeton NJ 08544, USA\relax                                                                                                                                                                                                       \label{inst:0036}
\and LESIA, Observatoire de Paris, Universit\'{e} PSL, CNRS, Sorbonne Universit\'{e}, Universit\'{e} de Paris, 5 Place Jules Janssen, 92190 Meudon, France\relax                                                                                                                                                 \label{inst:0037}
\and naXys, University of Namur, Rempart de la Vierge, 5000 Namur, Belgium\relax                                                                                                                                                                                                                                 \label{inst:0038}
}

   \date{ }

% \abstract{}{}{}{}{} 
% 5 {} token are mandatory
 
\abstract
  {\textit{Gaia} Early Data Release 3 (\textit{Gaia} EDR3) contains results for 1.812~billion sources 
  in the magnitude range $G=3$ to 21 based on observations collected by the European Space
  Agency \textit{Gaia} satellite during the first 34~months of its operational phase.}  
  {We describe the input data, the models, and the processing used for the astrometric content
  of \textit{Gaia} EDR3, as well as the validation of these results performed within the astrometry task.}  
  {The processing broadly followed the same procedures as for \textit{Gaia} DR2, but with significant 
  improvements to the modelling of observations. 
  For the first time in the \textit{Gaia} data processing, colour-dependent calibrations of the line- and 
  point-spread functions have been used for sources with well-determined colours from DR2. 
  In the astrometric 
  processing these sources obtained five-parameter solutions, whereas other sources were processed 
  using a special calibration that allowed a pseudocolour to be estimated as the sixth astrometric parameter. 
  Compared with DR2, the astrometric calibration models have been extended, and the spin-related 
  distortion model includes a self-consistent determination of basic-angle variations, improving the 
  global parallax zero point.}  
  {\textit{Gaia} EDR3 gives full astrometric data (positions at epoch J2016.0, parallaxes, and proper 
  motions) for 1.468~billion sources (585~million with five-parameter solutions, 882~million with
  six parameters), and mean positions at J2016.0 for an additional 344~million. 
  Solutions with five parameters are generally more accurate than six-parameter solutions, and are 
  available for 93\% of the sources brighter than the 17th magnitude. 
  The median uncertainty in parallax and annual proper motion is 0.02--0.03~mas at magnitude $G=9$ 
  to 14, and around 0.5~mas at  $G=20$. Extensive characterisation of the statistical properties of the
  solutions is provided, including the estimated angular power spectrum of parallax bias from the quasars.}
  {}


   \titlerunning{\textit{Gaia} Early Data Release 3 -- Astrometric solution} 
   \authorrunning{Lindegren et al.}

   \maketitle

%
%________________________________________________________________

\section{Introduction} 
\label{sec:intro}

\textit{Gaia} Early Data Release 3 (EDR3; \citeads{EDR3-DPACP-130}) contains provisional 
astrometric and photometric data for more than 1.8~billion ($1.8\times 10^9$) sources
based on the first 34~months of observations made by the European Space Agency's 
\textit{Gaia} mission (\citeads{2016A&A...595A...1G}) since the start of the nominal 
operations in July 2014. The astrometric data in EDR3 include the five astrometric 
parameters (position, parallax, and proper motion) for 1.468~billion sources, and the 
approximate positions at epoch J2016.0 for an additional 344~million mostly faint 
sources. All sources have magnitudes in \textit{Gaia}'s unfiltered photometric 
passband $G$, and 1.544~billion have two-colour photometry in the passbands 
$G_\text{BP}$ and $G_\text{RP}$ defined by the blue and red photometers 
(BP and RP; \citealt{EDR3-DPACP-117}). The magnitudes of the well-observed sources range 
from $G=6$ to 21. All data are publicly available in the online \textit{Gaia} Archive at 
\href{https://archives.esac.esa.int/gaia}{\tt https://archives.esac.esa.int/gaia}.

The EDR3 is a subset of the full \textit{Gaia} Data Release 3 (DR3), planned for the 
first half of 2022. The full release will provide a much wider set of data, including
detailed spectrophotometric and variability information, additional astrometric 
data on non-single and extended objects, and the radial velocities, object classification, 
and astrophysical parameters for many sources. However, the basic astrometric 
information on the \textit{Gaia} DR3 sources, obtained by treating all of them as 
single stars, has already been provided in EDR3 and will not change for DR3.
  
This paper gives an overview of the processing leading up to the EDR3 astrometry,
as well as of the main characteristics of the astrometric results. Further details are provided 
in the online documentation of the \textit{Gaia} Archive and in specialised papers. In 
particular, the celestial reference frame of \textit{Gaia} (E)DR3 is described in 
\citet{EDR3-DPACP-133}, the parallax bias (zero point) is discussed in
\citet{EDR3-DPACP-132}, and the overall properties of the release are reviewed in
\citet{EDR3-DPACP-126}. A general description of the \textit{Gaia} mission can be found
in \citetads{2016A&A...595A...1G}.

The core astrometric solution for \textit{Gaia}, known as AGIS (astrometric global 
iterative solution), was comprehensively described in the pre-launch paper by 
\citetads{2012A&A...538A..78L}. This remains a useful general reference for AGIS 
in spite of the many modifications and improvements introduced since 2012. 
We also refer frequently to \citetads{2018A&A...616A...2L},
which describes the astrometric solution for \textit{Gaia} DR2.

\begin{figure*}
\center
%\sidecaption
%  \includegraphics[width=16cm]{Figures/timeline.pdf}   % data segment = "S"
  \includegraphics[width=16cm]{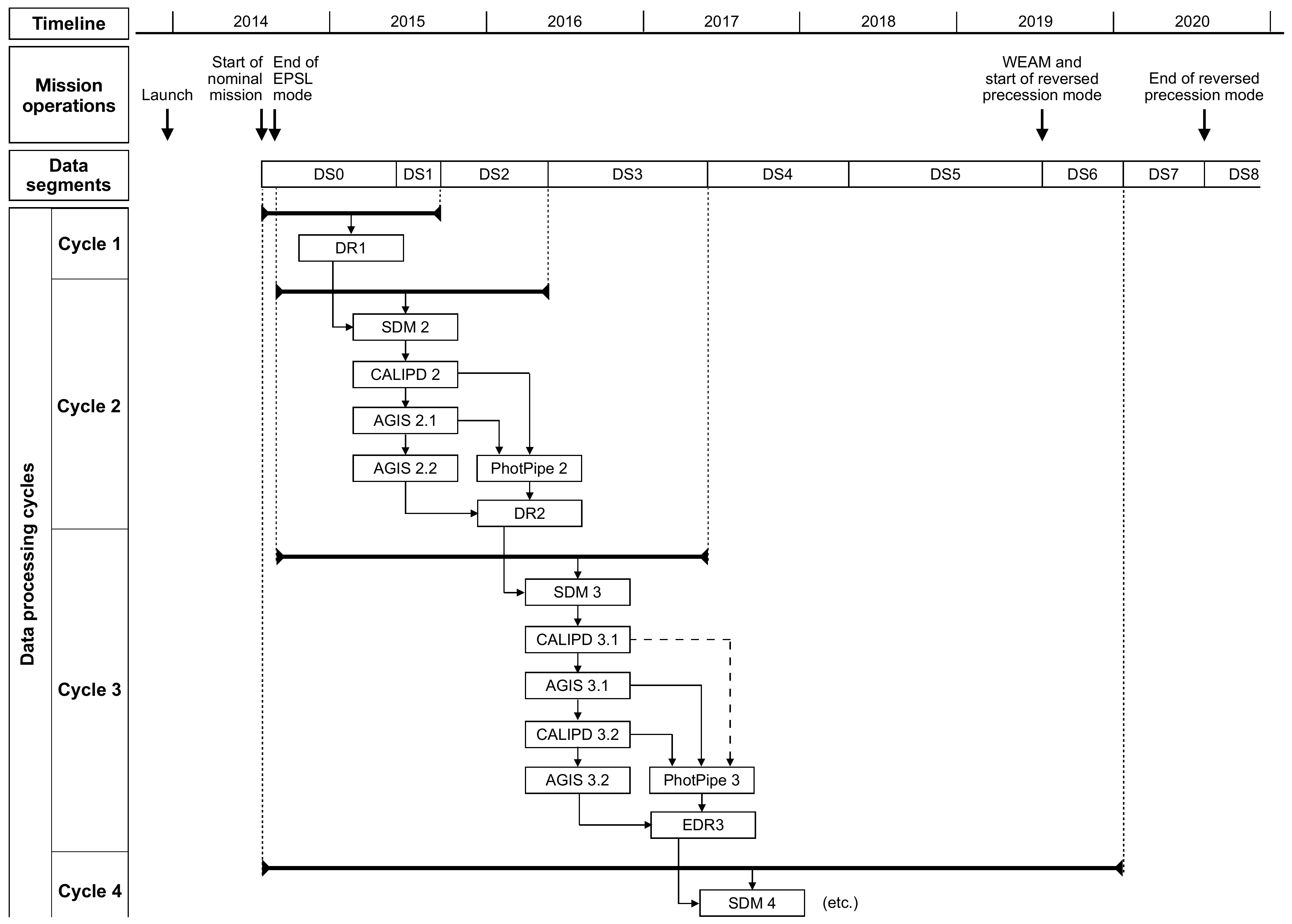}  %{Figures/timeline_DS.pdf}  % data segment = "DS"
    \caption{Main steps of the EDR3 astrometry processing and their place in the
    cyclic processing scheme of DPAC. \textit{Gaia} EDR3 (and DR3) are generated in
    the third processing cycle (cycle~3). The stretches of observational data processed 
    in the different cycles are indicated by thick horizontal lines. The boxes connected by arrows 
    show the sequence of processing steps and their interdependencies, but they are not placed
    chronologically on the timeline. Only steps directly relevant for the astrometry are shown,
    leaving out most of the complexities of the full DPAC processing. No details are given for DR1.
    The first month of 
    the nominal mission, with observations made in the ecliptic pole scanning law (EPSL) mode, 
    was not used for the astrometry in DR2 and EDR3, but may be incorporated in later releases. 
    The Whitehead eclipse avoidance manoeuvre (WEAM) on 16 July 2019 marks the beginning
    of the extended mission. In the first year of the extended mission (data segments DS6 
    and DS7), scanning was made in the reversed precession mode (Sect.~\ref{sec:acRate}). 
    The processes SDM, CALIPD, AGIS, and PhotPipe are explained in Sect.~\ref{sec:tasks}.
    }
    \label{fig:timeline}
\end{figure*}

\section{Overview of the astrometric processing}
\label{sec:overview}

\subsection{Main processing tasks}
\label{sec:tasks}

In the cyclic processing scheme adopted by the \textit{Gaia} Data Processing and Analysis 
Consortium (DPAC; \citeads{2016A&A...595A...1G}), EDR3 and DR3 are products of the 
third processing cycle, using observations in the first four data segments called DS0--DS3
in Fig.~\ref{fig:timeline}. The data segments are just a convenient, but essentially
arbitrary division of the raw data by acquisition time. As suggested in the figure, the cycles 
treat successively larger chunks of the raw data by including additional segments, but in 
every cycle the old segments are always reprocessed together with the new ones. This 
iterative reprocessing of earlier data segments is necessary in order to achieve the 
uniformly best treatment of all the data, and a consistent assignment of source 
identifiers to the on-board detections.

In Fig.~\ref{fig:timeline}, the boxes labelled PhotPipe represent the complex photometric 
processing described elsewhere \citep{EDR3-DPACP-117,DPACP-118,DPACP-119,DPACP-120}.
This is not part of the astrometric processing as such but is included in the diagram because 
the photometric information, and in particular the colour information encoded in the 
effective wavenumbers ($\nu_\text{eff}$; Sect.~\ref{sec:nuEff}) calculated in PhotPipe, are 
needed for calibrating the colour-dependent line-spread and point-spread functions (LSF 
and PSF) of the astrometric instrument. Because PhotPipe runs essentially in parallel with 
the astrometric solution (AGIS), this implies that the astrometric processing in cycle~$N$ 
must use photometric information from cycle~$N{-}1$.

The boxes in Fig.~\ref{fig:timeline} labelled SDM, CALIPD, and AGIS represent the three 
main stages in the processing of the raw CCD (charge-coupled device) data that are of 
immediate relevance for the astrometry:

In the first stage, the SDM (source, detection-classifier, and cross-match) aims to identify all on-board
detections belonging to the same source and assign a unique source identifier (\gacs{source\_id}) 
to each such cluster of detections \citep{EDR3-DPACP-124}. An important part of the process is the 
identification of spurious detections, created for example by the diffraction spikes of bright stars 
\citepads{2016A&A...595A...3F}. 
Because the updated source list and table of links to the (genuine) detections created by the 
SDM is used by all subsequent processes, this is one of the first tasks to be executed in a cycle.
The source list from the previous cycle is a starting point for the task, but the new data and 
improved reconstruction of the satellite attitude (a key element in translating observed 
transit times into positions) unavoidably require some of the old sources to be split or merged,
in addition to creating entirely new ones. For example, EDR3 contains many pairs of sources 
(most of which are genuine binaries) that are separated by less than 0.4~arcsec, where DR2 had only one. 
Such cases could lead to the assignment of new source identifiers for both components.
The auxiliary table \gacs{dr2\_neighbourhood} helps to trace the evolution of source identifiers.

The second stage, CALIPD, consists of two parts, calibration (CAL) and image parameter 
determination (IPD). In CAL, the LSF (for one-dimensional observations) and PSF 
(for two-dimensional observations) are calibrated as functions of time, colour, 
and several other variables in order to take into account the optical imperfections of the 
instrument and their temporal evolution \citep{EDR3-DPACP-73}. The LSF and PSF 
describe the shape of the image profile for a point source as well as the small 
displacement caused by chromatic effects (Sect.~\ref{sec:nuEff}).
In IPD, the LSF or PSF relevant for a particular observation is fitted to the sampled CCD image, 
yielding precise estimates of its one- or two-dimensional location in the pixel stream 
and of the total flux of the image in the $G$ band \citepads{2016A&A...595A...3F}. 
The resulting image locations constitute the main input data for the astrometric solution,
while the flux estimates are used for the determination of $G$ magnitudes in PhotPipe.
Whereas CAL only uses a small fraction of the available observations for the LSF and PSF
calibrations, IPD is applied to all observations in the skymapper (SM) and
astrometric field (AF).
 
In the third stage, AGIS performs a simultaneous least-squares
estimation of the attitude, instrument calibration, and the five astrometric parameters for
a subset of well-behaved primary sources (about 14.3~million in cycle~3). The calibration
includes corrections for effects that are not accounted for in the CALIPD, or only partially
corrected at that stage. The comprehensive pre-launch description of AGIS in 
\citetads{2012A&A...538A..78L} is complemented by specifics of the
current models in Sect.~\ref{sec:models}.

SDM and CALIPD belong to the intermediate data update (IDU) system, which includes 
several additional tasks such as astrophysical background estimation 
\citepads{2016A&A...595A...3F} and electronic calibrations \citepads{2018A&A...616A..15H}.
Compared with DR2, several major improvement of the IDU have been introduced with
cycle~3. In SDM the treatment of high-proper motion stars, very bright stars, variable 
sources, and close pairs has been much improved \citep{EDR3-DPACP-124}. In CALIPD the 
image profiles (LSF and PSF) are no longer assumed to be independent of time and
colour, as was the case in cycle~2, and a much more realistic two-dimensional model (PSF) is 
used \citep{EDR3-DPACP-73}. Moreover, as described in Sect.~\ref{sec:nuEff} and
Fig.~\ref{fig:timeline}, the CALIPD and AGIS tasks are for the first time iterated in 
order that CALIPD may benefit from the improved astrometry, attitude, and instrument 
calibration obtained by including the new data segment (DS3) in AGIS.

\subsection{Observations used}
\label{sec:data}

\textit{Gaia} EDR3 is based on data collected from the start of the nominal observations 
on 25 July 2014 (10:30 UTC) until 28 May 2017 (08:45 UTC), or 1038~days
(data segments DS0--DS3 in Fig.~\ref{fig:timeline}).
Similarly to the astrometric solution for DR2 \citepads{2018A&A...616A...2L}, this
solution did not use the observations in the first month of the operational phase, when
the special ecliptic pole scanning law (EPSL) was employed. The data for the astrometry 
therefore start on 22 August 2014 (21:00 UTC) and cover 1009~days or 2.76~yr, with some 
interruptions mentioned below. 

The time coverage for this solution is therefore about one year longer than the astrometric 
solution for \textit{Gaia} DR2, which covered 640~days or 1.75~year. The expected 
improvement from the added data and longer time baseline scales as $T^{-1/2}$ for the
parallaxes and positions at the mean epoch of observation, and as $T^{-3/2}$ for the
proper motions; thus uncertainties should be smaller by a factor 0.80 for the parallaxes
and positions, and by a factor 0.51 for the proper motions. As shown in Sect.~\ref{sec:form}, the
median ratios of the formal uncertainties are slightly better than this thanks to additional
improvements in the instrument and attitude modelling. The reference epoch J2016.0 used 
for the astrometry in \textit{Gaia} EDR3 (Sect.~\ref{sec:source}) is close to the mid-point of 
the observations.

The on-board mission timeline (OBMT) is conveniently used to label on-board events; it is
expressed as the number of nominal revolutions of exactly 21\,600~s (6~h) on-board time from 
an arbitrary origin.%
\footnote{The rubidium atomic clock on board of \textit{Gaia} does not count SI seconds because 
it is a free-running oscillator with some (very small) time-dependent frequency error. This is 
calibrated in a special part of the data processing (see Sect.~\ref{sec:aux}), but ignored when giving 
intervals in OBMT.}
The approximate relation between OBMT (in revolutions) and barycentric
coordinate time (TCB, in Julian years) at \textit{Gaia} is
\begin{equation}\label{eq:obmt}
\text{TCB} \simeq \text{J}2015.0 + (\text{OBMT} - 1717.6256~\text{rev})/(1461~\text{rev~yr}^{-1}) \, ,
\end{equation} 
or as a Julian Date,
\begin{equation}\label{eq:obmtJD}
\text{JD}2457023.75 + (\text{OBMT} - 1717.6256~\text{rev})/(4~\text{rev~d}^{-1}) \, .
\end{equation} 
The nominal observations start at OBMT 1078.38~rev. The astrometric solution used
data in the interval OBMT 1192.13--5230.09~rev (J2014.64032--J2017.40415), with 
major gaps as listed in Table~\ref{tab:gaps}.

\begin{table}
\caption{Major gaps and events affecting the astrometric solution.
\label{tab:gaps}}
\small
\begin{tabular}{ccrl}
\hline\hline
\noalign{\smallskip}
$t_\text{beg}$ & $t_\text{end}$ & length & description\\
\noalign{\smallskip}
\hline
\noalign{\smallskip}
%1088.120 & 1090.200 & 2.080 & VPU reset \\
1220.400 & 1225.200 & 4.800 & VPU reset \\
1316.490 & 1389.113 & 72.623 & decontamination {\#}4 \\
1443.800 & 1444.200 & 0.400 & refocus (following FoV) \\
1653.800 & 1660.000 & 6.200 & PAA anomaly \\
1820.900 & 1830.000 & 9.100 & PAA anomaly \\
2094.000 & 2099.000 & 5.000 & PDHU anomaly \\
2179.125 & 2191.000 & 11.875 & VPU software update \\
2192.252 & 2195.218 & 2.967 & observation gap\\
2238.000 & 2242.000 & 4.000 & unknown \\
2322.300 &	2401.559 & 79.259 & decontamination {\#}5 \\
2405.967 & 2408.643 & 2.676 & observation gap\\
2408.935 & 2409.968 & 1.033 & observation gap\\
2574.640 &	2575.400 & 0.760 & refocus (preceding FoV) \\
2954.200 & 2958.000 & 3.800 & moon eclipse \\
3045.133 & 3049.000 & 3.867 & PDHU anomaly\\
3603.250 & 3605.227 & 1.976 & observation gap\\
3646.800 & 3650.000 & 3.200 & moon eclipse \\
3663.700 & 3667.000 & 3.300 & moon eclipse \\
4074.210 & 4076.063 & 1.853 & observation gap\\
4112.463 & 4180.000 & 67.537 & decontamination {\#}6 \\
4271.753 & 4275.200 & 3.445 & PAA anomaly \\
4477.441 & 4481.000 & 3.550 & PAA anomaly \\
4512.502 & 4515.000 & 2.498 & PAA anomaly \\
4545.144 & 4548.000 & 2.856 & PAA anomaly \\
5078.547 & 5080.600 & 2.053 & STR anomaly \\
\noalign{\smallskip}
\hline
\end{tabular}
\tablefoot{The table lists gaps longer than 1~revolution (0.25~day) and some other events of relevance
for the calibration model. $t_\text{beg}$ and $t_\text{end}$ are the start and end times of the
gap in OBMT revolutions (see Eq.~\ref{eq:obmt}). The third column is the length of the gap in 
revolutions. Abbreviations: 
VPU = video processing unit,
FoV = field of view,
PAA = phased array antenna,
PDHU = payload data handling unit,
STR = star tracker.
}
\end{table}

\subsection{Use of colour information in CALIPD and AGIS}
\label{sec:nuEff}

In the focal plane of an all-reflecting telescope, free of wavefront aberrations, the 
point-spread function (PSF) is completely symmetric. Although the width 
of the PSF increases with wavelength, because of diffraction, its position does 
not change and is consequently
independent of the spectral composition of the light (achromatic). This is no 
longer true for a real instrument like \textit{Gaia}. Inevitable coma-like wavefront 
errors produce asymmetric PSFs, in which both the shape and location 
depend on the spectrum. Subtle wavelength-dependent effects can also be 
introduced by the CCD detector itself. We use `chromaticity' as a generic term 
for these several effects, but especially for the variation of the PSF location 
with colour. Chromaticity creates colour-dependent biases in the astrometric
results, unless it is properly calibrated and corrected for in the processing.

Chromaticity should ideally be completely eliminated already in CALIPD, so that
the astrometric solution (AGIS) would not need to care about the sources having 
different colours. This requires (i) that in CAL both the shape and location of the 
LSF or PSF are accurately calibrated as functions of the spectral energy 
distribution (multiplied by the wavelength passband); and (ii) that in IPD the location 
and flux of the image are estimated using the correct profile, depending on 
the actual spectrum of the source in each observation. The astrometric parameters 
determined in the subsequent AGIS solution will then be free from chromatic 
biases. There is a certain circularity here: To achieve (i), CAL must be able to identify 
the point in the image profile that corresponds to the achromatic centre of the source, 
and this can only be done by means of the (achromatic) astrometric parameters 
determined by AGIS. This strong interdependency between CALIPD and AGIS is
dealt with by executing the two tasks alternately, which motivates the sequence 
CALIPD~3.1, AGIS~3.1, CALIPD~3.2, AGIS~3.2 in Fig.~\ref{fig:timeline}. The 
CALIPD/AGIS sequence should ideally be iterated until convergence, but in cycle~3
only two iterations (3.1 and 3.2) were made. This appears to be sufficient in practice,
because AGIS is able to eliminate most of the chromatic effects left uncorrected in IPD
via the colour-dependent terms in the AGIS calibration model (Sect.~\ref{sec:cal}).

In cycle~3 two simplifying assumptions are made, both of which may be relaxed at
some future time. The first is that the spectral information needed for the chromaticity
correction is fully encoded
in the effective wavenumber, defined as $\nu_\text{eff}=\langle\lambda^{-1}\rangle$. 
Here $\lambda$ is the wavelength, and angular brackets denote a mean value 
weighted by the detected photon flux per unit wavelength interval. 
This quantity was chosen, in preference to (say) the effective wavelength or 
colour index, based on pre-launch studies using the properties of the \textit{Gaia} 
instrument as expected at the time. According to these studies, the effective wavenumber 
provides a good one-dimensional parametrisation of chromaticity for ordinary 
stellar spectra, but may not be enough to describe shifts at the few $\mu$as 
level in atypical cases such as quasar spectra. Thus, more complex dependencies on 
the source spectrum may have to be considered in the future, but for the time being 
we use $\nu_\text{eff}$ as defined.

The second assumption is that the spectrum (or effective wavenumber) is the 
same in all observations of a given source. Although this is a sufficiently good 
approximation for most sources, it may prevent us from reaching the full potential 
of \textit{Gaia} for some variable objects. The remedy is simple in principle, namely
to use the actual colour of the source at each observation, but this may require 
an additional iteration over PhotPipe and the variability analysis 
\citepads{2018A&A...618A..30H}.

Even with the simplifications mentioned above, the conditions for eliminating 
chromaticity in CALIPD are not fully met in cycle~3. The main obstacle is that many 
sources do not have reliable colour information that can be used to select the 
appropriate image profiles for the IPD. The effective wavenumbers used in 
CALIPD~3.1 and 3.2 were calculated in PhotPipe~2 directly from the sampled and 
calibrated mean BP and RP spectra, and are given in EDR3 as 
\gacs{nu\_eff\_used\_in\_astrometry}. The analysis of the BP and RP spectra 
is very challenging in crowded areas and at the faintest magnitudes, owing to 
the blending of overlapping spectra and the difficulty to estimate the background 
accurately \citep{EDR3-DPACP-118}. A strict filtering on the quality of 
$\nu_\text{eff}$ was adopted in order to avoid that biases in the photometric
colour might propagate into the astrometry. Of particular concern was the BP+RP 
flux excess issue \citepads{2018A&A...616A...4E}, which in DR2 tended to make faint 
sources in crowded areas too blue. As a result of the adopted filtering, about
two thirds of the sources in EDR3 do not have a valid \gacs{nu\_eff\_used\_in\_astrometry}.
The situation is more favourable for brighter sources, where, for example, only 
12\% of the sources with $G<18$~mag lack a valid $\nu_\text{eff}$.

For the many sources without a valid $\nu_\text{eff}$, special procedures were 
used both in IPD and AGIS. In the IPD, image parameters 
were estimated by fitting the calibrated LSF or PSF for the default wavenumber
$\nu_\text{eff}^\text{\,def}=1.43~\mu$m$^{-1}$. This value was chosen to be
close to the mean $\nu_\text{eff}$ of faint sources, for which the default value 
is mostly used; thus, the averaged error introduced by the procedure is minimised. 
In AGIS, a six-parameter solution was computed for these sources, where the sixth
unknown, after the standard five astrometric parameters, is the pseudocolour.
This quantity, denoted $\hat{\nu}_\text{eff}$, is an astrometric estimate of the 
effective wavenumber $\nu_\text{eff}$. In order to estimate the pseudocolour, 
it is assumed that the chromatic shift of the image location caused by using the 
wrong colour (that is, the default colour) in IPD is linearly proportional to 
${\nu}_\text{eff}-\nu_\text{eff}^\text{\,def}$. The constant of proportionality 
is a property of the instrument that can be determined in a special AGIS calibration 
solution using sources for which $\nu_\text{eff}$ is known (step~\ref{step:Cprime}
in Sect.~\ref{sec:steps}). 

The sources with six-parameter solutions in EDR3 are identified by the flag  
$\gacs{astrometric\_params\_solved}=95$. The estimated $\hat{\nu}_\text{eff}$ 
(expressed in $\mu$m$^{-1}$) is given as \gacs{pseudocolour}; like the other 
astrometric parameters it comes with a formal uncertainty (\gacs{pseudocolour\_error}) 
and correlation coefficients (\gacs{ra\_pseudocolour\_corr}, etc.). The full $6\times 6$ 
covariance matrix can thus be reconstructed, which makes it possible to compute 
improved estimates of the astrometric parameters if a better estimate of the colour  
than the pseudocolour is available (see Appendix~\ref{sec:col6p}). We note that 
\gacs{nu\_eff\_used\_in\_astrometry} is not given for the sources with six-parameter 
solutions.

Conversely, sources with a standard five-parameter solution 
($\gacs{astrometric\_params\_solved}=31$) have the field
\gacs{nu\_eff\_used\_in\_astrometry} set, but no \gacs{pseudocolour}. 
Neither colour field is set for sources that have only a position in EDR3
($\gacs{astrometric\_params\_solved}=3$). 

\begin{figure}
\centering
  \includegraphics[width=0.9\hsize]{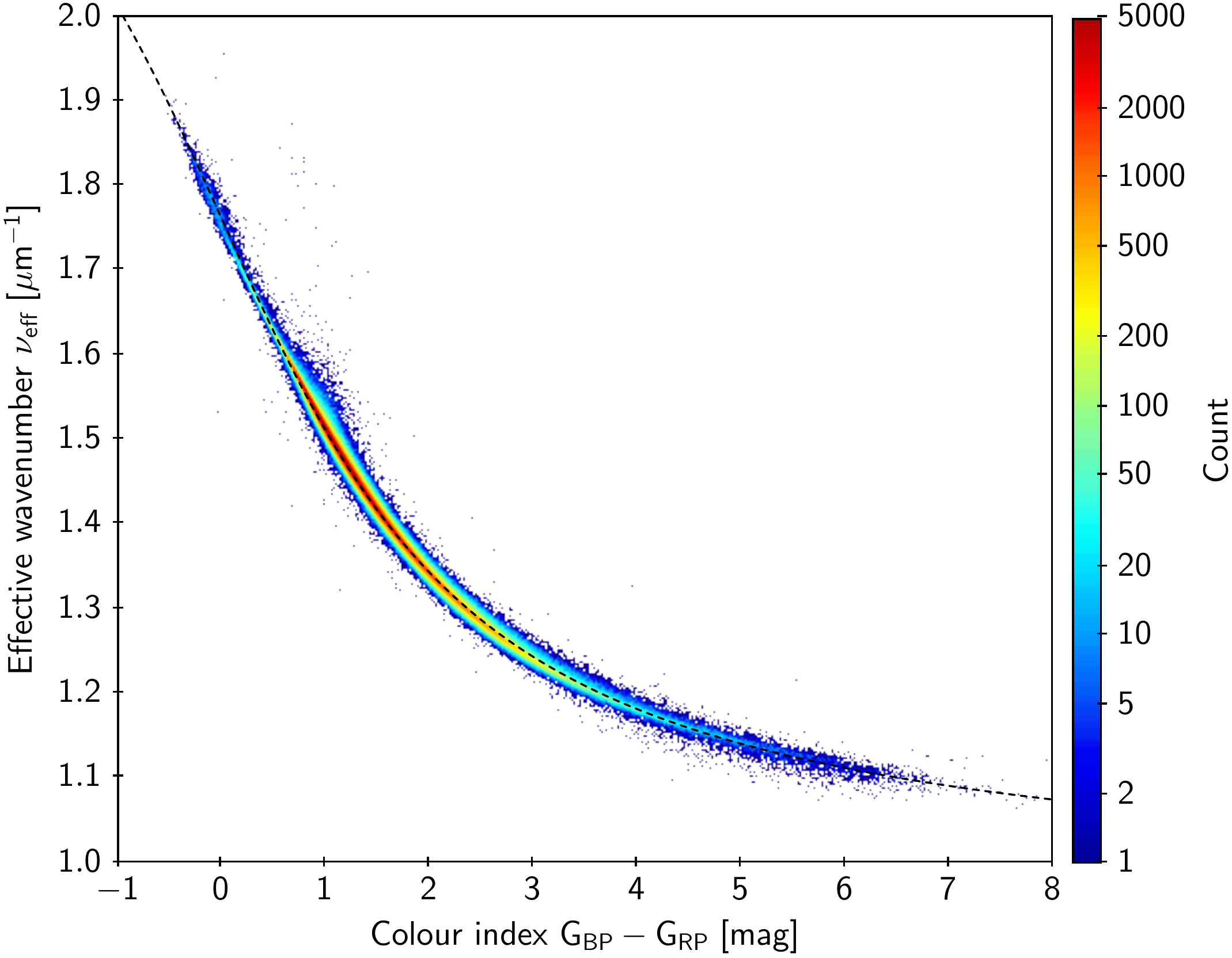}  %{Figures/nuEffVsBpRp1.pdf}
    \caption{Relation between the colour index and effective wavenumber for a
    random sample of 1.5~million sources in EDR3 brighter than $G=18$. The
    dashed curve is the approximate mean relation in Eq.~(\ref{eq:nuEff1}) or (\ref{eq:nuEff2}).}
    \label{fig:nuEffVsBpRp}
\end{figure}

The relation between the colour index  $G_\text{BP}-G_\text{RP}$ (\gacs{bp\_rp})
and effective wavenumber $\nu_\text{eff}$ (\gacs{nu\_eff\_used\_in\_astrometry}) in EDR3
is illustrated in Fig.~\ref{fig:nuEffVsBpRp}. As shown by the diagram, there is no unique 
one-to-one relation between the two colour parameters. One reason is that $\nu_\text{eff}$ 
for cycle~3 was computed in the previous cycle (by PhotPipe~2), and is therefore not 
completely consistent with other photometric data in EDR3, including the colour indices. 
But the main reason for the scatter is the very different methods of computation 
($\nu_\text{eff}$ as a weighted sum over the sampled BP and RP spectra, 
$G_\text{BP}-G_\text{RP}$ from the integrated BP and RP fluxes), which give slightly 
different results depending on the detailed spectra. When an approximate relation
is needed, the following analytical formulae may be useful:
\begin{gather}\label{eq:nuEff1}
\nu_\text{eff} \simeq 1.76 - \frac{1.61}{\pi}\,\text{atan}\,
\Bigl(0.531(G_\text{BP}-G_\text{RP})\Bigr)\quad \mu\text{m}^{-1}\, , \\
G_\text{BP}-G_\text{RP} \simeq \frac{1}{0.531}\,\text{tan}\,
\Biggl(\frac{\pi}{1.61}\,(1.76-\nu_\text{eff})\Biggr)\quad \text{mag} \, . \label{eq:nuEff2}
\end{gather}
For $-0.5\le G_\text{BP}-G_\text{RP}\le 7$ they represent the mean relation for stellar
objects to within $\pm 0.007~\mu$m$^{-1}$ in the effective wavenumber. 
The atan/tan functions conveniently describe the non-linear relation to a useful approximation,
and has the additional advantage that $\nu_\text{eff}$ is restricted to the physically plausible
interval $[0.955,~2.565]~\mu$m$^{-1}$ for arbitrarily large (positive or negative) colour 
indices.

\subsection{Auxiliary data}
\label{sec:aux}

The processing of \textit{Gaia} data aims at producing the most accurate astrometric catalogue 
consistent with the observations, using a minimum of external auxiliary data. Some external data 
are nevertheless needed, for example to align the catalogue with the celestial reference system 
and correct for stellar aberration. The main auxiliary data used in the processing are described below. 

\paragraph{Reference frame} The orientation of the axes of the International Celestial Reference System 
(ICRS) is conventionally defined by means of the accurate positions for extragalactic radio sources observed
by very long baseline interferometry. As of 1~January 2019, the defining list is the third realisation of 
the International Celestial Reference Frame (ICRF3; \citeads{2020A&A...ICRF3}) containing 4588
radio sources. The orientation of \textit{Gaia}-CRF3, the celestial reference frame of \textit{Gaia} EDR3
\citep{EDR3-DPACP-133}, was fixed by means of 2269 ICRF3 S/X sources, for which optical counterparts have 
been identified in EDR3 and which have a valid colour information \gacs{nu\_eff\_used\_in\_astrometry}.
In order to correct a specific problem identified with the bright reference frame of EDR3 
(Sect.~\ref{sec:adHoc}) we also make use of the positional reference frame of \textsc{Hipparcos} at 
epoch J1991.25 as defined by the revised \textsc{Hipparcos} catalogue \citepads{2007ASSL..350.....V}.  

\paragraph{Ephemerides} Accurate barycentric ephemerides of \textit{Gaia} and of all the major bodies 
in the solar system, as well as for some moons and minor planets, are needed in order to interpret the
directions observed by \textit{Gaia} in terms of astrometric parameters defined in the 
barycentric system. The solar system ephemeris used for EDR3 
is the INPOP10e provided by the IMCCE \citepads{2016NSTIM.104.....F}. 
The orbit of \textit{Gaia} was determined at the Mission Operations Centre (MOC)
located at ESOC (Darmstadt, Germany), using conventional Doppler and range tracking 
as well as Delta-Differential One-way Range (Delta-DOR) measurements, the latter
using two tracking stations and calibrated by simultaneous observations of a quasar 
with known position.

The elementary along-scan (AL) astrometric observation is the precise time, 
$t_\text{obs}$, when the centre of a
stellar image crosses the calibrated fiducial line on the CCD. This time is initially given as an on-board
time (OBT), that is the number of nanoseconds counted by the on-board rubidium clock from an arbitrary 
origin, but must be transformed to the coordinate time (TCB) of the event before it can be used in the
astrometric solution. This transformation, known as the time ephemeris, is derived from an analysis of 
time couples (the OBT of a signal generated on board and the reading of the ground-station clocks
when it was received at the ground station), using a sophisticated model that takes into account \textit{Gaia}'s position relative to the Earth, Earth orientation parameters, relativistic effects in the signal propagation,
the influence of the Earth's troposphere, differences between the ground-station clocks and UTC,  
etc.\ \citepads{2017SSRv..212.1423K}. 

\paragraph{Basic-angle corrector} 
The basic angle monitor (BAM) is an interferometric device measuring short-term
($\lesssim 1$~day) variations of the basic angle at $\mu$as precision 
\citepads{2016SPIE.9904E..2DM}. BAM measurements are available since before the start of
nominal operations and throughout the entire period of observations used for EDR3. They
were processed off-line using the same methods as for DR2 
(Sect.~2.4 in \citeads{2018A&A...616A...2L}), resulting in a table of basic-angle
jumps (with the estimated time and amplitude of each jump) and, in between the jumps, 
a continuous function of time represented by a spline. 
The jumps and spline together define the function $\Delta\Gamma(t)$ in Eq.~(\ref{eq:calAL}).

\section{Models}
\label{sec:models}

\subsection{Source model}
\label{sec:source}

The astrometric processing for \textit{Gaia} EDR3 is based on a consistent 
theory of relativistic astronomical reference systems \citepads{2003AJ....126.2687S}. 
The primary coordinate system is the Barycentric Celestial Reference System 
(BCRS) with origin at the solar system barycentre and axes aligned with the 
International Celestial Reference System (ICRS). The time-like coordinate of 
the BCRS is the barycentric coordinate time (TCB). The
\textit{Gaia} relativity model (\citeads{2003AJ....125.1580K}; \citeads{2004PhRvD..69l4001K})
provides a rigorous general-relativistic modelling of astrometric observations.

For the purpose of deriving the main astrometric results in EDR3, it is assumed that
all sources outside of the solar system move with uniform velocity relative to the solar 
system barycentre. Thus, non-linear motions caused by binarity and other perturbations 
are presently ignored, but will be taken into account in future \textit{Gaia} releases.
In the present model, which we refer to as the standard model of stellar motion 
(\citeads{1997ESASP1200.....E}; \citeads{2020A&A...633A...1L}), the motion of the source is 
completely specified by six kinematic
parameters, conventionally taken to be the standard five astrometric parameters
($\alpha$, $\delta$, $\varpi$, $\mu_{\alpha*}{\,=\,}\mu_\alpha\cos\delta$, $\mu_\delta$)
and the radial velocity ($v_r$). All parameters refer to the adopted reference epoch, 
which for the EDR3 astrometry is
$\text{J2016.0}=\text{JD~2457\,389.0 (TCB)}=\text{1 January 2016, 12:00:00}$ (TCB).
This is exactly 0.5~Julian year (182.625~days) later than the reference epoch 
J2015.5 adopted for \textit{Gaia} DR2.

In spite of the well-known fact that a large fraction of the stars in the solar neighbourhood are members
of double and multiple systems, the standard model of stellar motion is very often a good 
model for the observed motions of stars in our Galaxy, and further away, at least over the
relatively short time span covered  by \textit{Gaia}'s observations. In practice, only 
$\sim$10\% of the stars may have proper motions that are noticeably non-linear over 
a few years (cf.\ \citeads{2005ESASP.576...97S}). One reason for this is the extremely
wide range of periods in physical systems, which means that most of them either have
too long periods to show significant curvature over a short time, or they are so close and 
have such short periods that their photocentric wiggles are small and average out over a 
few years. The standard model is also very often an excellent approximation for extragalactic
sources such as active galactic nuclei (AGNs) or quasars. The astrometric solution for 
\textit{Gaia} relies heavily on the lucky circumstance that the motions of most point-like 
sources in the sky can be accurately represented by this simple model. 

The standard model takes into account perspective acceleration through terms depending 
on the radial velocity $v_r$. In \textit{Gaia} DR2 this effect was only considered for some 
50 nearby \textsc{Hipparcos} sources; for EDR3 it is taken into account whenever possible, 
using radial-velocity data from \textit{Gaia}'s radial-velocity spectrometer 
(RVS; \citealt{2018A&A...616A...6S}) as provided in \textit{Gaia} DR2. For a small number of nearby 
stars (mainly white dwarfs), this was complemented with radial velocities from the literature.
Apart from the change in reference epoch and the more frequent use of radial velocity
data, the source model for EDR3 is exactly the same as was used for DR2.

In the standard model, the radial velocity is needed, in addition to the usual five astrometric parameters,
for a complete specification of the six-dimensional phase space vector of a nearby star. 
Because of this, $v_r$ (or $\mu_r= v_r\varpi/A_\text{u}$, where $A_\text{u}$ is the astronomical
unit) is sometimes called the sixth astrometric parameter. This is potentially confusing
in connection with the six-parameter solutions discussed in Sect.~\ref{sec:nuEff} and
elsewhere, where the sixth parameter is the pseudocolour $\hat{\nu}_\text{eff}$, that is 
the astrometrically estimated effective wavenumber (colour) of the source. In contrast
to the pseudocolour, the radial velocity is never estimated from \textit{Gaia} data
in any of the solutions discussed here, although it will be possible in the future for a small 
number of nearby high-velocity stars \citepads{1999A&A...348.1040D}.

\subsection{Attitude model}
\label{sec:attitude}

The attitude model for \textit{Gaia} EDR3 is the same as was used for DR2,
except that AL observations made in window class WC0b (see Sect.~\ref{sec:cal})
were not used for the attitude determination. The attitude model includes a 
pre-computed AL corrective attitude that removes much of the rapid attitude 
irregularities created by micro-clanks and high-frequency thruster noise. 
We refer to Sect.~3.2 of \citetads{2018A&A...616A...2L} for a description
of the DR2 model.

\begin{figure*}
\center
  \includegraphics[width=16cm]{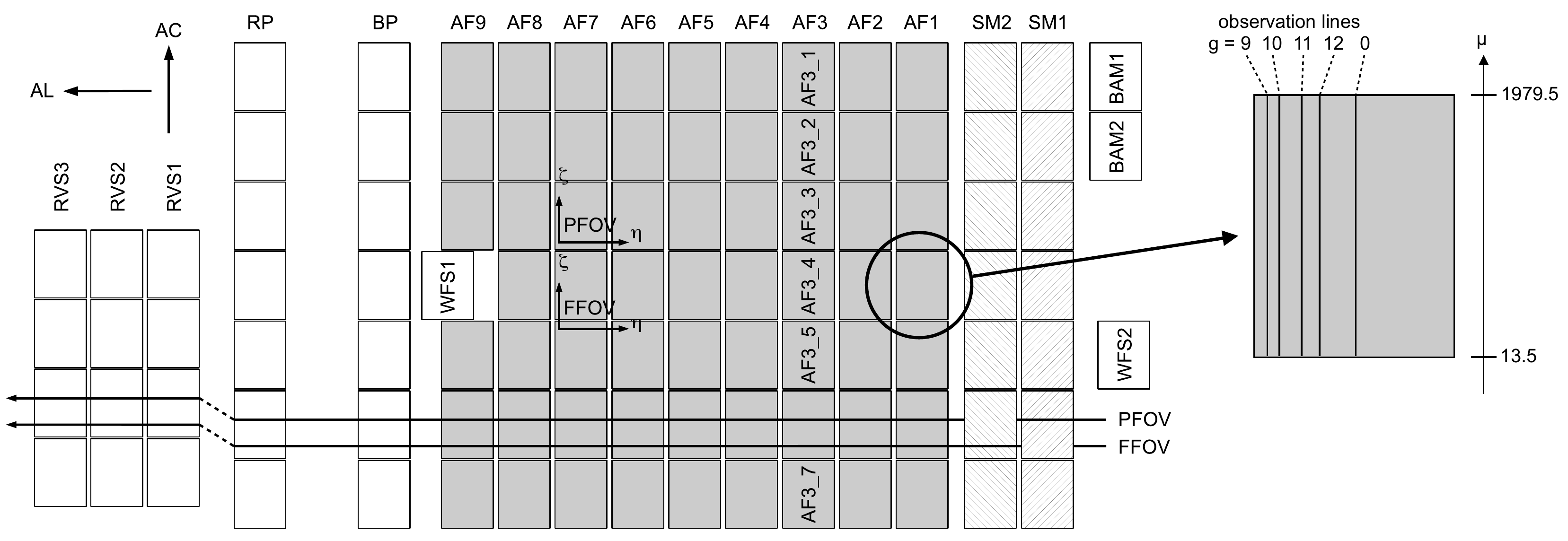}  %{Figures/fpa.pdf}
\caption{Layout of CCDs in \textit{Gaia}'s focal plane. Star images move 
from right to left in the diagram, as indicated in the lower part of the drawing by the 
nominal paths of two images, one in the preceding FoV (PFoV) and one in the 
following FoV (FFoV). The along-scan (AL) and across-scan (AC) directions are 
indicated in the top left corner. To the right, one of the CCDs is shown magnified, with 
the fiducial observation lines indicated for selected gates ($g$). Also indicated is the
AC pixel coordinate $\mu$, running from 13.5 to 1979.5 across the image area of
each CCD. The skymappers (SM1, SM2) provide source image detection and FoV 
discrimination, but their measurements are not used in the astrometric solution. The 
astrometric field (AF1--AF9) provides accurate AL measurements and (for two-dimensional 
windows) AC positions. Other CCDs are used for the blue and red photometers 
(BP, RP), the radial-velocity spectrometer (RVS), wavefront sensing (WFS), and
basic-angle monitoring (BAM). One of the CCD strips (AF3) illustrates the system for 
labelling individual CCDs by strip and row index. The origin of the field angles 
$(\eta,\zeta)$ is at different physical locations on the CCDs in the two fields.
(Adapted from \citeads{2012A&A...538A..78L}.)\label{fig:fpa}}
\end{figure*}

\begin{table*}[t]
\caption{Summary of the astrometric calibration model and number of calibration
parameters in the astrometric solution for \textit{Gaia} EDR3.\label{tab:cal}}
\small
\begin{tabular}{rllcccccccccccccr}
\hline\hline
\noalign{\smallskip}
&& \multicolumn{1}{l}{Basis functions} 
& \multicolumn{13}{c}{---------------------- Multiplicity of dependencies ----------------------} & 
\multicolumn{1}{c}{Number of} \\ 
\multicolumn{2}{l}{Effect ($i$) and brief description} & 
\multicolumn{1}{l}{$K_{lm}(\tilde{\mu},\tilde{t})$} & \multicolumn{1}{c}{$K_{lm}$} 
& \multicolumn{1}{c}{$j$} & $f$ & \multicolumn{1}{c}{$n$} & 
$g$ & $b$ & $w$ & \multicolumn{1}{c}{$\nu_\text{eff}$} & $G$ & $S$ & $\phi$ 
& \multicolumn{1}{c}{$\Delta t$} & $\dot{\zeta}$ &
\multicolumn{1}{c}{parameters} \\
\noalign{\smallskip}
\hline
\noalign{\smallskip}
%1& SM-AL large scale, window class  & $lm=00,01$ & 2 & 310 & 2 & 7 & -- & -- & 2 & -- & -- & -- &  17\,360\\ 
1& AL large-scale geometric & $lm=00,10,20,01$ & 4 & 310 & 2 & 62 & -- & -- & 4 & -- & -- & -- & -- & -- & -- &  615\,040\\ 
2& AL medium-scale gate & $lm=00,10$ & 2 & 19 & 2 & 62 & 8 & 9 & -- & -- & -- & -- & -- & -- & -- & 339\,264 \\ 
3& AL large-scale colour & $lm=00,01$ & 2 & 19 & 2 & 62 & -- & -- & 4 & 1 & -- & -- & -- & -- & -- & 18\,848 \\ 
4& AL large-scale saturation & $lm=00$ & 1 & 19 & 2 & 62 & -- & -- & 2 & -- & -- & 1 & -- & -- & -- & 4\,712 \\ 
5& AL large-scale subpixel & $lm=00$ & 1 & 19 & 2 & 62 & -- & -- & 4 & -- & -- & -- & 2 & -- & -- & 18\,848 \\ 
6& AL large-scale CTI & $lm=00$ & 1 & 19 & 2 & 62 & -- & -- & 4 & -- & -- & -- & -- & 4 & -- &  37\,696\\ 
7& AL large-scale AC rate & $lm=00$ & 1 & 19 & 2 & 62 & -- & -- & 1 & -- & -- & -- & -- & -- & 1 &  2\,356\\ 
\noalign{\smallskip}
\hline
\noalign{\smallskip}
8& AC large-scale geometric & $lm=00,10,20,01$ & 4 & 19 & 2 & 62 & -- & -- & 2 & -- & -- & -- & -- & -- & -- &  18\,848\\ 
9& AC large-scale gate & $lm=00$ & 1 & 19 & 2 & 62 & 8 & -- & -- & -- & -- & -- & -- & -- & -- & 18\,848 \\ 
10& AC large-scale colour & $lm=00$ & 1 & 19 & 2 & 62 & -- & -- & 2 & 1 & -- & -- & -- & -- & -- & 4\,712 \\ 
11& AC large-scale magnitude & $lm=00$ & 1 & 19 & 2 & 62 & -- & -- & 2 & -- & 1 & -- & -- & -- & -- & 4\,712 \\ 
12& AC large-scale saturation & $lm=00$ & 1 & 19 & 2 & 62 & -- & -- & 2 & -- & -- & 1 & -- & -- & -- & 4\,712 \\ 
\noalign{\smallskip}
\hline
\end{tabular}
\tablefoot{The column Basis functions lists the combinations of indices $l$ and $m$
used to model variations with AC coordinate on a CCD ($\tilde{\mu}$) and with time within 
a time granule ($\tilde{t}$). Multiplicity of dependencies gives the number of distinct 
functions or values for each dependency, or a dash if there is no dependency: 
basis functions ($K_{lm}$, Eq.~\ref{eq:Klm}), granule index ($j$), 
field index ($f$), CCD index ($n$), gate ($g$), stitch block ($b$), window class ($w$), 
effective wavenumber ($\nu_\text{eff}$), magnitude ($G$), saturation ($S$), 
subpixel phase ($\phi$), time since last charge
injection ($\Delta t$), and AC scan rate ($\dot{\zeta}$). The last column 
is the product of multiplicities, equal to the number of calibration parameters for the effect.
The SMs are not considered here.}
\end{table*}

\subsection{Calibration model}
\label{sec:cal}

The astrometric calibration model for \textit{Gaia} EDR3 is similar to the one
used for DR2, as described in Sect.~3.3 of \citetads{2018A&A...616A...2L}, but
with additional dependencies described below. 
The general principles of the calibration model are described in Sect.~3.4 of 
\citetads{2012A&A...538A..78L}, and only a few basic concepts are recalled here.
At any time, the attitude represents a solid-body rotation from the celestial reference
system to \textit{Gaia}'s scanning reference system (SRS), nominally fixed with respect
to the CCDs as viewed through the two FoVs (preceding and following).
Within a FoV, directions with respect to the SRS are usually expressed by means of the
field angles $(\eta,\zeta)$, with origin at the nominal centre of the FoV (Fig.~\ref{fig:fpa}). 
According to the scanning law, stellar images traverse the FoV in the direction of decreasing 
$\eta$ (at the AL rate of approximately 60~arcsec~s$^{-1}$) and at approximately constant 
$\zeta$ (the AC rate is at most $\pm 0.18$~arcsec~s$^{-1}$). The fundamental AL measurement 
used for the astrometry is the precise time when an image transits across a fiducial 
`observation line' line fixed to the CCD (Fig.~\ref{fig:fpa}, right). The astrometric calibration 
of the instrument (as opposed to the LSF and PSF calibrations by CAL) is essentially a specification 
of the location of the observation line in field angles, that is of the functions $\eta(\mu)$ and 
$\zeta(\mu)$, where $\mu$ is the AC pixel coordinate.  

More precisely, the AL and AC calibration functions are written as the 
sums of the nominal calibrations and several `effects' that describe the dependence on 
various quantities, such as time, CCD, and FoV (see Eqs.~\ref{eq:calAL} and \ref{eq:calAC}). 
The effects used in the EDR3 calibration model in the AF
are listed in Table~\ref{tab:cal}.
The skymappers (SM1 and SM2 in Fig.~\ref{fig:fpa}) obtain a similar, but simpler, 
calibration. However, the SM observations are not at all used in the astrometric solution, 
and their calibration is not discussed in this paper.

\begin{figure}
\centering
  \includegraphics[width=8cm]{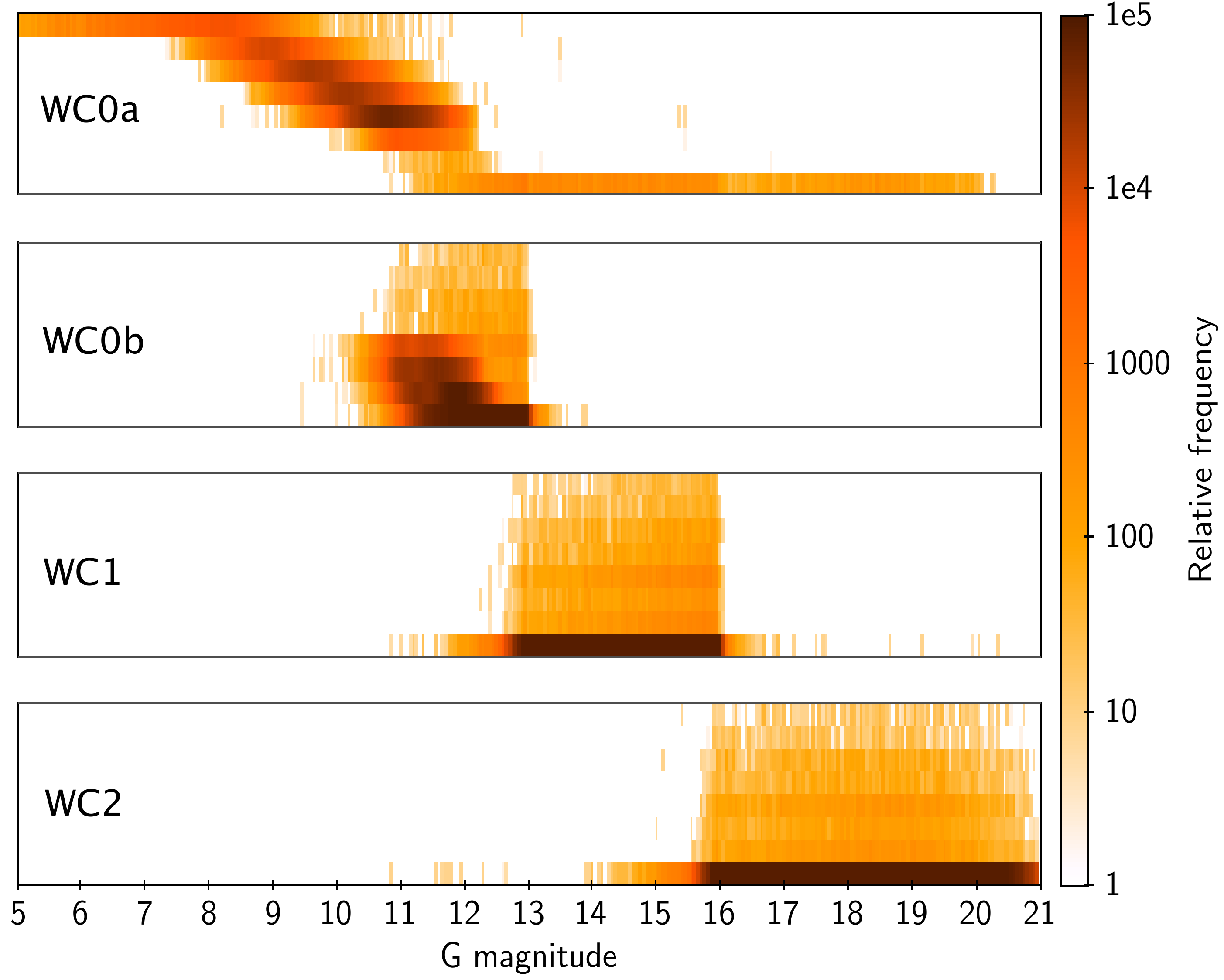} %{Figures/freqPerGclassGateAnn.pdf}
  \caption{Relative frequency of observations in the various
  combinations of window class and gate, as a function of magnitude.
  The four blocks represent the four window classes (WC); within each WC
  the eight stripes represent (from top to bottom) gate number 4, 7, 8, 9, 10, 
  11, 12, and 0. The graph was constructed from a random 1\% sample of
  the AF observations of the primary sources. The faint sources observed in 
  WC0a at gate~0 are the Calibration Faint Stars, a small fraction of faint 
  observations receiving full-pixel resolution windows for calibration purposes
  \citepads{2016A&A...595A...1G}.}
  \label{fig:obsFreqVsWcGate}
\end{figure}

Compared with the corresponding table for the 
DR2 model (Table~2 in \citeads{2018A&A...616A...2L}), Table~\ref{tab:cal} contains effects with
several new dependencies ($S$, $\phi$, $\Delta t$, $\dot{\zeta}$). Their introduction
in the model was motivated by systematic trends seen in the residuals from preliminary 
solutions, in which the calibration model did not include the effects. 
The complete set of dependencies is as follows.
\begin{itemize}
\item
AC pixel coordinate $\mu$ on the CCD, which is a continuous value running
from 13.5 to 1979.5 across the AC extent of the CCD image area (Fig.~\ref{fig:fpa}).
The offset by 13~pixels allows for the presence of pre-scan pixel data.
\item
Time $t$, divided into granules such that $t_j\le t<t_{j+1}$ in the granule indexed 
by $j$. Two different time axes are used, with 310 and 19 granules spanning the length 
of the data; the typical duration of the granules is, respectively, about 3~d and 63~d. 
\item 
FoV index $f$, specifying preceding or following FoV. We use the
convention $f=+1$ in the PFoV and $f=-1$ in the FFoV.
\item
CCD $n$, with 62 different values in the astrometric field.
\item
Gate $g$, taking eight different values with $g=0$ for ungated observations
(Fig.~\ref{fig:obsFreqVsWcGate}). The number of active TDI lines is 4500 for $g=0$,
2900 ($g=12$), 2048 ($g=11$), 1024 ($g=10$), 512 ($g=9$), 256 ($g=8$), 
128 ($g=7$), and 16 ($g=4$). Gates 1--3, 5, and 6 are not used in normal operations.
\item
Stitch block $b$, with nine different values in the AC direction of a CCD. $b$ is uniquely
defined by the AC pixel coordinate through $b=\lfloor (\mu+128.5)/250\rfloor$,
where $\lfloor\,\rfloor$ is the floor function.
\item
Window class $w$, with four values. In the DR2 calibration model, three window classes 
WC0, WC1, and WC2 were used, approximately corresponding to magnitude ranges
$G\lesssim 13$, $13\lesssim G\lesssim 16$, and $16\lesssim G$, respectively. (The
WC represents the CCD sampling scheme chosen at detection time, depending 
mainly on the real-time estimate of the magnitude derived from the SM observation, 
but also on several other factors. There is consequently no strict relation between the 
mean calibrated $G$ magnitude, given in the catalogue, and the WC.) 
In the EDR3 model, WC0 was further 
subdivided into WC0a  (for $G\lesssim 11$) and WC0b (for $11\lesssim G\lesssim 13$),
see Fig.~\ref{fig:obsFreqVsWcGate}.
\item
Effective wavenumber $\nu_\text{eff}$ is the photon-weighted inverse wavelength, 
calculated from the BP and RP spectra in the photometric processing \citep{EDR3-DPACP-118} 
and expressed in $\mu\text{m}^{-1}$. The cyclic processing scheme 
adopted by DPAC implies that the
$\nu_\text{eff}$ used for the EDR3 astrometry was generated in the preceding cycle,
corresponding to DR2 photometry, and is sometimes missing or inconsistent with the 
EDR3 photometry. The actual values used for the astrometry (and IDU pre-processing)
is given in the \textit{Gaia} Archive as \gacs{nu\_eff\_used\_in\_astrometry}. For sources
without a reliable $\nu_\text{eff}$ a special calibration was employed (step~\ref{step:Cprime}
in Sect.~\ref{sec:steps}).   
\item
Magnitude $G$: Like the effective wavenumber, the magnitude used in the astrometric
processing was derived from DR2, but since the differences are generally small and only the
(less critical) AC calibration depends on $G$, the actual value used is not given in the 
Archive.  
\item
Saturation $S$: This is a flag produced by the IPD as part of the IDU pre-processing. 
It is set to 1 if the raw observed sample exceeds 
a pre-defined conservative threshold, as determined from early mission data, for the 
CCD column and sample binning; otherwise $S=0$. 
The astrometric effects of the saturation are only calibrated for WC0a and WC0b.
\item
Subpixel phase $\phi$: This is $2\pi$ times the fractional part of the precise observation 
time $t$, as determined by the IPD and expressed in TDI periods of on-board time. (The TDI 
period is the time it takes to shift the charges on the CCDs by one pixel AL, or approximately
0.982~ms.) Inaccuracies in the LSF and PSF calibrations used for the IPD may result in
systematic AL errors that are periodic functions of $\phi$.
\item
Time since the last charge injection $\Delta t$: To minimise the effects of charge transfer
inefficiency (CTI) in the CCDs, charge injections are made at regular time intervals
of 2000~TDI periods. CTI may cause systematic AL shifts of the image centroids, which
increase with $\Delta t$.    
\item
Across-scan (AC) rate $\dot{\zeta}$: The nominal scanning law of \textit{Gaia} 
\citepads{2016A&A...595A...1G} produces a quasi-periodic ($\simeq 6$~h period) variation 
of the AC rate, with an amplitude of approximately 173~mas~s$^{-1}$. Imperfections 
in the PSF modelling may result in systematic AL errors that depend on the AC rate.
This dependence is only calibrated for observations in WC0b using gates 11, 12, and 0
(that is, for a CCD exposure time of about 2.0, 2.8, or 4.4~s).
\end{itemize}
Within a time granule, the variation with $t$ and $\mu$ is modelled as a 
linear combination of basis functions
\begin{equation}\label{eq:Klm}
K_{lm}(\tilde{\mu},\tilde{t})=\tilde{P}_l(\tilde{\mu})\tilde{P}_m(\tilde{t}\,) \, ,
\end{equation}
where $\tilde{P}_l(x)$, $\tilde{P}_m(x)$ are the shifted Legendre polynomials%
\footnote{The shifted Legendre polynomials $\tilde{P}_n(x)$ are related to the 
(ordinary) Legendre polynomials $P_n(x)$ by $\tilde{P}_n(x)=P_n(2x-1)$.
Specifically, $\tilde{P}_0(x)=1$, $\tilde{P}_1(x)=2x-1$, and 
$\tilde{P}_2(x)=6x^2-6x+1$. The shifted Legendre polynomials are orthogonal 
on $0\le x\le 1$.\label{fn2}}
of degree $l$ and $m$,
$\tilde{\mu}=(\mu-13.5)/1966$ is the normalised AC pixel coordinate, and 
$\tilde{t}=(t-t_j)/(t_{j+1}-t_j)$ the normalised time within granule $j$.
The third and fourth columns in Table~\ref{tab:cal} list the combination of indices 
$l$ and $m$ used for a particular effect and the number of basis functions 
$K_{lm}$. Most of the effects only
use $lm=00$, meaning that the effect is modelled as constant with $t$ and $\mu$
for a given combination of the other indices and variables.

Each combination of indices $l$, $m$, $j$, $f$, $n$, $g$, $b$, and $w$ indicated in 
Table~\ref{tab:cal} is a `calibration unit' and receives an independent calibration.
Within a calibration unit, effect $i$ is a linear combination of products 
$\smash{K_{lm}(\tilde{\mu},\tilde{t})\,\Psi^{(i)}_k(x)}$, where $K_{lm}$ describes the dependence
on $t$ and $\mu$ according to Eq.~(\ref{eq:Klm}), and $\smash{\Psi^{(i)}_k(x)}$ ($k=0,~1,~\dots$) 
describe the dependence on some other variable $x$, which could be
$\nu_\text{eff}$, $G$, $S$, $\phi$, $\Delta t$, or $\dot{\zeta}$. The relevant
functions are:
\begin{gather}
\label{eq:PhiNuEff}
\Psi^{(3,\,10)}_0(\nu_\text{eff})=\nu_\text{eff}-1.43~\mu\text{m}^{-1} \, , \\
\label{eq:PhiG}
\Psi^{(11)}_0(G)=G-12.6\, , \\
\label{eq:PhiS}
\Psi^{(4,\,12)}_0(S)=S\, , \\ 
\label{eq:PhiPhi}
\Psi^{(5)}_0(\phi)=\cos\phi\, ,\quad \Psi^{(5)}_1(\phi)=\sin\phi\, , \\
\label{eq:PhiDeltaT}
\Psi^{(6)}_k(\Delta t) = a_k - \exp(-\Delta t/\tau_k) \, ,\quad k=0\dots 3\, , \\
\label{eq:PhiDotZeta}
\Psi^{(7)}_0(\dot{\zeta}) = |\,\dot{\zeta}\,|^2\, .
\end{gather}
The function $\Psi^{(11)}_0(G)$ in Eq.~(\ref{eq:PhiG}) is only used for ungated 
AC observations; otherwise, it is set to 1.
Equation~(\ref{eq:PhiPhi}) describes a periodic variation with subpixel phase $\phi$.
In Eq.~(\ref{eq:PhiDeltaT}), the variation with time since the last charge injection 
$\Delta t$ is assumed to be a linear combination of four exponentials,
with e-folding times $\tau_k=10$, 100, 500, and 2000~TDI periods. The constants 
$a_k=(\tau_k/2000)[1-\exp(-2000/\tau_k)]$ are such that the mean
value of $\smash{\Psi^{(6)}_k(\Delta t)}$ over $0\le\Delta t\le 2000$ is zero. 
This means that the mean displacement of the images caused by the CTI is 
not taken out by this calibration, only its variation with $\Delta t$. The resulting 
calibration parameters are thus mainly interesting as diagnostics of the effect 
(Fig.~\ref{fig:cti}). The quadratic dependence on the AC rate $\dot{\zeta}$ 
in Eq.~(\ref{eq:PhiDotZeta}) models a possible bias caused by the AC smearing of 
the PSF; this effect was not well modelled in the PSF calibration for EDR3 
(cf.\ Sect.~\ref{sec:acRate} and Appendix~\ref{sec:plxAc}). 
Formally, $\Psi^{(1,2,8,9)}=1$ for the effects that only depend on $t$ and $\mu$.

The complete AL calibration model is 
\begin{equation}\label{eq:calAL}
\begin{split}
\eta(\mu,t,\nu_\text{eff},\dots) &= \eta_{ng}^{(0)}(\mu) + \frac{f}{2}\Delta\Gamma(t)\\
&+ \sum_{i=1}^7 \sum_{l,\,m,\,k} \Delta\eta^{(i)}_{lmk} K_{lm}\,\Psi^{(i)}_k + \delta\eta(f,t,\eta,\zeta) \, ,
\end{split}
\end{equation}
where the first term is the nominal location of the fiducial observation line for CCD $n$,
gate $g$ 
(Eq.~14 in \citeads{2012A&A...538A..78L}); the second contains the basic angle correction
$\Delta\Gamma(t)$ derived from BAM data (Sect.~\ref{sec:aux}); the third is the sum of the 
seven effects $i$ in the upper part of Table~\ref{tab:cal}, with calibration parameters 
$\Delta\eta^{(i)}_{lmk}$; and the last term is the spin-related distortion model fitted as 
global parameters (Sect.~\ref{sec:glob}). 
For brevity, the dependences on $f$, $n$, $g$, $b$, and $w$ and the arguments of 
$K_{lm}$ and $\smash{\Psi^{(i)}_k}$ have been suppressed.
This gives a total of 1\,036\,764 AL parameters (not counting the spin-related distortion
parameters), which is more than three times as many as used for the DR2 calibration model
(see Table~1 in \citeads{2018A&A...616A...2L}). Besides the longer time interval 
covered by the data, this reflects the more complex modelling made necessary (and
possible) thanks to the generally improved quality of the input data and resulting
solution. Several new effects have been introduced (saturation, subpixel, CTI, and 
AC rate), and the AL large-scale geometric calibration now depends also on the 
window class. Some of these effects should eventually be taken out by the LSF and PSF 
calibrations, but that was not yet possible in the present cycle.

The AC calibration model is 
\begin{equation}\label{eq:calAC}
\begin{split}
\zeta(\mu,t,\nu_\text{eff},\dots) = \zeta_{fng}^{(0)}(\mu) 
&+ \sum_{i=8}^{12} \sum_{l,\,m,\,k} \Delta\zeta^{(i)}_{lmk} K_{lm}\,\Psi^{(i)}_k \\
\Bigl[~&+~\delta\zeta(f,t,\eta,\zeta)~\Bigr] \, ,
\end{split}
\end{equation}
where $\zeta^{(0)}_{fng}(\mu)$ is the nominal calibration and $\Delta\zeta^{(i)}_{lmk}$ 
the calibration parameter for the five effects in the bottom part of Table~\ref{tab:cal}.
The last term is the spin-related distortion in AC; but as explained in Sect.~\ref{sec:glob}
this is not used for EDR3, and is therefore put within square brackets in Eq.~(\ref{eq:calAC}).  
This gives a total of 51\,832 AC calibration parameters, which is 10\% smaller than 
in DR2, in spite of the longer time period covered. The main reason for this is that the 
dependences on time and AC coordinate were found to be overly complicated in DR2 
and have been simplified.

Selected results of the astrometric calibration are given in Appendix~\ref{sec:calData}.

\subsection{Spin-related distortion model}
\label{sec:glob}

Thermo-mechanical perturbations of the instrument over time
scales close to and below the rotational period of 6~h
present a special problem for the AGIS calibrations. Such variations
will be called `quick' below. It is known
that it is impossible to fully calibrate the quick variations of the
instrument \citepads{2017A&A...603A..45B}. The reason for this is 
a degeneracy between the source parameters, attitude parameters, 
and calibration parameters for the quick variations of the instrument. 
This is known as the VBAC degeneracy (Velocity error and
effective Basic Angle Calibration).%
\footnote{The mention of velocity here may seem surprising. For an astrometric 
satellite such as \textit{Gaia}, the observational effects of a 
small error in the translational velocity, as used in the modelling of stellar aberration, 
is found to be indistinguishable from a certain combination of errors in the attitude and in the 
basic angle. This aspect of the data processing plays no role for EDR3, where velocity 
is taken to be known (Sect.~\ref{sec:aux}).} 
A special case of this is the well-known degeneracy between the global parallax zero 
point and a specific form of basic-angle and attitude variations \citepads{2017A&A...603A..45B},
but the VBAC degeneracy is much more general: Any time-dependent distortion of 
the celestial positions is observationally indistinguishable from some specific 
combination of attitude errors and quick instrument variations. In the design 
considerations for \textit{Gaia} it was indeed a fundamental requirement that 
the instrument must either be extremely stable on time scales shorter than a few times 
the spin period, or have the means to monitor the variations continuously to a very high 
precision.

However, the VBAC degeneracy does not imply that arbitrary quick instrument variations
are degenerate with the source and attitude parameters. On the contrary, most such variations 
are not degenerate, and therefore in principle possible to calibrate from the astrometric 
observations themselves, that is, without the need for special procedures or metrology 
devices like the BAM. Over the last decade, 
a considerable effort has been put into investigating how, and to what extent, quick 
variations of the \textit{Gaia} instrument can be calibrated from the astrometric observations.
The spin-related distortion model presented here is a limited version of more general 
models that may be used for future \textit{Gaia} releases. It nevertheless represents 
a significant advance over the DR2 model (Sect.~3.4 in \citeads{2018A&A...616A...2L}).
Although the distortion model logically belongs to the instrument calibration model,
it is fitted as part of the global block in AGIS for purely implementation-technical reasons. 

\subsubsection{General model}
\label{sec:vbacfoc}

We begin by formulating the spin-related distortion model in its most general form.
At any moment of time an arbitrary distortion of the AL and AC
field angles $(\eta,\zeta)$ for a source in a given FoV $f$
can be represented as a two-dimensional expansion over a family of
orthogonal functions $\Phi_{lm}(\eta,\zeta)$:
\begin{eqnarray}\label{expansion-general-eta}
\delta\eta(f,t,\eta,\zeta)
=\sum\limits_{l\ge0}\sum\limits_{m\ge0}\;\delta\eta_f^{lm}(t)\;\Phi_{lm}(\eta,\zeta)\,,\\
\label{expansion-general-zeta}
\delta\zeta(f,t,\eta,\zeta)
=\sum\limits_{l\ge0}\sum\limits_{m\ge0}\;\delta\zeta_f^{lm}(t)\;\Phi_{lm}(\eta,\zeta)\,.
\end{eqnarray}
\noindent
For the nominally rectangular AF of \textit{Gaia}, a convenient set of orthogonal 
functions are the products of the Legendre polynomials $P_k(x)$ for the AL and AC coordinates: 
\begin{equation}\label{Phi-lm-Gaia}
\Phi_{lm}(\eta,\zeta)=P_l(\tilde\eta)\,P_m(\tilde\zeta)\,,
\end{equation}
where $\tilde\eta$ and $\tilde\zeta$ are the field angles $\eta$
and $\zeta$ linearly scaled to the interval $[-1,1]$. 
(Although not apparent in these equations, the scaling of $\zeta$ is actually different in the
two FoVs, owing to the offset of the AC origins indicated in Fig.~\ref{fig:fpa}.)
%The Zernike polynomials, well-known from optics (e.g.\ \citeads{1999poet.book.....B}), 
%are not useful here as they are orthogonal on a circular area. 

We use Eqs.~(\ref{expansion-general-eta}) and (\ref{expansion-general-zeta}) to model the
spin-related distortion terms in Eqs.~(\ref{eq:calAL}) and (\ref{eq:calAC}). Their time-variations
are defined by the functions $\smash{\delta\eta_f^{lm}(t)}$ and $\smash{\delta\zeta_f^{lm}(t)}$
to be specified below. 

Full-scale simulations of the AGIS solution have shown that the terms in Eqs.~(\ref{expansion-general-eta}) 
and (\ref{expansion-general-zeta}) with $l+m\ge1$ are not degenerate with the source parameters and attitude, 
and can therefore safely be determined from the observations. For example, simulated variations 
in which $\smash{\delta\eta_f^{lm}(t)}$ and $\smash{\delta\zeta_f^{lm}(t)}$ for $1\le l+m\le 5$ were 
represented as B-splines, with a knot interval of 10~min and random coefficients, could be completely 
recovered with no rank-deficiency and only a moderate slowdown of the convergence of the 
iterative solution. The terms with $l=m=0$, on the other hand, involve the VBAC degeneracy,
and the time variation of these terms need to be chosen with care in order not to jeopardise 
the source parameters. It is therefore natural to split the further specification of the general 
spin-related distortion model in two parts, corresponding to VBAC (for $l=m=0$) and 
FOC (for $l+m\ge 1$). Here, FOC stands for Focal length and Optical distortion Calibration, 
since the calibration for $l+m\ge 1$ obviously covers also a variation of the focal length 
of the instrument.

The general model in Eqs.~(\ref{expansion-general-eta}) and (\ref{expansion-general-zeta})
can also describe slow variations of the instrument and is in principle degenerate with certain 
parts of the calibration model described in Sect.~\ref{sec:cal}. As this degeneracy does 
not involve the source parameters it is harmless for the astrometry, but since it could slow down 
the convergence of the AGIS iterations it should nevertheless be avoided if practically feasible. 
One such case is the near-degeneracy mentioned below between effect~7 in Table~\ref{tab:cal} 
and the Fourier terms of order $p=2$ in Eqs.~(\ref{eq:calVF3b}) and (\ref{eq:calVF3c}).

%FOC
\subsubsection{FOC}
\label{sec:foc}

Because the FOC calibration ($l+m\ge 1$) has no degeneracy with the source and attitude 
parameters, we are quite free to choose the maximum degree $l+m$ of the basis functions
$\Phi_{lm}(\eta,\zeta)$ and the parametrisation of the time dependency of the functions
$\smash{\delta\eta_f^{lm}(t)}$ and $\smash{\delta\zeta_f^{lm}(t)}$. Of course, it is always
necessary to limit the degree and time resolution so that the number of parameters is reasonable
in relation to the number of observations, keeping the overall solution numerically and 
practically tractable.
For example, because the AF has nine strips of CCDs (see Fig.~\ref{fig:fpa}),
it is not numerically feasible to have the AL degree $l>9$. 
Numerous test solutions using EDR3 data were made to explore some 
of the many possible options, and the configuration finally adopted for AGIS~3.2 is in some 
sense the best one found in the limited time available. 

One conclusion from the test solutions was that 
the FOC correction in AC does not bring any improvement at this stage and it is therefore not
used for EDR3; hence the bracketed term in Eq.~(\ref{eq:calAC}). It was also found that the
polynomials in Eq.~(\ref{expansion-general-eta}) can be restricted to $1\le l+m\le 3$, which
gives 18 coefficients $\smash{\delta\eta_f^{lm}(t)}$ to be considered (nine per FoV). Similarly to the AL
calibration model in Sect.~\ref{sec:cal}, it was found necessary to determine FOC separately
for each window class. For WC0a, WC1, and WC2, the ten coefficients with $1\le l+m\le 2$ 
were determined as cubic splines with a knot interval of 20~min. The remaining eight 
coefficients with $l+m=3$ were fitted as Fourier polynomials
\begin{equation}\label{eq:calVF3b}
\delta\eta_f^{lm}(t) = d(t)^{-2} \sum_{p=1}^8 \Bigl(c_{f\!plm}\cos p\Omega(t) 
+ s_{f\!plm}\sin p\Omega(t)\Bigr)\, ,
\end{equation}
where $d(t)$ is the Sun--\textit{Gaia} distance in au, $\Omega(t)$ is
the heliotropic spin phase \citepads{2018A&A...616A...2L}, and 
$c_{f\!plm}$ and $s_{f\!plm}$ are the free parameters fitted to the data.
The scaling by $d(t)^{-2}$ here and in the following equations accounts for the 
variation in solar irradiance.
For WC0b, all 18 coefficients $\smash{\eta_f^{lm}(t)}$ were fitted as
Fourier polynomials, as in Eq.~(\ref{eq:calVF3b}), but omitting the terms with
$p=2$ to avoid the near-degeneracy with effect~7 in Table~\ref{tab:cal}.
Furthermore, all Fourier polynomials for FOC were fitted independently for
the two time intervals before and after OBMT~4513~rev.  That moment of
time (one of the gaps in Table~\ref{tab:gaps}) was found to be a boundary 
between slightly different behaviours of the residuals in test solutions; 
ultimately, this behaviour can be traced back to a particular change of 
LSF and PSF models in CALIPD 3.1 at that moment.
The resulting FOC model has a total of 2\,033\,184 parameters.

%VBAC
\subsubsection{VBAC}
\label{sec:vbac}

The terms in Eqs.~(\ref{expansion-general-eta}) and (\ref{expansion-general-zeta}) with 
$l=m=0$ represent the distortion averaged over each FoV. It is readily seen that this is 
equivalent to a combination of four time dependent variations, namely,
(i) of the AL attitude by 
$\smash{\frac{1}{2}\bigl(\delta\eta_{+1}^{0,0}+\delta\eta_{-1}^{0,0}\bigr)}$;
(ii) of the AC attitude in the PFoV by
$\smash{\delta\zeta_{+1}^{0,0}}$;
(iii) of the AC attitude in the FFoV by
$\smash{\delta\zeta_{-1}^{0,0}}$; and
(iv) of the basic angle by
$\smash{\delta\Gamma=\delta\eta_{+1}^{0,0}-\delta\eta_{-1}^{0,0}}$.
The flexibility of the attitude modelling means that the first three variations are completely
degenerate with the attitude determination, and should not be further considered in the 
VBAC model. Therefore, the only variation to consider for $l=m=0$ is a time-dependent 
basic angle variation, $\delta\Gamma(t)$. 

$\delta\Gamma(t)$ can be regarded as an additive correction to the basic angle variation 
$\Delta\Gamma(t)$ in Eq.~(\ref{eq:calAL}) that comes from the analysis of BAM data 
(Sect. \ref{sec:aux}). 
It should be recalled that $\Delta\Gamma(t)$ includes both basic angle jumps (due to
sudden structural changes in the optics) and a smooth representation of the basic angle 
variations between jumps, including a very good approximation of the quick variations.
However, because the CCDs for the BAM are located outside of the AF
(Fig.~\ref{fig:fpa}), we cannot assume that the variations measured with the BAM are 
fully representative for the whole FoV -- indeed, in the presence of FOC distortion this
is not to be expected. Moreover, the BAM device itself may be subject to perturbations that
are not relevant for the astrometric observations. For these reasons it is highly desirable
to calibrate as much as possible of the basic angle variations directly from the astrometric
observations, which can be done with VBAC. Owing to the VBAC degeneracy there are nevertheless 
components of the basic angle variations that cannot be determined from the observations,
and the BAM signal remains indispensable as the only handle we may have on those components.

For EDR3, the same representation of $\delta\Gamma(t)$ was 
used as for DR2 (Eq.~10 in \citeads{2018A&A...616A...2L}), but split in two parts, 
\begin{equation}\label{eq:deltaGamma-split}
\delta\Gamma(t)=\delta\Gamma_\text{A}(t)+\delta\Gamma_\text{B}(t)\,,
\end{equation}
with
\begin{multline}\label{eq:calVF3c}
\delta\Gamma_\text{A}(t) = d(t)^{-2} 
\sum_{p=2}^8\, \Bigl[\delta C_{p,0}+(t-t_\text{ep})\delta C_{p,1}\Bigr]\cos p\Omega(t) \\
+ d(t)^{-2} 
\sum_{p=1}^8\, \Bigl[\delta S_{p,0}+(t-t_\text{ep})\delta S_{p,1}\Bigr]\sin p\Omega(t) \,  
\end{multline}
and
\begin{equation}\label{eq:deltaGamma-C10}
\delta\Gamma_\text{B}(t) = d(t)^{-2} \Bigl[\delta C_{1,0}+(t-t_\text{ep})\delta C_{1,1}\Bigr]\cos\Omega(t)\,.
\end{equation}
Here $d(t)$ and $\Omega(t)$ have the same meaning as in Eq.~(\ref{eq:calVF3b}),
$t_\text{ep}=\text{J}2016.0$, and $\delta C_{p,q}$, $\delta S_{p,q}$ are the constant 
coefficients determined from the data. The split in Eq.~(\ref{eq:deltaGamma-split}) 
is motivated by the near-degeneracy of $\delta C_{1,0}$ with a global parallax shift
(\citeads{2017A&A...603A..45B}; \citeads{2018A&A...616A...2L}),
which necessitates a special treatment of this term; this is deferred till Sect.~\ref{sec:C10}.
The parameter $\delta C_{1,1}$ is included in $\delta\Gamma_\text{B}(t)$ only because it
naturally belongs together with $\delta C_{1,0}$; it is not strongly correlated with other
parameters and could instead have been fitted with the other VBAC parameters in 
Eq.~(\ref{eq:calVF3c}).

The representation of $\delta\Gamma_\text{A}(t)$ in
Eq.~(\ref{eq:calVF3c}) contains 30 parameters describing linear
variations of the $d(t)^{-2}$-scaled Fourier coefficients in
$\Omega(t)$. Analysis of the test AGIS solutions and their residuals has shown
that the effective basic angle variations obtained with this model are substantially 
different for the different window classes. A separate set of Fourier coefficients was
therefore fitted for each window class.
Moreover, similarly as for the Fourier coefficients
in the FOC model, separate fits were made for the time intervals before and after 
OBMT~4513~rev, and the coefficients for $p=2$ were omitted for WC0b.
The resulting model for $\delta\Gamma_\text{A}(t)$ has a total of 232 parameters.

% C10
\subsubsection{Treatment of the near-degeneracy with parallax ($\delta C_{1,0}$)}
\label{sec:C10}

Here we consider the VBAC correction $\delta\Gamma_\text{B}(t)$ in 
Eq.~(\ref{eq:deltaGamma-C10}), containing the two additional parameters
$\delta C_{1,0}$ and $\delta C_{1,1}$. Unlike the parameters in $\delta\Gamma_\text{A}(t)$,
which were fitted per window class and separately before and after OBMT~4513~rev, 
there is only a single set of these two parameters. They are fitted using all data except WC0b.

As already mentioned, the parameter $\delta C_{1,0}$ cannot be easily fitted in an
iterative solution like AGIS because it is highly correlated to a global shift of all parallaxes
\citepads{2017A&A...603A..45B}. However, this also means that if the correction $\Delta\Gamma(t)$ 
to the basic angle derived from BAM data has an error described by $\delta C_{1,0}$, there will be
a global shift of the parallaxes. Owing to the profound scientific importance of the parallax
zero point, every effort should be made to avoid such an error. To this end a
method has been developed to calibrate $\delta C_{1,0}$ directly from the astrometric 
observations of \textit{Gaia}. The method was thoroughly tested in a series of detailed
end-to-end simulations of the iterative solution, which demonstrated the feasibility of the 
method and probed the limits of its applicability. It was tested with cycle~2 data (but 
not used in the solution for \textit{Gaia} DR2), and finally employed in the primary astrometric 
solution for EDR3. Full details of the method will be published elsewhere; here we describe 
only its most important elements.

The possibility to fit $\delta C_{1,0}$ 
is based on the small but not completely negligible differences between
heliocentric and barycentric quantities, and the fact that the solar irradiance and parallax
factor scale differently with the varying distance from the Sun or solar system barycentre.
Based on physical considerations, the model for the basic angle variations in 
Eqs.~(\ref{eq:calVF3c}) and (\ref{eq:deltaGamma-C10}) scales as $d(t)^{-2}$ and is
periodic in $\Omega(t)$, where $d(t)$ and $\Omega(t)$ are heliocentric, that is, reckoned 
with respect to the Sun. On the other hand, the AL parallax effect depends on the 
corresponding barycentric quantities $d_\text{b}(t)$ and $\Omega_\text{b}(t)$ measured
relative to the solar system barycentre. More precisely, 
the ability to determine absolute parallaxes depends on the AL parallax 
factor being different in the two FoVs (see Fig.~2 in \citeads{2016A&A...595A...1G}). 
Relevant for the parallax zero point is therefore the 
differential AL parallax factor $d_\text{b}(t)\sin\xi_\text{b}(t)\cos\Omega_\text{b}(t)$
(see footnote~\ref{fn:Omega} in Appendix~\ref{sec:plxAc}), where $\xi_\text{b}(t)$ 
is the angle between \textit{Gaia}'s spin axis and the direction to the barycentre.
Thus, in our model for $\delta\Gamma(t)$, only the term proportional to 
$\cos\Omega(t)$, that is the one containing $\delta C_{1,0}$, has a strong 
correlation with the parallax zero point.
The differences between the heliocentric and barycentric 
quantities, of the order of 0.01~au and 0.01~rad, and the annual variations in
$d$ and $d_\text{b}$, by about $\pm 1.7$\%, all contribute 
towards a decorrelation of the parallax zero point from $\delta C_{1,0}$. 

For the actual cycle~3 data, the correlation coefficient between $\delta C_{1,0}$ 
and the parallax zero point is $\simeq\,$0.99992. Such a high degree of correlation 
(collinearity) in a least-squares estimation problem would normally be considered 
crippling, but it need not be so if the number of observations is very high, which it is 
in this case, and the modelling is sufficiently accurate, which we strive for. 
If all the $\sim$85~million unknowns in the primary astrometric solution could be 
obtained by direct solution of the normal equations, a valid solution for $\delta C_{1,0}$ 
would be obtained because the full normal equations take 
into account the correlations among all parameters. However, we are forced to 
use iterative solution methods, and it turns out that the introduction of $\delta C_{1,0}$ 
in AGIS effectively prevents the convergence of the block-iterative solution in its
original form. The non-convergence is however not caused by the strong correlation 
itself, but by the circumstance that the correlated parameters are in different blocks.
In AGIS the different blocks of source, attitude, calibration, and global parameters are 
treated as independent least-squares problems in a given iteration, thus ignoring 
correlations between, for example, the global block (containing $\delta C_{1,0}$)
and the source blocks (containing the parallaxes) when updates for the next iteration 
are computed.

We nevertheless found a way to obtain a converged solution including $\delta C_{1,0}$, by
using a special option in the global block of AGIS, called `consider parameters'. This device
was originally introduced for a different purpose,%
\footnote{The term `consider parameter' has various meanings in
the literature. Here we refer to a parameter that is included in the
estimation of updates to the current parameter values, but for which
no actual update is applied to the parameter. The consider parameter thus remains
at its original value (in this case zero), but the solution computes updates to the
other parameters, and uncertainties and correlations among all parameters,
exactly as if the consider parameter had been included in the fit. As the name suggests,
consider parameters are intended to help the researcher decide whether a particular
signal, modelled by the consider parameters, exists in the data, and how the covariance
of the solution would be affected if they were included in the fit.}
but here it is used to allow the AGIS iterations to converge in a reasonable time.
This particular use of consider parameters has been thoroughly tested in simulations,
and we are therefore confident in its fundamental correctness. Briefly, here is how it
works. In the global block, we introduce three more unknowns (consider parameters) 
that are strongly correlated with $\delta C_{1,0}$, namely one additive constant to all 
parallaxes, and two parameters for certain variations of the attitude (see Eq.~15 in 
\citeads{2017A&A...603A..45B}). All three consider parameters are fully degenerate with 
the parallaxes and the ordinary attitude parameters, but because their updates are ignored
in each iteration they remain at their initial zero values and do not affect the computation
of the right-hand side of the observation equations (the residuals). Their inclusion in
the left-hand side of the global block does however modify the updates to the regular
global parameters, including $\delta C_{1,0}$, and this is what allows the iterations to
converge. Because the three consider parameters remain at zero, it does not matter that
they are degenerate with other parameters, and the solution, after convergence, must be 
the same as a solution without them -- if such a solution could be obtained by some 
different algorithm. 
The role of the consider parameters in the block-iterative primary AGIS solution 
can formally be understood as a modification of the preconditioner of the adjustment 
scheme (e.g.\ \citealt{doi:10.1137/1.9780898718003}; \citeads{2012A&A...538A..77B}).

The condition number of the normal matrix for the fit of $\delta C_{1,0}$ 
is about $10^5$, so its inversion using normal 64-bit arithmetic is 
quite accurate. Although somewhat delicate, the fit works in practice and 
delivers a reasonably stable value of $\delta C_{1,0}$ after a number of 
AGIS iterations. The formal uncertainty of $\delta C_{1,0}$ from the fit 
is about 1~$\mu$as.
However, the fragile character of the fitting of $\delta C_{1,0}$ necessitates 
certain precautions: (i) $\delta C_{1,0}$ and the consider parameters 
should only be introduced at the very last stage of the AGIS iterations 
(see Table~\ref{tab:iterations}); (ii) only AL data should be used in the fit; 
(iii) in EDR3, the observations in window class WC0b have larger systematics
and were therefore omitted from the fit. 

In the primary AGIS solution for EDR3, $\delta C_{1,0}$ shifted the parallax zero 
point by about $+20~\mu$as compared with the same solution without $\delta C_{1,0}$, 
and by about $+10~\mu$as compared with DR2. The global parallax zero point of 
EDR3 is about $-17$~$\mu$as \citep{EDR3-DPACP-132}.
Although the inclusion of $\delta C_{1,0}$ in the global model for EDR3 did not bring
the global parallax zero point to zero, the partial success of the method is very 
encouraging and fosters the hope that the zero point issue can be resolved, at the level
of a few $\mu$as, in future releases that will benefit from much improved calibration models.

\section{Astrometric solutions}
\label{sec:agis}

\subsection{Main steps of the solutions}
\label{sec:steps}

The tasks labelled AGIS~3.1 and AGIS~3.2 in Fig.~\ref{fig:timeline} each consists of
several steps, the most important ones being:
\begin{enumerate} 
\item
Preprocess the input data (transits) from IDU: This includes filtering (removing transits
that are unmatched or of poor quality according to IPD flags, or outside the specified
time interval) and sorting the transits by position. Sorting uses the healpix index
\citepads{2005ApJ...622..759G} encoded in the \gacs{source\_id}.
\item
Select a set of primary sources to ensure a sufficient density of well-behaved sources with
a good coverage in magnitude and colours.
\item
Fit an initial attitude for the required time interval, using source parameters from a 
previous cycle or phase; also define data gaps where transits are missing or of poor quality.
\item
Generate the corrective attitude from rate data as described in Sect.~3.2 of 
\citetads{2018A&A...616A...2L}.
\item\label{step:prim}
Calculate a primary solution by simultaneously estimating source (S), attitude (A),
calibration (C), and global (G) parameters in an iterative least-squares solution involving
only the primary sources. See Sect.~\ref{sec:prim} for a brief explanation.
\item\label{step:Cprime}
Compute a separate set of calibration parameters (C$'$) for sources where
IPD used the default colour $\nu_\text{eff}^\text{\,def}=1.43~\mu$m$^{-1}$. This
calibration is based on a subset of the primary sources where image parameters 
were determined by IPD using both the actual colours and the default value.
\item
Calculate secondary solutions for all sources (Sect.~\ref{sec:sec}). 
The computation is equivalent to the S block in step~\ref{step:prim},
except that sources with default colour obtain six-parameter solutions
using calibration C$'$. In this step the acceptance criteria detailed in Sect.~\ref{sec:accept}
are checked and, if necessary, a fallback solution computed. 
\item
Postprocess the results: This includes calculating various statistics such as the
renormalised unit weight error (RUWE).
\item\label{step:CoMaRa}
Regenerate attitude and calibration data for use by downstream processes such as 
PhotPipe. This fills some time gaps and intervals (including the EPSL) that were excluded 
for the astrometry, but where the observations may still be useful for other processes.
The skymapper (SM) geometry is also calibrated at this point. Although the SM observations 
are not used in the astrometric solution, they are needed in downstream processes.
\item
Regenerate attitude and astrometric calibration data for the LSF and PSF calibrations in the  
CALIPD of the next processing cycle or phase. This uses the same calibration model as for 
the primary solution, but including only the purely geometric part of the model, that is the 
effects numbered 1, 2, 8, and 9 in Table~\ref{tab:cal}. This is known as the `NoCoMaRa' 
calibration: no dependency on colour, magnitude, or rate (as opposed to the normal,
`CoMaRa', calibration including all the effects). The rationale for this is that all 
dependencies on colour, magnitude, AC rate, saturation, subpixel phase, and CTI
effects should ultimately be accounted for by the LSF and PSF calibrations, so that 
AGIS can be a purely geometric solution. This goal will never be reached if the AGIS 
calibration used for the LSF and PSF calibrations already removes (part of) the dependencies. 
By using NoCoMaRa for the LSF and PSF calibrations, the latter processes see the full extent 
of the dependencies. (In principle the attitude generated in steps~\ref{step:prim} and
\ref{step:CoMaRa} is already purely geometric, but owing to the non-orthogonality 
of some CoMaRa and NoCoMaRa effects, the best-fitting geometric attitude is slightly 
different for the two calibrations.) The NoCoMaRa calibration and attitude are not used 
by any downstream processes, only by IDU.
\item
Export all results to the main database, making them available to other processes.
\end{enumerate}
The same steps were executed in AGIS~3.1 and 3.2, but with many differences
in the details. In particular, the selection of primary sources and the calibration models
were different in the two phases, and numerous improvements and bug fixes were 
implemented in between. The models described in Sect.~\ref{sec:models}, and all other 
details given hereafter, refer to AGIS~3.2.

\subsection{Primary solution for AGIS~3.2}
\label{sec:prim}

Although all steps listed in the previous section are needed for a successful 
astrometric solution, the primary solution (step~\ref{step:prim}) is by far the most 
important and difficult one. As described elsewhere
\citepads{2012A&A...538A..78L}, the primary solution iteratively updates the four
kinds of unknowns (source, attitude, calibration, and global parameters). The algorithm
can be described in terms of four separate blocks, designated S, A, C, and G.
In S the astrometric parameters of the primary sources are updated based on current 
values for the other unknowns; in A the attitude parameters are updated based on
current source, calibration, and global parameters; and so forth. The blocks are
normally executed in a cyclic manner, for example S-A-C-G-S-A-C-G-S-A-$\dots$, 
where S-A-C-G constitutes one iteration. (For specific purposes, some of the blocks 
may be left out, meaning that the corresponding unknowns are kept fixed.)
In the simple iteration (SI) algorithm, there is no memory of the updates in previous
iterations that can be used to optimise the next update; this algorithm 
reliably converges in all relevant cases
and is numerically very stable, but may require many iterations for complete convergence. 
The conjugate gradient (CG) algorithm \citepads{2012A&A...538A..77B} speeds up 
convergence considerably, but is less stable and (unlike SI) does not allow observation 
weights to be changed from one iteration to the next. Weight adjustment is necessary 
for a good treatment of outliers and for estimating the excess noise. Most often
AGIS employs a hybrid scheme consisting of three SI iterations during which the weights 
are adjusted, followed by three CG iterations with fixed weights; this sequence is 
then repeated as many times as required. A complete run typically ends with a 
sequence of simple iterations, confirming that the solution is sufficiently converged.

The primary solution for AGIS~3.2 processed about 6.5~billion ($6.5\times 10^9$) 
CCD observations for 14.3~million primary sources. The solution determined 
71.5~million source parameters together with 10.7~million attitude parameters, 
1.1~million calibration parameters, and 2.0~million global parameters; the redundancy 
factor (mean number of observations per unknown) is $\simeq\,$76.
(This does not count the corrective attitude, for which a largely different set of
observations was used to estimate some 17~million parameters.)

The sequence of iterations executed in the primary solution for AGIS~3.2 is 
detailed in Table~\ref{tab:iterations}. After a warm-up run to obtain good starting
values, a total of 165 iterations were made using the global model described in 
Sect.~\ref{sec:glob}. In the last 39 iterations the global parameter $\delta C_{1,0}$
was also adjusted, resulting in a significant reduction of the (negative)
parallax bias (Sect.~\ref{sec:C10}).

\begin{table}
\caption{Iteration sequences for the primary solutions of AGIS~3.2.  
\label{tab:iterations}}
\small
\begin{tabular}{lllll}
\hline\hline
\noalign{\smallskip}
Iterations & Blocks active & Algorithm & Remark\\ % & Snapshots for validation \\ 
\noalign{\smallskip}
\hline
\noalign{\smallskip}
(1--27) & SAC & hybrid & warm up run\\ % & SO1 (I0) \\
1--89 & SACG & hybrid & $\delta C_{1,0}=0$ \\ % & SO4--SO9 (I36, I39, I51, I62, I71, I83) \\
90--126 & SACG & simple & $\delta C_{1,0}=0$ \\ %& SO9--SO11 (I95, I96, I97) \\
127--141 & SACG & hybrid & $\delta C_{1,0}$ free \\ %& SO12 (I141) \\
142--165 & SAG & simple & $\delta C_{1,0}$ free, ad hoc corr. \\ %& SO14--SO16 (I147, I156, I165)\\ 
166--181 & C & simple & step~6 (calibration C$'$) \\ 
\noalign{\smallskip}
\hline
\end{tabular}
\tablefoot{Columns~2 and 3 describe the AGIS configuration for the sequence of
iterations in the first column (see text).
$\delta C_{1,0}$ is the global parameter discussed in Sect.~\ref{sec:C10}. 
%SO1 to SO16 are extracts of the solution at selected iterations, used for validation and
%monitoring purposes (Sect.~\ref{sec:val}). 
Iterations~1--165 correspond to step~5 in Sect.~\ref{sec:steps}.
Between iterations 141 and 142 the ad hoc correction to the calibration parameters 
for the bright sources (Sect.~\ref{sec:adHoc}) was applied, after which the calibration
was not updated. Iterations 166--181 correspond to step~6, computing the special 
calibration C$'$ needed for sources with default colour in IPD.
}
\end{table}

\subsection{Secondary solutions}
\label{sec:sec}

In this step the astrometric parameters were computed for all sources, including the primary sources.
Depending on the colour information used in the IPD for a particular source, it received a 
five- or six-parameter solution as described in Sect.~\ref{sec:nuEff}, but otherwise the treatment
was identical. The five-parameter solutions used calibration C obtained in step~5 of 
Sect.~\ref{sec:steps}, while the six-parameter solutions used calibration C$'$ obtained in step~6. 
For sources with an insufficient number of observations, or where the astrometric results failed to meet
the acceptance criteria for a five- or six-parameter solution (Sect.~\ref{sec:accept}), only 
the mean position at the reference epoch (J2016.0) is published. 

The secondary solutions processed nearly 78~billion FoV transits, generating converged
solutions for 2.495~billion sources (of which 585~million five-parameter, 883~million six-parameter, 
and 1027~million two-parameter solutions). Subsequently some of the five- and six-parameter solutions 
and most of the two-parameters solutions were removed because they failed to meet the acceptance 
criteria (Sect.~\ref{sec:accept}). The final number of sources and other statistics are given in 
Sect.~\ref{sec:results}.

\subsection{Acceptance criteria and fallback (two-parameter) solutions}
\label{sec:accept}  

The decision whether a converged, non-duplicated secondary solution is accepted as a 
five- or six-parameter solution, or at all retained for publication, depends on the four
quantities $N_\text{tr, astr}$, $N_\text{vpu}$, $\sigma_\text{pos, max}$, and $\sigma_\text{5d, max}$
calculated in the course of the source update process. Here,
$N_\text{tr, astr}$ is the number of FoV transits (detections) used in the AGIS solution; in
the \textit{Gaia} Archive it is given as \gacs{astrometric\_matched\_transits}.
$N_\text{vpu}$ (\gacs{visibility\_periods\_used}) is the number of distinct observation epochs 
(visibility periods) used in the solution, where a visibility period is a group of observations 
separated from other groups by a gap of at least four days.
$\sigma_\text{pos, max}$ (not in the \textit{Gaia} Archive) is the semi-major axis of the 
error ellipse in position at the reference epoch J2016.0 (Eq.~B1 in \citeads{2018A&A...616A...2L}).
Finally, $\sigma_\text{5d, max}$ (\gacs{astrometric\_sigma5d\_max}) is the five-dimensional 
equivalent to $\sigma_\text{pos, max}$, calculated as described in Sect.~4.3 of 
\citetads{2018A&A...616A...2L} but with $T=2.76383$~yr for the time coverage of the data 
used in the solution and ignoring the pseudocolour for six-parameter solutions.

For every source, a solution with five or six parameters (depending on the colour information used in
IPD) was first tried. This was accepted if it converged and satisfied the
criterion
\begin{equation}\label{eq:critP56}
G\le 21.0~~~  \& ~~~ N_\text{vpu}\ge 9 ~~~
\& ~~~ \sigma_\text{5d, max}<(1.2~\text{mas})\times\gamma(G)\, ,
\end{equation}
where $\gamma(G)=10^{0.2\max(6-G,~0,~G-18)}$.
This is similar to the DR2 criterion (Eq.~11 in \citeads{2018A&A...616A...2L}), except that the minimum
$N_\text{vpu}$ is higher and the upper limit on $\sigma_\text{5d, max}$ was increased for $G<6$ 
to accommodate the sharply rising uncertainty for the brightest sources (Fig.~\ref{fig:uncVsG}).
The present threshold on $N_\text{vpu}$ removes most cases where a lower threshold might produce 
spurious solutions, like the ones found in DR2 with very large (positive or negative) parallaxes.
We note that the $G$ used in Eq.~(\ref{eq:critP56}) is not the EDR3 value, which was unavailable 
at the time, but the value from DR2, or the real-time magnitude estimate from the on-board 
object detection if the source was not in DR2. In EDR3 there are 143\,546 sources with 
five- or six-parameter solutions and EDR3 magnitude $G > 21$, and conversely some sources with 
two-parameter solutions that would have passed Eq.~(\ref{eq:critP56}) if the EDR3 magnitude
had been used. (Elsewhere in this paper $G$ stands for the EDR3 value \gacs{phot\_g\_mean\_mag}.)

If the five- or six-parameter solution did not converge, 
or failed to satisfy Eq.~(\ref{eq:critP56}), prior information
on the parallax and proper motion was added, based on the Galactic model described in 
\citetads{2015A&A...583A..68M} and Sect.~4.3 of \citetads{2018A&A...616A...2L}. In such
cases only the position parameters ($\alpha$, $\delta$) at epoch J2016.0 and their covariances 
were retained out of the full
five- or six-parameter solution. As explained in \citetads{2015A&A...583A..68M}, the purpose of 
the Galactic prior is to provide more realistic uncertaintainties for the positions of sources with
a very small number of observations, by making some reasonable assumption about the
sizes of their parallaxes and proper motions. The resulting position is called a two-parameter 
solution ($\gacs{astrometric\_params\_solved}=3$), although 
in reality all five or six parameters are estimated. A two-parameter solution was accepted for 
publication if it satisfied the criterion
\begin{equation}\label{eq:crit}
N_\text{tr, astr}\ge 5 \quad \& \quad \sigma_\text{pos, max}<100~\text{mas} \, .
\end{equation}
It can be noted that any solution that satisfies Eq.~(\ref{eq:critP56}) also satisfies Eq.~(\ref{eq:crit}), 
which therefore holds for all sources in EDR3. In contrast to the corresponding criterion for 
\textit{Gaia} DR2 (Eq.~12 in \citeads{2018A&A...616A...2L}), Eq.~(\ref{eq:crit}) puts no upper limit 
on the \gacs{astrometric\_excess\_noise}, because it was found that such a limit rejects many 
partially resolved binaries that should be retained in the catalogue for completeness,
even though they do not have full astrometric data.

Finally, all sources, irrespective of the kind of solution, must be solitary in the sense that
there is no other source within a radius of 0.18~arcsec, as calculated from the position 
parameters ($\alpha$, $\delta$) at the reference epoch. If multiple sources are found at 
smaller separations, only one source is kept, namely, in order of precedence:
(i) a source previously identified as relevant for the extragalactic reference frame; 
(ii) the five- or six-parameter solution with the smallest $\sigma_\text{5d, max}$; or 
(iii) the two-parameter solution with the smallest $\sigma_\text{5d, max}$. 
In such cases the retained source has the flag \gacs{duplicated\_source} set in the
\textit{Gaia} Archive.

\subsection{Ad hoc correction of WC0 calibration}
\label{sec:adHoc}   

Between iterations 141 and 142 an ad hoc correction was applied to the WC0
calibration parameters in order to mitigate a known problem with the bright 
($G\lesssim 13$) reference frame. The motivation and procedure for this 
correction, which should not be needed in future processing cycles, are briefly
as follows.

During the internal validation of the AGIS~2.2 solutions, carried out by the
astrometry team prior to the publication of \textit{Gaia} DR2, it was found that
the reference frame of the bright sources ($G\lesssim 12$--13) in DR2 was 
rotating, relative to the frame defined by the fainter quasars, at a rate of 
about 0.15~mas~yr$^{-1}$ (Sect.~5.1 of \citeads{2018A&A...616A...2L}).
The problem was confirmed by \citetads{2018ApJS..239...31B} in a 
comparison with proper motions calculated from the position differences 
between the DR2 and \textsc{Hipparcos} catalogues, and by
\citetads{2020A&A...633A...1L,2020A&A...637C...5L} in a comparison
with radio-interferometric (VLBI) observations of bright radio stars.
The likely cause of the effect is explained in Appendix~B 
of \citetads{2020A&A...633A...1L}.

A similar effect was seen during the production of AGIS~3.1. A major concern
then was that these systematics, if left uncorrected in AGIS~3.1, would propagate 
into the time-dependent LSF and PSF calibrations of WC0 sources in CALIPD~3.2,
only to appear again in AGIS~3.2. It was therefore decided to implement an 
ad hoc correction to the calibration parameters of WC0 in AGIS~3.1,
counteracting the effect. This procedure successfully fixed the bright reference 
frame in AGIS~3.1, but the problem nevertheless reappeared in AGIS~3.2,
albeit with different values. 
A similar ad hoc correction was therefore made after iteration 141 in the 
AGIS~3.2 iteration sequence (Table~\ref{tab:iterations}). Because the calibration
was not updated in the subsequent iterations, the correction remained effective
in the final results.

To explain the correction it is useful to consider how the AL astrometric measurements 
are affected by a change in the source positions corresponding to a small error in the 
celestial reference frame. The orientation error at a certain time is given by the 
(numerically small) rotation vector $\vec{\varepsilon}$, such that the change in the 
unit vector $\vec{u}$ towards a source is $\Delta\vec{u}=\vec{\varepsilon}\times\vec{u}$. 
Let $\vec{z}$ be the unit vector, at the same instant, along the nominal spin axis of 
\textit{Gaia}  (more precisely, $\vec{z}$ is the third axis of the scanning reference 
system SRS; e.g.\ Fig.~2 in \citeads{2012A&A...538A..78L}).
$\vec{z}$ and $\vec{u}$ must be nearly orthogonal for the source to be observed 
in one of the FoVs, and for simplicity we assume $\vec{z}'\vec{u}=0$. The 
tangent vector of the AL field angle $\eta$ at the source is then the unit vector 
$\vec{z}\times\vec{u}$, and the component of $\Delta\vec{u}$ in the AL direction is 
$\Delta\eta=(\vec{z}\times\vec{u})'\Delta\vec{u} = 
\vec{z}'\vec{\varepsilon}-(\vec{z}'\vec{u})(\vec{u}'\vec{\varepsilon})=\vec{z}'\vec{\varepsilon}$.
Both $\vec{\varepsilon}$ and $\vec{z}$ are functions of time, thus
$\Delta\eta(t) = \vec{z}(t)'\vec{\varepsilon}(t)$. Here $\vec{z}(t)$ is set by the 
scanning law, while the standard model of stellar motion (Sect.~\ref{sec:source}) 
requires that the frame orientation error is a linear function of time, 
$\vec{\varepsilon}(t)=\vec{\varepsilon}(t_\text{ep})+(t-t_\text{ep})\vec{\omega}$.
The function $\Delta\eta(t)$ therefore has six degrees of freedom corresponding to the 
components of the vectors $\vec{\varepsilon}(t_\text{ep})$ and $\vec{\omega}$. The
important conclusion from this brief discussion is that only very specific forms of
time-dependent AL displacements in the calibration model could be mistaken for  
a reference frame error. 

In the astrometric calibration
model (Sect.~\ref{sec:cal}), the AL large-scale calibration for WC1 ($G\simeq 13$ to 16) has 
a fixed origin, when averaged over both FoVs, but for WC0 and WC2 it is necessary
to permit time-dependent displacements relative to WC1. This means that each WC could 
in principle have its own reference frame, namely if the relative displacement between their
calibrations can be described in the form of the function $\Delta\eta(t)$ introduced above
for some vectors $\vec{\varepsilon}(t_\text{ep})$ and $\vec{\omega}$. 
In practice this should not be a problem, because many primary 
sources around magnitude 13 and 16 are not always observed in the same WC, and they
will only obtain consistent solutions if the reference frame is the same in all WC. This
mechanism apparently works as expected for the transition between WC1 and WC2 around 
$G=16$, but not for the transition between WC1 and WC0 around $G=13$. The probable
reason for this is the generally problematic calibrations in WC0, both in CALIPD and AGIS.

The ad hoc correction amounts to adding the time-dependent correction 
$\Delta\eta(t)=(t-t_\text{ep})\vec{z}(t)'\vec{\omega}$ to the WC0a and WC0b calibrations, where 
$\vec{\omega}=[-0.0166,~-0.0950,~+0.0283]'$~mas~yr$^{-1}$ was estimated from 
a comparison of the proper motions of \textsc{Hipparcos} stars, as obtained in iteration 
141 from the \textit{Gaia} observations, and as derived from the positional differences 
between \textit{Gaia} DR2 and the \textsc{Hipparcos} catalogue \citepads{2007ASSL..350.....V}.
For lack of better information, it was necessary to assume that the positional
systems of the different window classes agreed at the reference epoch, in other words that
$\vec{\varepsilon}(t_\text{ep})=\vec{0}$. In effect, the applied correction implies that the 
bright reference frame of EDR3, when extrapolated to the \textsc{Hipparcos} epoch 
J1991.25, agrees with the \textsc{Hipparcos} reference frame. The uncertainty in the 
alignment of the \textsc{Hipparcos} reference frame to the ICRS at epoch J1991.25 
was $\pm 0.6$~mas in each axis \citepads{1997A&A...323..620K}, which gives a systematic 
uncertainty of at least $(0.6~\text{mas})/(24.75~\text{yr})\simeq 0.024$~mas~yr$^{-1}$ 
per axis in the spin of the bright reference frame of \textit{Gaia}~EDR3.

It is important to note that the ad~hoc correction does not adjust the proper motions individually
for agreement with the \textsc{Hipparcos} positions (as was done for the \textit{Gaia} DR1 TGAS 
solution; \citeads{2016A&A...595A...2G}); only the reference frame is adjusted via the WC0
calibration parameters. Nevertheless, resorting to this procedure is very unsatisfactory
and hopefully exceptional: Improved calibrations in CALIPD and AGIS for the WC0 observations 
should eliminate the need for it in future releases. However, it highlights the need for 
independent means to verify the consistency of the \textit{Gaia} reference frame over the full 
range of magnitudes, for example by means of VLBI observations of radio stars 
\citepads{2020A&A...633A...1L}.

\begin{figure}
\center
  \resizebox{0.9\hsize}{!}{\includegraphics{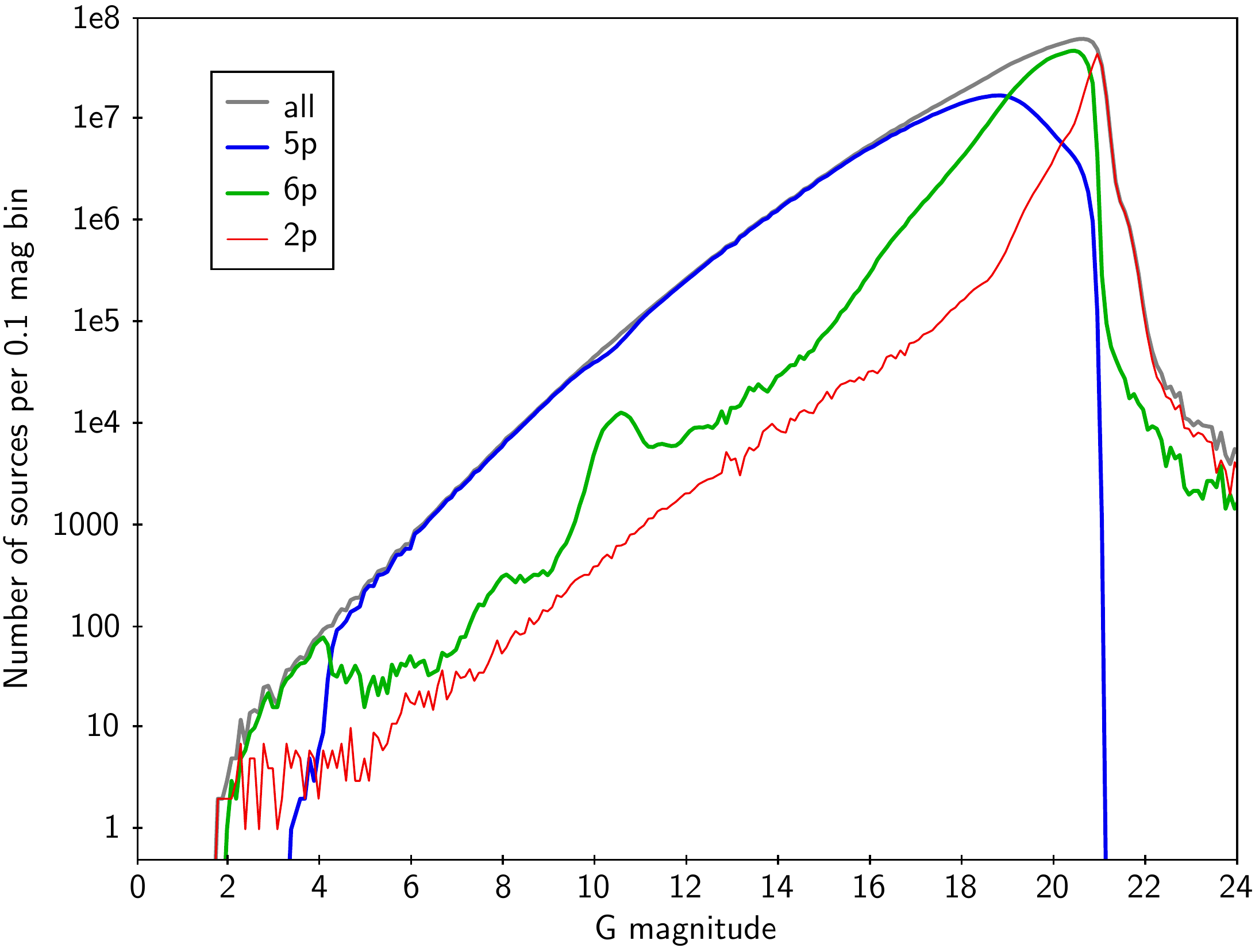}} %{Figures/histNsrc.pdf}}
    \caption{Magnitude distribution of sources in \textit{Gaia} EDR3. 
	The grey denotes all sources, the blue denotes sources with five-parameter solutions, 
	the green denotes sources with six-parameter solutions, and the red denotes sources 
	with two-parameter solutions.}
    \label{fig:histNsrc}
\end{figure}

\section{Results: Astrometric properties of EDR3}
\label{sec:results}

\subsection{Overview of the data}
\label{sec:overviewData}

%The main table of \textit{Gaia} EDR3, \gacs{gaia\_source}, gives
%astrometric data for more than 1.8~billion sources. The exact numbers
%with five-, six-, and two-parameter solutions are:
%\begin{itemize}
%\item 5p: \phantom{1\,}585\,416\,709\quad ($\gacs{astrom\_params\_solved}=31$)
%\item 6p: \phantom{1\,}882\,328\,109\quad ($\gacs{astrom\_params\_solved}=95$)
%\item 2p: \phantom{1\,}343\,964\,953\quad ($\gacs{astrom\_params\_solved}=3$)
%\item all: 1\,811\,709\,771
%\end{itemize}
The main table of \textit{Gaia} EDR3, \gacs{gaia\_source}, gives
astrometric data for more than 1.8~billion sources. The exact numbers are
585\,416\,709 sources with five-parameter solutions ($\gacs{astrom\_params\_solved}=31$),
882\,328\,109 with six-parameter solutions ($\gacs{astrom\_params\_solved}=95$), and
343\,964\,953 with two-parameter solutions ($\gacs{astrom\_params\_solved}=3$).
In total there are 1\,811\,709\,771 sources. 
Their distribution in $G$ magnitude (\gacs{photometric\_g\_mean\_mag})
is shown in Fig.~\ref{fig:histNsrc}.

In the following we give statistics related to the quantities 
listed below with their brief explanations.
\begin{itemize}
\item \gacs{ra\_error} = standard uncertainty in right ascension at epoch J2016.0, 
$\sigma_{\alpha*}=\sigma_\alpha\cos\delta$
\item  \gacs{dec\_error} = standard uncertainty in declination at epoch J2016.0, 
$\sigma_{\delta}$
\item  \gacs{parallax\_error} = standard uncertainty in parallax, 
$\sigma_{\varpi}$
\item \gacs{pmra\_error} = standard uncertainty of proper motion in right ascension, 
$\sigma_{\mu\alpha*}=\sigma_{\mu\alpha}\cos\delta$
\item \gacs{pmdec\_error} = standard uncertainty of proper motion in declination, 
$\sigma_{\mu\delta}$
\item \gacs{pseudocolour\_error} = standard uncertainty of the pseudocolour,
$\sigma_{\hat{\nu}\text{eff}}$ 
\item semi-major axis of error ellipse in position at epoch J2016.0, $\sigma_\text{pos,max}$
(Eq.~B.1 in \citeads{2018A&A...616A...2L})
\item semi-major axis of error ellipse in proper motion, $\sigma_\text{pm,max}$
(Eq.~B.2 in \citeads{2018A&A...616A...2L})
\item \gacs{ruwe} = renormalised unit weight error (RUWE). The unit weight error (UWE)
is the square root of the normalised chi-square of the astrometric fit to the AL observations, 
$\text{UWE}=[\,\chi^2/(n-n_\text{p})]^{1/2}$, where $n$ is the number of 
good CCD observations of the source (see below) and $n_p=5$ or 6 the number of
parameters fitted. $\text{UWE}\simeq 1.0$ is expected for a well-behaved source, but
that is often not the case owing to calibration errors. The RUWE is calculated by
empirical scaling of the UWE, depending on $G$ and $\nu_\text{eff}$ or 
$\hat{\nu}_\text{eff}$, such that $\text{RUWE}\simeq 1.0$ for well-behaved sources
(see also Sect.~\ref{sec:gof}). This statistic is not given for two-parameter solutions.
\item \gacs{astrometric\_excess\_noise} = excess source noise, $\epsilon_i$: This
is the extra noise per observation that must be postulated to explain the scatter
of residuals in the astrometric solution for the source (see also Sect.~\ref{sec:gof}). 
The excess source noise is considered to be statistically significant if 
$\gacs{astrometric\_excess\_noise\_sig}>2$
\item  \gacs{visibility\_periods\_used} = number of visibility periods of the source,
that is, groups of observations separated by at least four days
\item  \gacs{astrometric\_matched\_observations} = number of FoV
transits of the source used in the astrometric solution
\item \gacs{astrometric\_n\_good\_obs\_al} = number of good
CCD observations AL of the source used in the astrometric solution 
\item fraction of outliers (bad CCD observations AL of the source) = 
$\gacs{astrometric\_n\_bad\_obs\_al}/\gacs{astrometric\_n\_obs\_al}$
\item \gacs{ipd\_gof\_harmonic\_amplitude} = amplitude of the natural logarithm of the
goodness-of-fit obtained in the IPD versus position angle of scan
\item \gacs{ipd\_frac\_multi\_peak} = fraction of CCD observations where IPD detected 
more than one peak
\item \gacs{ipd\_frac\_odd\_win} = fraction of FoV transits with truncated windows 
or multiple gates
\end{itemize}
The last three statistics were generated at the image parameter determination (IPD) stage
prior to the astrometric solution, and may include transits that were not used for the
astrometry. They are listed here because they provide information on (potentially) 
problematic sources, complementary to what is obtained from the astrometric 
fit (see Sect.~\ref{sec:gof}). We refer to the \textit{Gaia} Archive 
on-line documentation for further explanation of these statistics.

Although both the five- and six-parameter solution provides estimates of the five 
astrometric parameters (position, parallax, and proper motion), the six-parameter 
solution (with pseudocolour as the sixth parameter) is intrinsically less accurate 
because the default colour had to be used for the IPD. Moreover, the six-parameter 
solution is normally only used for sources that are problematic in some respect, 
for example in very crowded areas, which tends to reduce its accuracy even more. 
It is therefore usually relevant to give separate statistics for the two kinds of 
solution. In Tables~\ref{tab:statP5}--\ref{tab:statP2} we report the mean or median 
values of most of the statistics listed above, as functions of magnitude and separated 
by the kind of solution. The median is used for quantities that have a long-tailed 
distribution, for which the mean value might be less representative.

\begin{figure}
\centering
  \includegraphics[width=0.90\hsize]{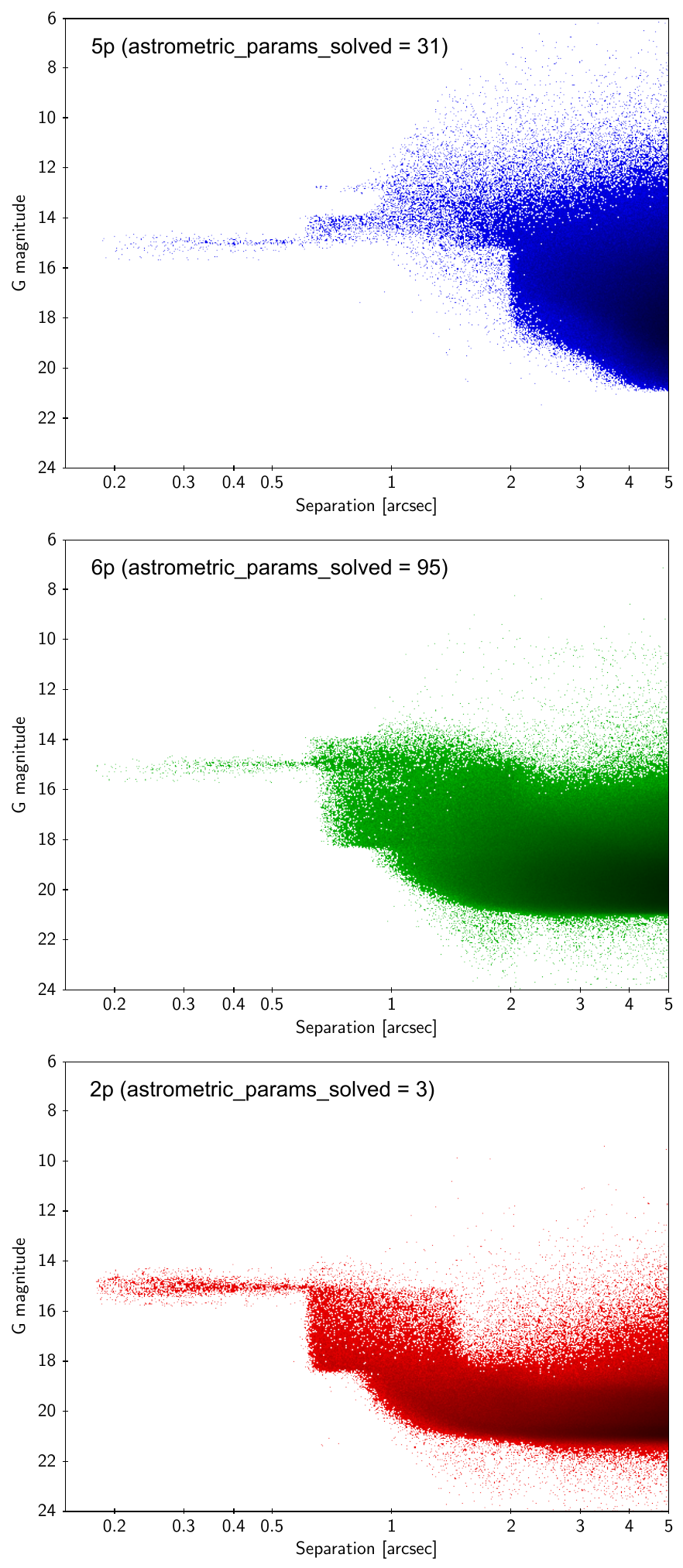}  %{Figures/resolutionG15.pdf}
  \caption{Neighbourhood of 15th magnitude sources in \textit{Gaia} EDR3.
  The diagrams show all sources within 5~arcsec from any one of the 2.8~million sources
  in EDR3 with $14.95<G<15.05$~mag.
    \textit{Top:} Five-parameter solutions.
    \textit{Middle:} Six-parameter solutions.
    \textit{Bottom:} Two-parameter solutions.}
  \label{fig:resG15}
\end{figure}

\subsection{Angular resolution}
\label{sec:resolution}

Resolution here refers to the minimum angular separation between distinct sources
in \textit{Gaia} EDR3, that is between objects with different \gacs{source\_id}. As
explained in Sect.~\ref{sec:accept}, the separation will by construction never be
smaller than 0.18~arcsec, but relatively few sources are found with separations less
than about 
0.6~arcsec owing to other limitations. In a given situation, the effective resolution 
(however it is defined) depends on many different factors such as the magnitudes 
of both components in a pair, their relative orientation on the sky, and the 
kinds of solutions involved (with five, six, or two parameters). The complex situation
is illustrated in Fig.~\ref{fig:resG15}, which shows the neighbourhood of all
sources in EDR3 of magnitude $G\simeq 15$, subdivided by the kind of solution.
Clearly most neighbours at separation 0.18--0.6~arcsec only have two-parameter
solutions, while neighbours with five-parameter solutions are usually either the
brighter of the two sources or more distant than 2~arcsec. The six-parameter 
solutions partly fill the gap for separations between 0.6 and 2~arcsec. Additional
statistics on the small-scale completeness of EDR3 are given in \citet{EDR3-DPACP-126}.

\subsection{Goodness-of-fit statistics}
\label{sec:gof}

Several of the statistics listed in Sect.~\ref{sec:overviewData} quantify the goodness-of-fit
of the single-star model to the observations, either at the image parameter determination
(IPD), where a model LSF or PSF is fitted to the CCD samples, or in the subsequent astrometric 
solution, where the standard model of stellar motion (Sect.~\ref{sec:source}) is fitted to the 
resulting image locations. A few remarks should be made concerning the interpretation of
these statistics and their interrelations.

For the user, the most relevant goodness-of-fit statistics from the IPD are 
\gacs{ipd\_gof\_harmonic\_amplitude}, \gacs{ipd\_frac\_multi\_peak}, and 
\gacs{ipd\_frac\_odd\_win}; and from the astrometric fit, the RUWE and the 
excess source noise with its significance. (The fraction of outliers is probably 
less useful: The outlier detection is designed to remove occasional large deviations, 
caused by temporary perturbations that are usually unrelated to the source.)
All of them describe (real or spurious) deviations from the simplest possible point-source 
model, but they are sensitive to different kinds of modelling errors, and all of them are more
or less sensitive to calibration errors. The sensitivity is usually a strong function 
of the magnitude of the source, and may also depend on geometric factors such as the 
distribution of scans across the source. All of this complicates the interpretation of the 
statistics. For example, there is no simple way to convert them into $p$-values, using the 
single-star model as a null hypothesis; instead, the relevant distributions must be 
determined empirically, if it is at all possible.
 
The IPD statistics may be quite powerful for detecting certain kinds of binaries.
\gacs{ipd\_gof\_harmonic\_amplitude} is normally small but could become large
for sources that have elongated images, such as partially resolved binaries, provided
that their position angles are relatively fixed. In such cases \gacs{ipd\_gof\_harmonic\_phase} 
indicates the position angle of the major axis modulo $180^\circ$. 
\gacs{ipd\_frac\_multi\_peak} is sensitive to resolved binaries that in some scan
directions produce more than one peak in the window. Finally, \gacs{ipd\_frac\_odd\_win} 
is sensitive to the presence of another (usually brighter) source causing the window to be 
`odd', that is truncated or with multiple gating modes. The source causing the gating
could however be quite far away on the CCD, or even in the other FoV. Transits with 
odd windows were not used in the astrometric solution for EDR3.
From Tables~\ref{tab:statP5}--\ref{tab:statP2} it is seen that sources with 
five-parameter solutions are usually very clean, as indicated by the IPD statistics,
while the six-parameter solutions have higher fractions of observations with 
multiple peaks or odd windows (which partially explains why they did not
have sufficiently good BP and RP photometry for the calculation of $\nu_\text{eff}$), 
and the two-parameter solutions are even worse. Towards the faint magnitudes the 
\gacs{ipd\_frac\_multi\_peak} is always decreasing, because the diminishing 
signal-to-noise ratio makes the detection of secondary peaks increasingly difficult.

The astrometric goodness-of-fit measures RUWE and excess source noise 
(\gacs{astrometric\_excess\_noise}) quantify the same thing, namely how much 
the motion of the image centre (as determined by the IPD) deviates from 
the standard model of stellar motion fitted in the astrometric solution. However, while
the \gacs{astrometric\_excess\_noise} gives the discrepancy in angular measure (mas) 
per AL observation (ideally $=0$ for a good fit), the RUWE gives the discrepancy as a 
dimensionless factor (ideally $=1.0$). The RUWE was obtained from the unit weight 
error $\text{UWE}=[\,\chi^2/(n-n_\text{p})]^{1/2}$ by applying an empirical scaling 
factor to compensate for calibration errors, which tend to increase the UWE for bright, 
blue, and very red sources. A corresponding correction was not applied to the 
\gacs{astrometric\_excess\_noise}, which must therefore be interpreted with 
some caution for sources with $G\lesssim 13$, $\nu_\text{eff}\gtrsim 1.65$ 
($G_\text{BP}{-}G_\text{RP}\lesssim 0.4$), or $\nu_\text{eff}\lesssim 1.24$ 
($G_\text{BP}{-}G_\text{RP}\gtrsim 3.0$). The significance (signal-to-noise ratio) 
of the excess source noise is given by \gacs{astrometric\_excess\_noise\_sig}: 
\gacs{astrometric\_excess\_noise} should be regarded as insignificant (that is, 
effectively zero) if $\gacs{astrometric\_excess\_noise\_sig}\lesssim 2$. Alternatively, 
$\gacs{astrometric\_excess\_noise}/\gacs{astrometric\_excess\_noise\_sig}$ may
be taken as an estimate of the uncertainty of the excess source noise.
The RUWE and excess source noise are sensitive to the photocentric motions of 
unresolved objects, such as astrometric binaries, which are not revealed by the
IPD statistics, and therefore complement the latter in binary detection.

The standard uncertainties given in EDR3 have been adjusted to take into account 
the excess noise, whether it represents an astrometric mismatch or a calibration 
issue; they should not be further inflated based on the goodness-of-fit 
statistics.

\begin{table*}[t]
\caption{Summary statistics for the 585~million sources in \textit{Gaia} EDR3 
with five-parameter solutions.
\label{tab:statP5}}
\small   \setlength{\tabcolsep}{4.5pt}
\begin{tabular}{lrrrrrrrrrrl}
\hline\hline
\noalign{\smallskip}
& \multicolumn{10}{c}{Value at $G=$} \\
Quantity & $9$--$12$& $13$ & $14$ & $15$ & $16$ & $17$ & $18$ & $19$ & $20$ & $21$ & Unit \\
\noalign{\smallskip}\hline\noalign{\smallskip}
Fraction of sources with 5-param.\ solution & 92.9& 97.0& 97.1& 96.7& 93.8& 87.5& 76.7& 49.7& 14.5&  1.4& \% \\
Median standard uncertainty in $\alpha$ ($\sigma_{\alpha*}$) at J2016.0 & 0.014& 0.012& 0.015& 0.022& 0.034& 0.054& 0.094& 0.175& 0.374& 1.006& mas \\
Median standard uncertainty in $\delta$ ($\sigma_{\delta}$) at J2016.0 & 0.013& 0.011& 0.013& 0.019& 0.030& 0.049& 0.086& 0.161& 0.335& 0.977& mas \\
Median standard uncertainty in $\varpi$ ($\sigma_{\varpi}$) & 0.018& 0.015& 0.019& 0.027& 0.042& 0.069& 0.120& 0.221& 0.459& 1.320& mas \\
Median standard uncertainty in $\mu_{\alpha*}$ ($\sigma_{\mu\alpha*}$) & 0.018& 0.016& 0.019& 0.028& 0.044& 0.071& 0.123& 0.229& 0.487& 1.445& mas~yr$^{-1}$ \\
Median standard uncertainty in $\mu_{\delta}$ ($\sigma_{\mu\delta}$) & 0.016& 0.014& 0.017& 0.024& 0.038& 0.063& 0.111& 0.208& 0.428& 1.401& mas~yr$^{-1}$ \\
Median renormalised unit weight error (RUWE) & 1.039& 1.023& 1.016& 1.013& 1.012& 1.011& 1.010& 1.010& 1.010& 1.016&   \\
Fraction with significant excess source noise & 97.3& 59.4& 19.2& 16.8& 13.8& 10.6&  7.5&  5.0&  5.0& 11.7& \% \\
Median excess source noise (when significant) & 0.113& 0.089& 0.188& 0.224& 0.285& 0.409& 0.606& 0.976& 1.801& 3.668& mas \\
Mean number of visibility periods used & 20.9& 21.0& 20.9& 20.8& 20.7& 20.7& 20.7& 20.5& 19.2& 13.1&   \\
Mean number of FoV transits used & 43.7& 43.7& 43.6& 43.4& 43.2& 43.2& 43.4& 42.6& 38.7& 20.5&   \\
Mean number of good CCD observations AL & 378& 378& 377& 376& 374& 374& 376& 368& 334& 175&   \\
Mean fraction of bad CCD observations AL &  1.2&  0.4&  0.5&  0.5&  0.5&  0.4&  0.4&  0.4&  0.5&  0.6& \% \\
Median \gacs{ipd\_gof\_harmonic\_amplitude} & 0.018& 0.021& 0.020& 0.020& 0.020& 0.020& 0.020& 0.022& 0.029& 0.073&  \\
Mean \gacs{ipd\_frac\_multi\_peak} &  1.9&  1.5&  1.5&  1.7&  0.9&  0.6&  0.3&  0.2&  0.1&  0.9& \% \\
Mean \gacs{ipd\_frac\_odd\_win} &  0.0&  0.0&  0.0&  0.0&  0.0&  0.0&  0.1&  0.1&  0.2&  0.8& \% \\
\noalign{\smallskip}\hline
\end{tabular}
\end{table*}

\begin{table*}[t]
\caption{Summary statistics for the 882~million sources in \textit{Gaia} EDR3 
with six-parameter solutions.
\label{tab:statP6}}
\small   \setlength{\tabcolsep}{4.5pt}
\begin{tabular}{lrrrrrrrrrrl}
\hline\hline
\noalign{\smallskip}
& \multicolumn{10}{c}{Value at $G=$} \\
Quantity & $9$--$12$& $13$ & $14$ & $15$ & $16$ & $17$ & $18$ & $19$ & $20$ & $21$ & Unit \\
\noalign{\smallskip}\hline\noalign{\smallskip}
Fraction of sources with 6-param.\ solution &  6.2&  2.4&  2.2&  2.7&  5.6& 11.9& 22.4& 48.6& 77.7& 27.5& \% \\
Median standard uncertainty in $\alpha$ ($\sigma_{\alpha*}$) at J2016.0 & 0.017& 0.026& 0.025& 0.033& 0.052& 0.081& 0.131& 0.225& 0.430& 1.031& mas \\
Median standard uncertainty in $\delta$ ($\sigma_{\delta}$) at J2016.0 & 0.016& 0.024& 0.023& 0.030& 0.045& 0.069& 0.112& 0.194& 0.382& 1.025& mas \\
Median standard uncertainty in $\varpi$ ($\sigma_{\varpi}$) & 0.023& 0.033& 0.031& 0.040& 0.061& 0.093& 0.151& 0.266& 0.526& 1.390& mas \\
Median standard uncertainty in $\mu_{\alpha*}$ ($\sigma_{\mu\alpha*}$) & 0.023& 0.034& 0.033& 0.042& 0.065& 0.102& 0.167& 0.291& 0.565& 1.469& mas~yr$^{-1}$ \\
Median standard uncertainty in $\mu_{\delta}$ ($\sigma_{\mu\delta}$) & 0.021& 0.031& 0.029& 0.036& 0.051& 0.078& 0.132& 0.240& 0.490& 1.470& mas~yr$^{-1}$ \\
Median standard uncertainty in $\hat{\nu}_\text{eff}$ ($\sigma_{\hat{\nu}\text{eff}}$) & 0.004& 0.006& 0.006& 0.008& 0.012& 0.019& 0.031& 0.055& 0.109& 0.281& $\mu$m$^{-1}$ \\
Median renormalised unit weight error (RUWE) & 1.127& 1.362& 1.167& 1.104& 1.102& 1.085& 1.053& 1.033& 1.029& 1.048&   \\
Fraction with significant excess source noise & 99.6& 95.1& 75.4& 68.3& 64.0& 58.3& 45.5& 27.8& 17.7& 19.6& \% \\
Median excess source noise (when significant) & 0.169& 0.275& 0.387& 0.405& 0.471& 0.569& 0.809& 1.343& 2.153& 3.838& mas \\
Mean number of visibility periods used & 21.4& 21.1& 20.7& 20.1& 18.9& 18.2& 18.4& 18.5& 18.1& 13.3&   \\
Mean number of FoV transits used & 46.1& 44.6& 44.0& 42.6& 39.3& 38.2& 37.9& 37.8& 36.6& 21.1&   \\
Mean number of good CCD observations AL & 400& 384& 379& 366& 339& 331& 328& 327& 315& 181&   \\
Mean fraction of bad CCD observations AL &  1.3&  1.5&  1.9&  2.1&  1.5&  1.0&  0.7&  0.6&  0.6&  0.6& \% \\
Median \gacs{ipd\_gof\_harmonic\_amplitude} & 0.020& 0.033& 0.032& 0.030& 0.035& 0.034& 0.032& 0.030& 0.031& 0.071&  \\
Mean \gacs{ipd\_frac\_multi\_peak} &  9.4& 25.4& 21.6& 18.3& 10.7&  7.0&  5.5&  3.2&  1.0&  0.5& \% \\
Mean \gacs{ipd\_frac\_odd\_win} &  0.0&  4.8&  7.9&  8.1&  6.3&  5.2&  4.4&  4.2&  3.8&  2.1& \% \\
\noalign{\smallskip}\hline
\end{tabular}
\end{table*}

\begin{table*}[t]
\caption{Summary statistics for the 344~million sources in \textit{Gaia} EDR3 with 
two-parameter solutions.
\label{tab:statP2}}
\small   \setlength{\tabcolsep}{4.5pt}
\begin{tabular}{lrrrrrrrrrrl}
\hline\hline
\noalign{\smallskip}
& \multicolumn{10}{c}{Value at $G=$} \\
Quantity & $9$--$12$& $13$ & $14$ & $15$ & $16$ & $17$ & $18$ & $19$ & $20$ & $21$ & Unit \\
\noalign{\smallskip}\hline\noalign{\smallskip}
Fraction of sources with 2-param.\ solution &  0.8&  0.7&  0.7&  0.6&  0.6&  0.6&  0.9&  1.7&  7.8& 71.1& \% \\
Median standard uncertainty in $\alpha$ ($\sigma_{\alpha*}$) at J2016.0 & 1.771& 1.260& 1.127& 1.116& 1.060& 1.008& 0.979& 1.436& 2.579& 3.250& mas \\
Median standard uncertainty in $\delta$ ($\sigma_{\delta}$) at J2016.0 & 1.703& 1.211& 1.078& 1.067& 0.994& 0.925& 0.888& 1.252& 2.106& 2.847& mas \\
Mean number of visibility periods used & 17.9& 18.4& 18.2& 18.0& 17.7& 17.0& 15.7& 10.9&  8.0&  6.9&   \\
Mean number of FoV transits used & 35.2& 36.3& 35.9& 35.4& 34.7& 33.6& 30.7& 19.7& 13.3& 10.0&   \\
Mean number of good CCD observations AL & 299& 306& 304& 300& 294& 285& 261& 167& 113& 85&   \\
Mean fraction of bad CCD observations AL &  4.3&  3.5&  3.5&  3.5&  3.5&  3.4&  3.1&  2.5&  1.6&  0.9& \% \\
Median \gacs{ipd\_gof\_harmonic\_amplitude} & 0.217& 0.251& 0.258& 0.261& 0.279& 0.273& 0.256& 0.259& 0.175& 0.096&  \\
Mean \gacs{ipd\_frac\_multi\_peak} & 59.2& 53.0& 44.8& 43.6& 36.0& 27.8& 20.8& 14.2&  6.8&  0.3& \% \\
Mean \gacs{ipd\_frac\_odd\_win} &  0.0& 12.5& 19.2& 19.8& 27.2& 31.1& 33.3& 35.6& 25.2&  3.1& \% \\
\noalign{\smallskip}\hline
\end{tabular}
\end{table*}

\begin{figure*}
\centering
  \includegraphics[width=0.95\hsize]{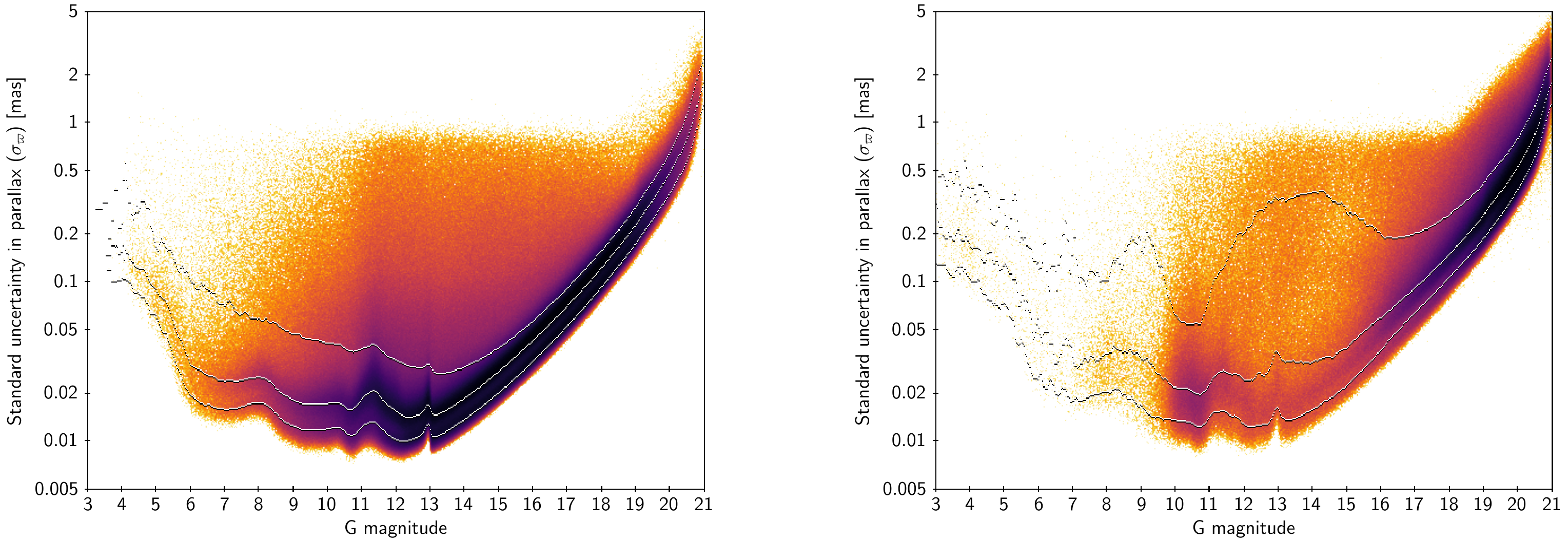}
  \caption{Uncertainty in parallax versus magnitude. \textit{Left:} Five-parameter solutions.
  \textit{Right:} Six-parameter solutions. The plots include all sources
  with $G<11.5$ and a geometrically decreasing random fraction of the fainter sources,
  so as to give a roughly constant number of sources per magnitude interval. The colour
  scale from yellow to black indicates an increasing density of data points in the diagram. 
  The curves show the 10th, 50th, and 90th percentiles of the distribution at a
  given magnitude.}
  \label{fig:uncVsG}
\end{figure*}

\begin{figure}
\centering
  \includegraphics[width=0.90\hsize]{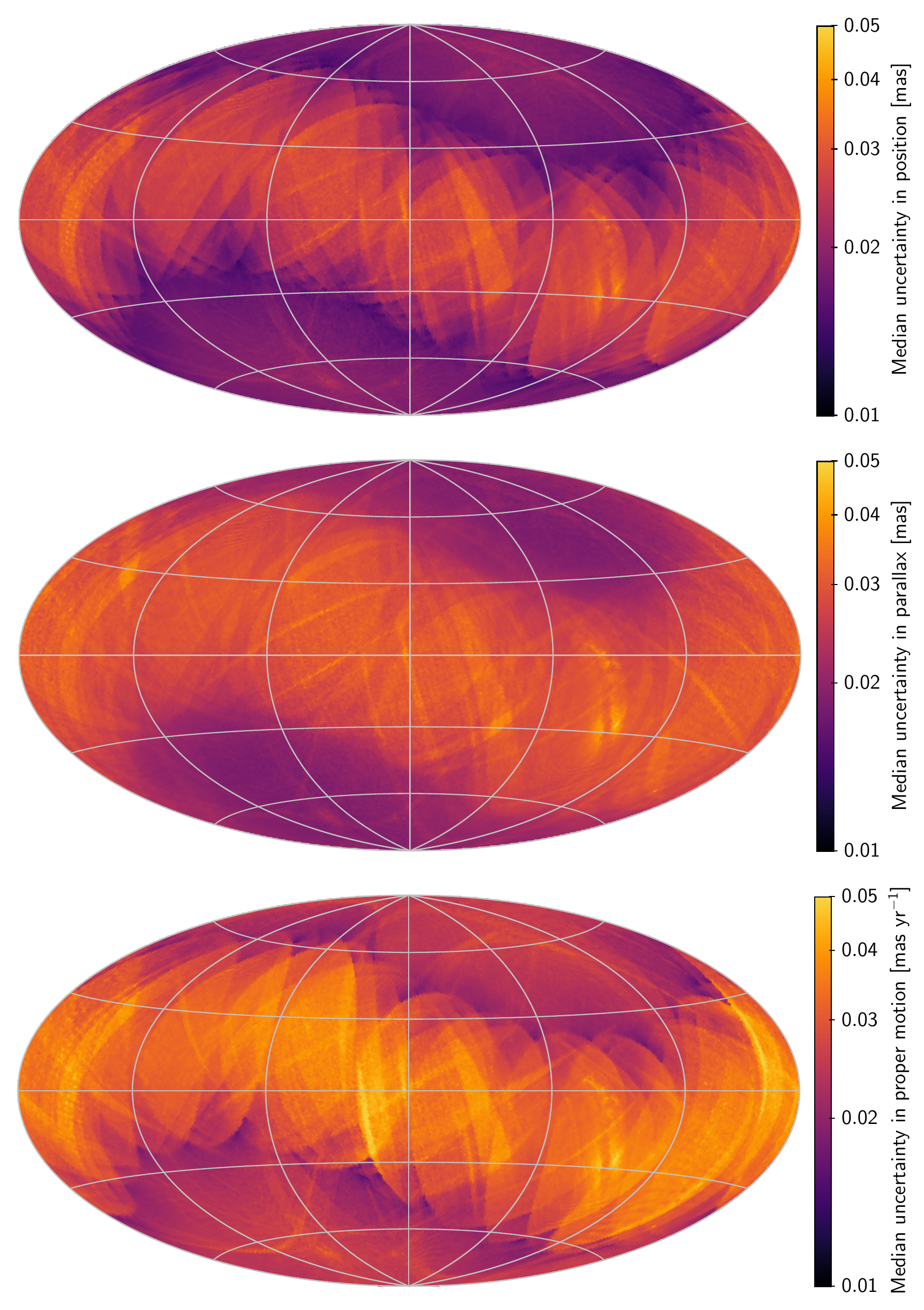}
  \caption{Formal uncertainties at $G\simeq 15$ for sources with a five-parameter 
  solution in EDR3.
    \textit{Top:} Semi-major axis of the error ellipse in position at epoch J2016.0.
    \textit{Middle:} Standard deviation in parallax.
    \textit{Bottom:} Semi-major axis of the error ellipse in proper motion.
    These and all other full-sky maps in the paper except Fig.~\ref{fig:mapsCorrEcl}
    use a Hammer--Aitoff projection in equatorial (ICRS) coordinates with 
    $\alpha=\delta=0$ at the centre, north up, and $\alpha$ increasing from right to left.}
  \label{fig:mapsUnc}
\end{figure}

\begin{figure}
\centering
  \includegraphics[width=0.90\hsize]{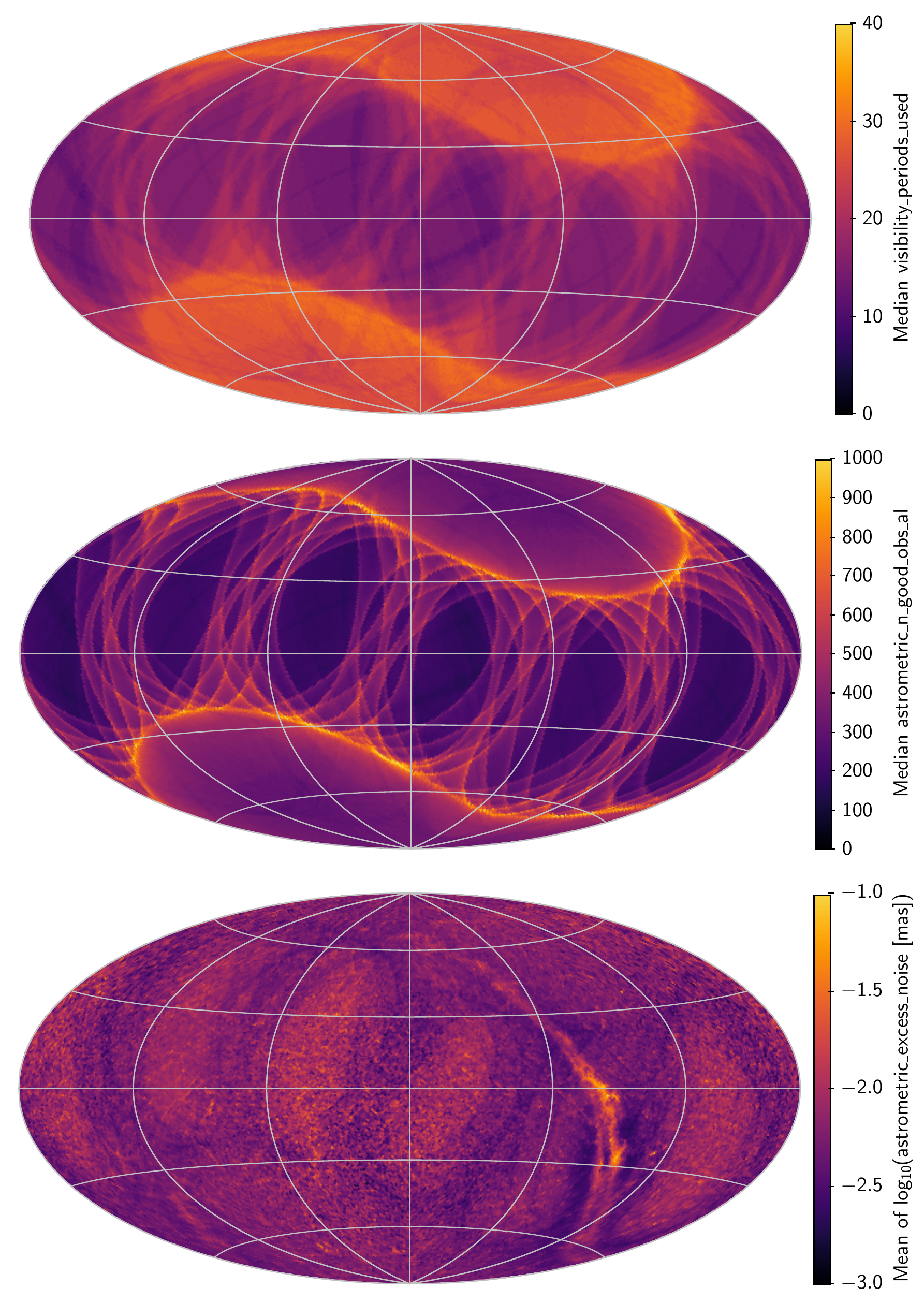}
  \caption{Selected observation statistics at $G\simeq 15$ for sources with
    a five-parameter solution in EDR3. These statistics are main factors
    governing the formal uncertainties of the astrometric data.
    \textit{Top:} Number of visibility periods used.
    \textit{Middle:} Number of good CCD observations AL. (The number of FoV 
    transits used in the solution looks very similar, only with a factor nine smaller numbers.)
    \textit{Bottom:} Mean of $\log_{10}[\max(0.001,\,\gacs{astrometric\_excess\_noise})]$.}
  \label{fig:mapsVpu}
\end{figure}

\subsection{Formal uncertainties}
\label{sec:form}

Tables~\ref{tab:statP5}--\ref{tab:statP2} give the median uncertainties of the
astrometric parameters at selected magnitudes for the different kinds of solutions. 
No statistics are given for $G<9$ owing to the relatively few sources and rapidly 
declining quality at this end (see Fig.~\ref{fig:uncVsG}).
Compared with DR2 (Tables~B.1 and B.2 in \citeads{2018A&A...616A...2L}), the gain 
in median uncertainty at $G=15$ is a factor 0.71 for the positions and parallaxes,
and 0.44 for the proper motions. This is slightly better than the factors 0.80 and 0.51
expected purely from the increased length of the data included in the solutions
(Sect.~\ref{sec:data}). The extra gain comes mainly from the improved robustness 
and homogeneity of results made possible by the higher redundancy of observations 
in the EDR3 solutions. 

For $G=9$--12 the gain in median uncertainty from DR2 to EDR3 is even more 
impressive thanks to the improved calibrations, which are relatively more important 
for the bright sources (cf.\  Sect.~\ref{sec:resid}): The factor is 0.43 for the positions 
and parallaxes, and 0.27 for the proper motions.

The comparison between the two releases is however complicated by the circumstance 
that in EDR3 there are three kinds of solutions (five, six, and two parameters), while 
in DR2 there are only five- and two-parameter solutions. At $G=15$ the five-parameter 
solutions comprised 99.0\% of the sources in DR2 and 96.7\% of the sources in EDR3, 
and the comparison above used the statistics for these subsets. At $G=15$ the median 
uncertainties in EDR3 are a factor 1.5 higher for the 2.7\% of the sources with 
six-parameter solutions than for the 96.7\% with five-parameter solutions. This
large ratio in the uncertainties reflects the generally more problematic nature 
of the sources receiving six-parameter solutions, also seen in the various
goodness-of-fit statistics discussed in Sect.~\ref{sec:gof}.

The fraction of sources that receive five-parameter solutions is higher than 90\%
down to $G\simeq 17$, but decreases rapidly for fainter sources. The fraction with
six-parameter solutions correspondingly increases down to $G\simeq 20$, after 
which there is instead a steep increase in the fraction of two-parameter solutions. 

At any magnitude there is a considerable spread in the uncertainties caused by
variations in the number of observations and the properties of the scanning law. For 
$G\le 12$ there are additional variations depending on the window classes and gates 
used for a particular source, and the onset of saturation for
the brightest sources. The spread is illustrated in Fig.~\ref{fig:uncVsG} 
for the parallaxes in the five- and six-parameter solutions. The uncertainties in position
and proper motion follow similar distributions. 

Figure~\ref{fig:mapsUnc} shows the median uncertainties in position, parallax, and 
proper motions at $G\simeq 15$ as functions of position. For the position and proper
motion data the semi-major axes of the error ellipses $\sigma_\text{pos,max}$,
$\sigma_\text{pm,max}$ are plotted.
The patterns are very similar at other magnitudes, only scaled according to the
general dependence on $G$ in Table~\ref{tab:statP5} or Fig.~\ref{fig:uncVsG}.
These patterns are mainly set by variations in the number, direction, and temporal
distribution of the scans across a given position, as governed by the scanning law.
In very crowded areas, such as along the Galactic plane and in the general direction
of the Galactic Centre, the increased level of excess source noise from background 
sources gives a local rise in the median uncertainties, which becomes more 
important at fainter magnitudes. Some relevant statistics are shown in 
Fig.~\ref{fig:mapsVpu}. A comparison with the corresponding maps for DR2
(Figs.~B.3 and B.4 in \citeads{2018A&A...616A...2L}) clearly shows an 
improved homogeneity in the uncertainties in the ecliptic belt, and the
smaller importance of the excess noise in EDR3.

\begin{figure*}
\centering
  \includegraphics[width=0.79\hsize]{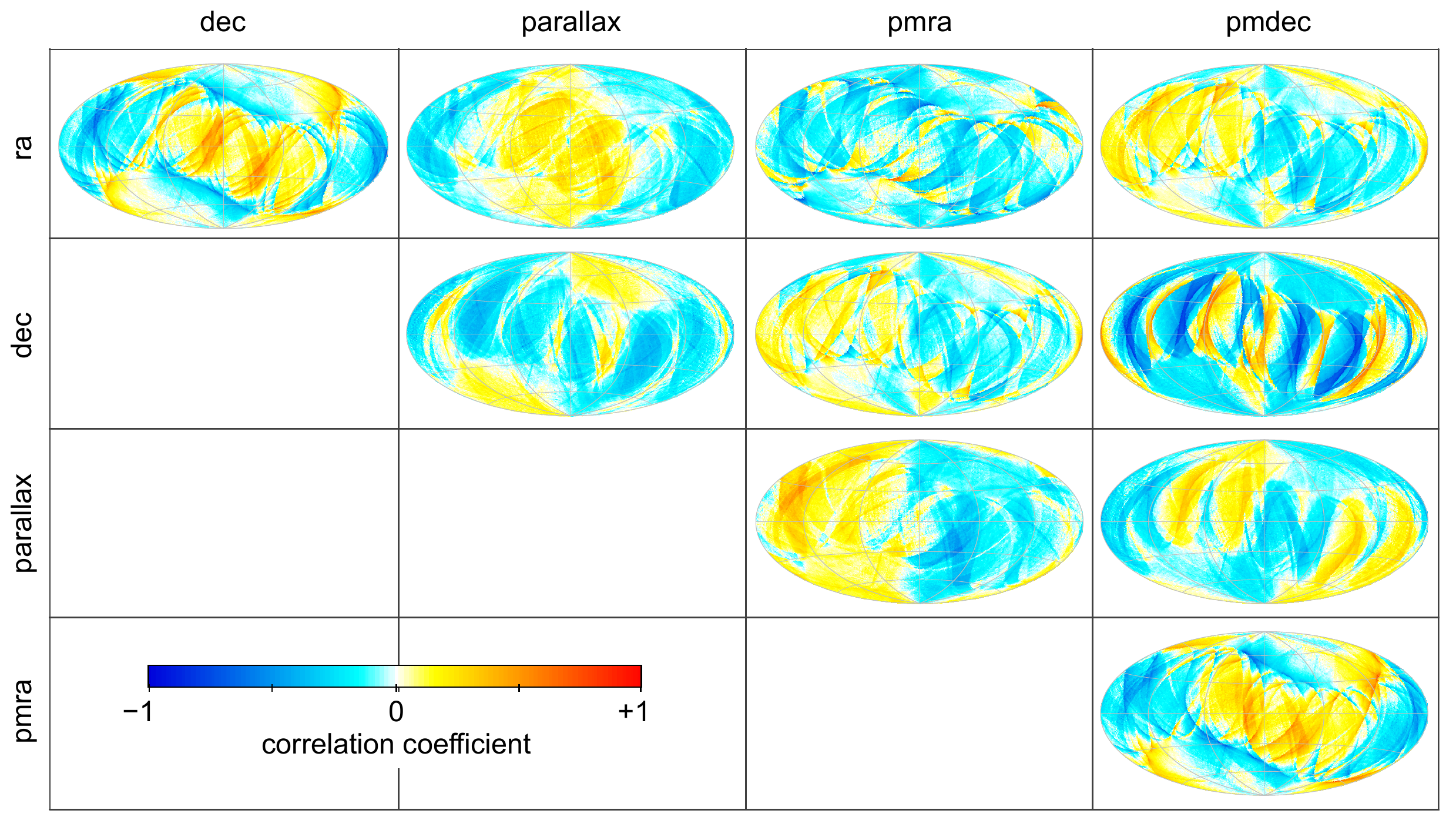}\hspace{0.19\hsize}
  \caption{Median correlation coefficients among the astrometric parameters for
  five-parameter solutions in \textit{Gaia} EDR3. The plots were made from a random
  selection of five-parameter sources with $G$ magnitude between 13 and 16. The correlations
  at other magnitudes are very similar.}
  \label{fig:mapsCorr5p}
\end{figure*}

\begin{figure*}
\centering
  \includegraphics[width=0.98\hsize]{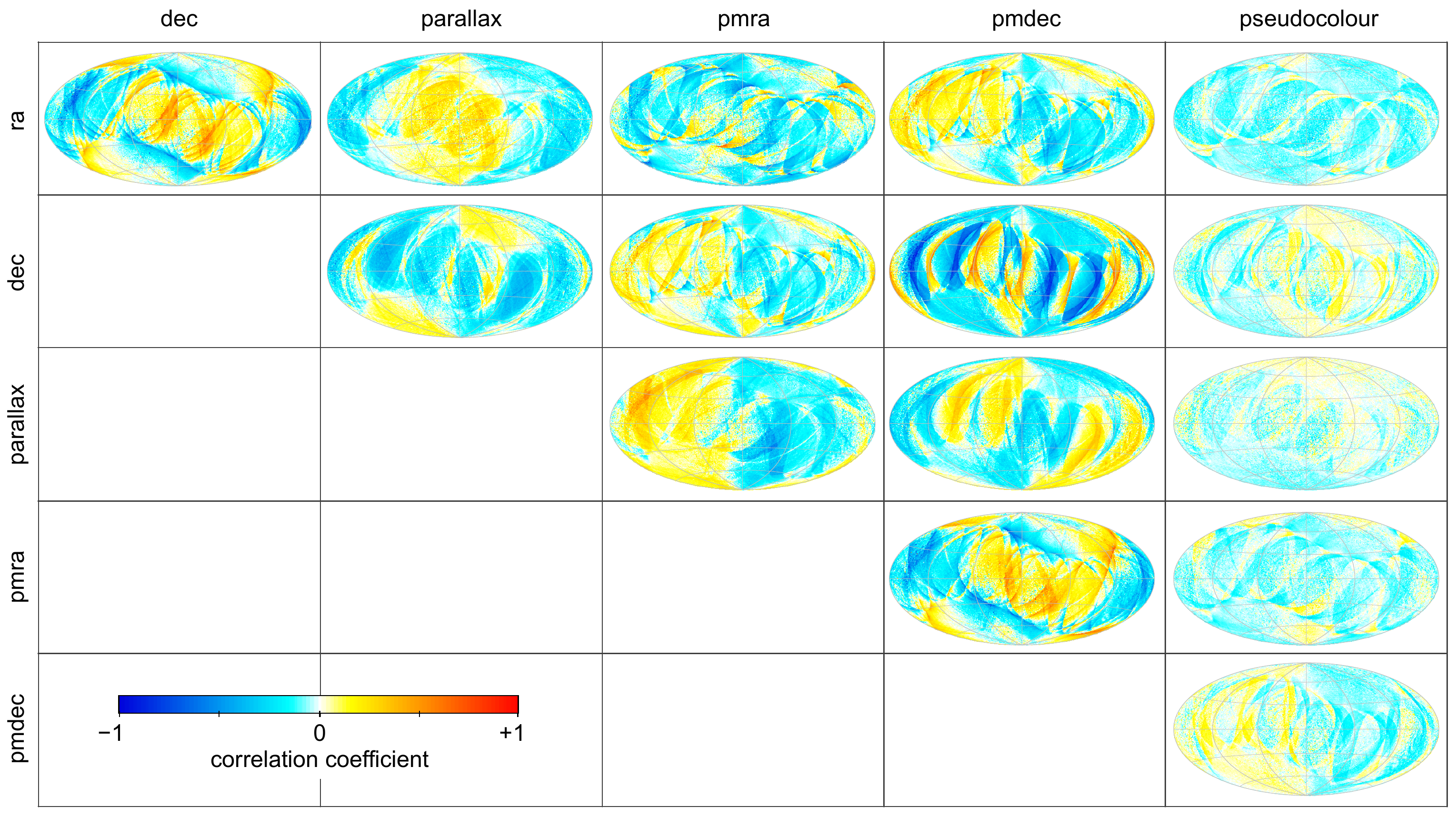}
  \caption{Median correlation coefficients among the astrometric parameters for
  six-parameter solutions in \textit{Gaia} EDR3. The plots were made using all 
  six-parameter sources with $G$ magnitude between 13 and 16. The correlations at other
  magnitudes are very similar.}
  \label{fig:mapsCorr6p}
\end{figure*}

\begin{figure*}
\sidecaption
  \includegraphics[width=120mm]{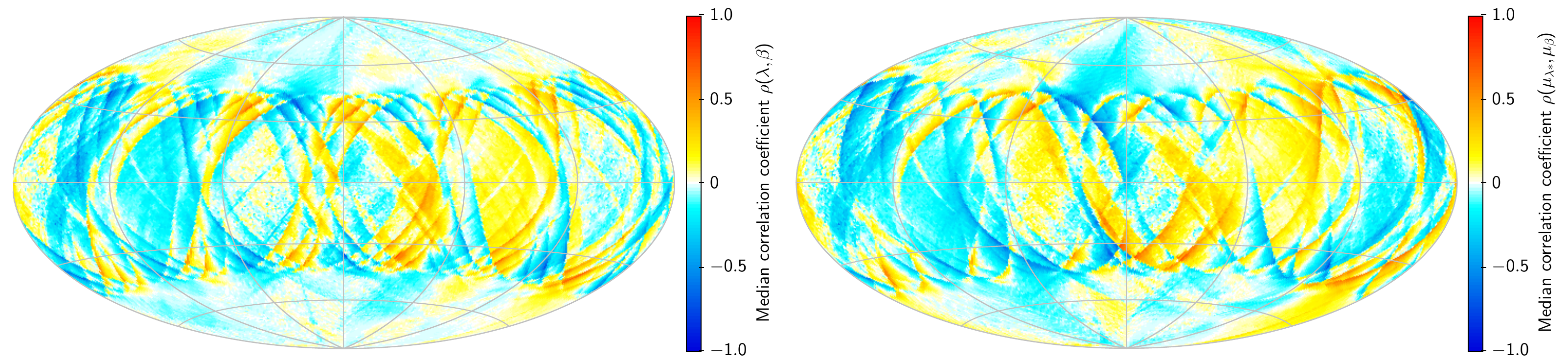}
  \caption{Correlations in ecliptic coordinates (five-parameter solutions, $G=13$--16). 
  \textit{Left:} Median correlation coefficient between errors in ecliptic longitude ($\lambda$)
  and latitude ($\beta$). 
    \textit{Right:} Same for the proper motion components ($\mu_{\lambda*}$, $\mu_\beta$).
  Contrary to other sky maps in the paper, these two use a projection in
  ecliptic coordinates with $\lambda=\beta=0$ in the centre.
  }
  \label{fig:mapsCorrEcl}
\end{figure*}

\subsection{Correlation coefficients}
\label{sec:corr}

\textit{Gaia} EDR3 gives the complete set of correlation coefficients $\rho$ between 
the astrometric parameters provided for a given source. For a source with
$n_\text{p}=5$, 6, or 2 parameters, we thus have  
$n_\text{p}(n_\text{p}-1)/2=10$, 15, or 1 non-redundant coefficients.
In the \textit{Gaia} Archive they are called \gacs{ra\_dec\_corr}, etc.; here we use the notation
$\rho(\alpha,\delta)$, etc. The correlations allow the elements of the 
$n_\text{p}\times n_\text{p}$ covariance matrix $\vec{K}$ to be reconstructed as
\begin{equation}\label{eq:cov}
\begin{split}
K_{00} = \sigma_{\alpha*}^2\,, \quad
&K_{11} = \sigma_\delta^2\,, \quad\dots,\\
&K_{01} = K_{10} = \rho(\alpha,\delta)\sigma_{\alpha*}\sigma_\delta\,, \quad\dots
\end{split} 
\end{equation}
where indices $0,\,1,\,\dots$ represent the parameters in the usual order, 
$\alpha$, $\delta$, $\varpi$, $\mu_{\alpha*}$, $\mu_\delta$, $\hat{\nu}_\text{eff}$.

The correlation coefficients for a given source are mainly determined by the distribution 
of scan directions and transit times among the observations of the source, which are
governed by the scanning law. The correlation coefficients are therefore practically
independent of magnitude, and we give here only statistics for sources with $G=13$
to 16~mag.
Figures~\ref{fig:mapsCorr5p} and \ref{fig:mapsCorr6p}
show the median correlation coefficients for five- and six-parameter solutions.
We note that the scanning law is (approximately) symmetric with respect to the ecliptic, 
which is reflected in many features depending on ecliptic latitude ($\beta$) rather than 
declination ($\delta$). Furthermore, the patterns are often distinctly different
for $|\,\beta\,|\lesssim 45^\circ$ (the ecliptic belt) and $|\,\beta\,|\gtrsim 45^\circ$ 
(the ecliptic caps).

Certain features of predominantly positive or negative correlations are caused by
the choice of ICRS (equatorial) coordinates for the position and proper motion 
parameters, and are much less pronounced if ecliptic coordinates are used. This 
is the case, for example, with the mainly positive correlations $\rho(\alpha,\delta)$
and $\rho(\mu_{\alpha*},\mu_\delta)$ in the ecliptic belt for $\alpha=270^\circ$ to 
$90^\circ$, and their mainly negative values for $\alpha=90^\circ$ to $270^\circ$.
Geometrically, this can be understood in terms of the orientation of the error ellipses
in position and proper motion: In the ecliptic belt their major axes are approximately 
aligned with the ecliptic, and consequently tilted by up to $\pm 23.5^\circ$ with 
respect to the equator, corresponding to non-zero correlations in the equatorial
components. As shown in Fig.~\ref{fig:mapsCorrEcl}, $\rho(\lambda,\beta)$ and 
$\rho(\mu_{\lambda*},\mu_\beta)$ (here shown on an ecliptic projection) are generally
smaller than $\rho(\alpha,\delta)$ and $\rho(\mu_{\alpha*},\mu_\delta)$. 

The correlations $\rho(\alpha,\mu_{\alpha*})$ and $\rho(\delta,\mu_\delta)$ are 
related to the mean epoch of observations contributing to the different parameters. 
A mean epoch later than the reference epoch J2016.0 gives 
a positive correlation between the position and proper motion, and vice versa. 
This is especially pronounced for $\rho(\delta,\mu_\delta)$, where regions of positive 
and negative correlations alternate along the ecliptic.   
The correlations among $\alpha$, $\delta$, $\varpi$, $\mu_{\alpha*}$, and $\mu_\delta$
are almost the same for the five- and six-parameter solutions. In the six-parameter
solutions, the correlations between $\hat{\nu}_\text{eff}$ and the other five parameters are 
generally small (RMS values around 0.1) compared with the correlations among the five
parameters (RMS values of 0.2--0.3). 

The generally small correlations between pseudocolour and the other five parameters
is a consequence of the variation in chromaticity along the path of the stellar image 
through the AF, and between successive observations in either FoV
in a few revolutions; by contrast, the scan directions and observation epochs relevant 
for the other correlations do not change much over several revolutions. The sizes of the 
correlations with pseudocolour are important for the potential to improve the six-parameter 
solutions by incorporating external colour information (Appendix~\ref{sec:col6p}).

\begin{figure*}
\centering
  \includegraphics[width=0.90\hsize]{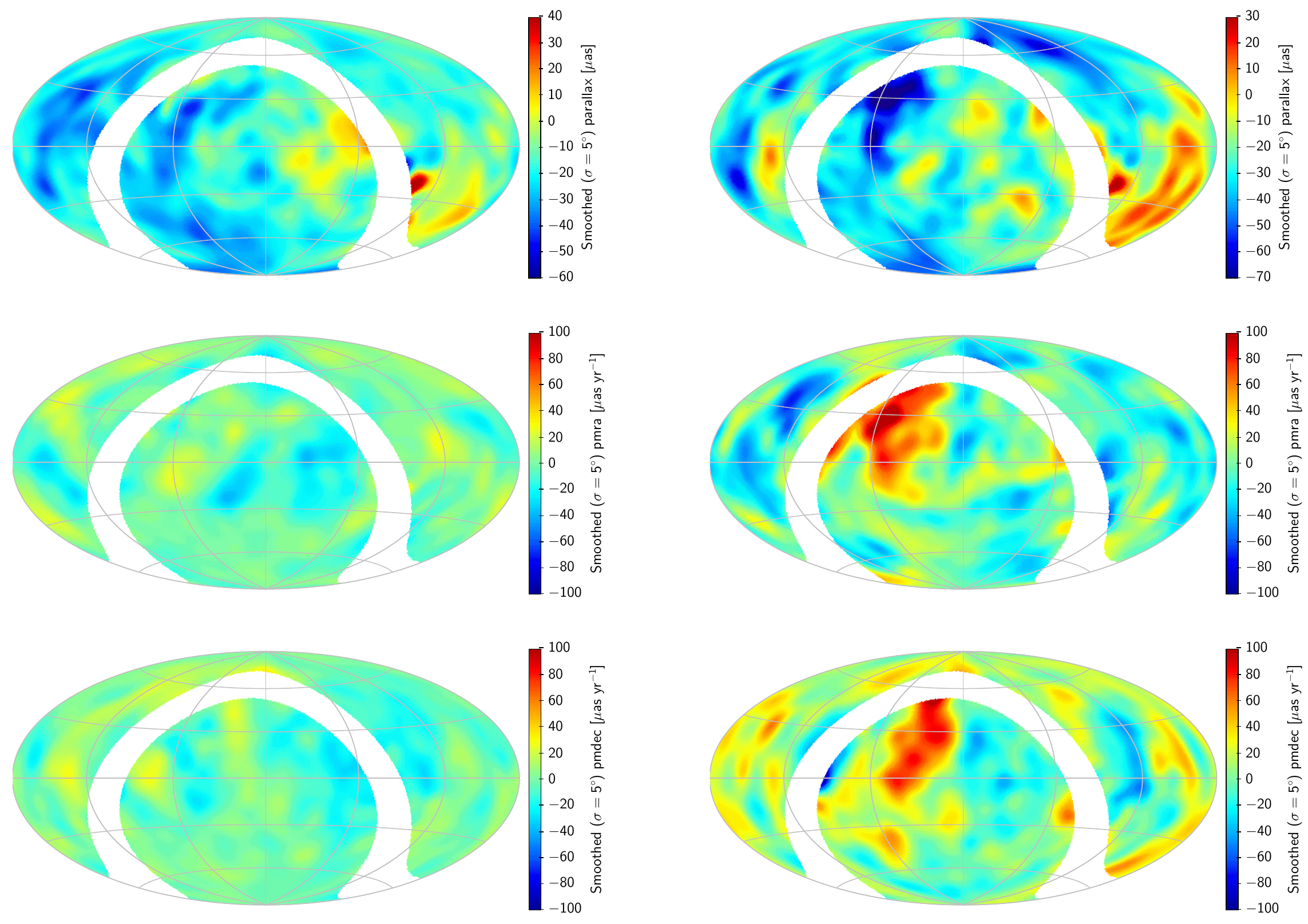}
  \caption{Smoothed maps of quasar parallaxes and proper motions.
  \textit{Left column:} \textit{Gaia} EDR3, using data for about 1.2~million quasars.
  \textit{Right column:} \textit{Gaia} DR2, using data for the 94\% of the quasars in the
  left column that have full astrometric solutions also in DR2.
  From top to bottom the maps show parallax, proper motion in right ascension, and
  proper motion in declination. The maps were smoothed using a Gaussian kernel with 
  standard deviation $5^\circ$.
    No data are shown for $|\,b\,|<10^\circ$, where $b$ is Galactic latitude.}
  \label{fig:mapsQso}
\end{figure*}

\begin{figure*}
\centering
  \includegraphics[width=0.93\hsize]{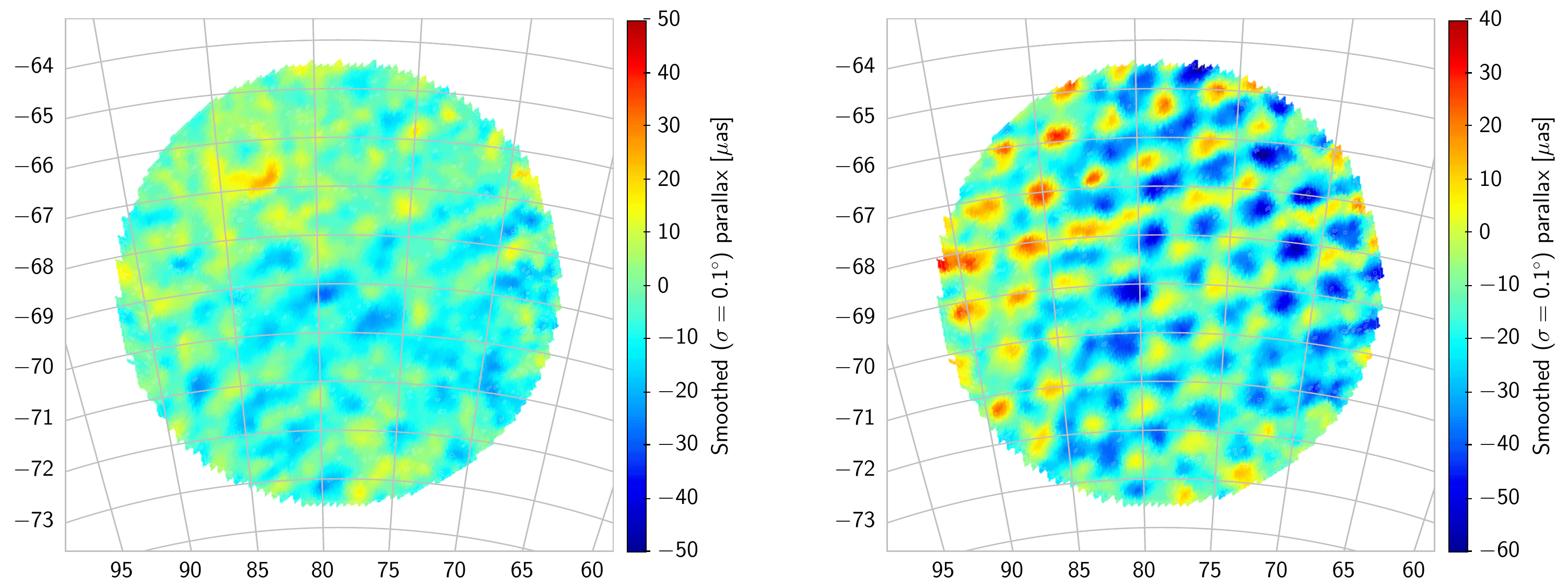}
  \caption{Smoothed maps of parallaxes in the LMC area, visualising small-scale 
  systematics (the `checkered pattern') in \textit{Gaia} EDR3 and DR2. 
  \textit{Left:} Smoothed parallaxes in EDR3 for sources in the magnitude 
  range $G=16$--18 (median $G=17.4$), kinematically selected as probable members 
  of the system (see Appendix~B in \citeads{EDR3-DPACP-132} for details).
  \textit{Right:} Smoothed parallaxes in DR2 for the same sample of sources.
  Both maps were smoothed using a Gaussian kernel with standard deviation $0.1^\circ$.
  While the sample includes about 730\,000 sources within $5^\circ$ radius of the 
  adopted centre, only smoothed points within a radius of $4.5^\circ$ are shown to 
  avoid unwanted edge effects. Comparison between the two diagrams is facilitated 
  by the use of the same colour scale, only shifted by $10~\mu$as to compensate for the 
  mean difference in parallax between DR2 and EDR3.}
  \label{fig:mapsLmcPlx}
\end{figure*}

\subsection{Systematic errors}
\label{sec:syst}

Several aspects of the systematic errors (biases) in the astrometric data are 
examined in the EDR3 catalogue validation paper \citep{EDR3-DPACP-126}.
The bias in parallax, and its variation with magnitude, colour, and ecliptic latitude, 
is extensively discussed in a separate paper \citep{EDR3-DPACP-132}. The global
properties of the system of positions and proper motions are discussed in the
EDR3 celestial reference frame paper \citep{EDR3-DPACP-133}. 

In this section we focus on the statistical variations of parallax and proper motion
biases on various angular scales, as revealed by the quasars and (for parallaxes on
scales $\lesssim 1^\circ$) by sources in the direction of the Large Magellanic Cloud
(LMC). We also illustrate the improvements achieved since DR2. The results presented here 
are derived using relatively faint sources ($G\simeq 16$--20), and little is known about 
small- and medium-scale variations at brighter magnitudes, in particular for $G<13$, 
where the sources in many respects behave differently from the fainter sources. Furthermore, 
we only give results for the five-parameters solutions, which are used for most 
sources brighter than $G\simeq 19$ (Fig.~\ref{fig:histNsrc}). 
In general the six-parameter solutions are probably worse than the five-parameter 
solutions in terms of systematics, but it is difficult to know whether this is an intrinsic 
property of the solutions or a consequence of the faintness and more problematic 
nature of most of the sources getting a six-parameter solution (Sect.~\ref{sec:nuEff}). 

Figure~\ref{fig:mapsQso} (left) shows smoothed maps of the parallaxes and proper motion
components for a sample of 1\,215\,942 quasars, namely the subset of sources in 
\textit{Gaia} EDR3 Archive table \gacs{agn\_cross\_id} with five-parameter 
solutions in \gacs{gaia\_source} (median $G=19.9$).
The selection of quasars in \gacs{agn\_cross\_id} is discussed in \citet{EDR3-DPACP-133}.
Smoothed values were computed using a Gaussian kernel of $5^\circ$ standard deviation.%
\footnote{More precisely, the smoothed value at a given point is computed as
the weighted median of the individual values within a radius of $15^\circ$, using
weights proportional to $\exp[-\frac{1}{2}(\theta/5^\circ)^2]$, where $\theta$ is 
the angle between the quasar and the smoothed point.}
The smoothed points in the Galactic zone ($|\,b\,|<10^\circ$) are not displayed, 
as they are dominated by noise from small-number statistics. The standard 
deviations of the smoothed maps (for $|\,b\,|>10^\circ$) are 10.8~$\mu$as in 
$\varpi$, 11.2~$\mu$as~yr$^{-1}$ in $\mu_{\alpha*}$, and 10.7~$\mu$as~yr$^{-1}$
in $\mu_\delta$.

For comparison, we show in the right column of Fig.~\ref{fig:mapsQso} the corresponding 
maps for \textit{Gaia} DR2 astrometry, calculated in the same manner for the 1\,141\,470 of the
sources in the EDR3 quasar sample that have full astrometric data also in DR2. To facilitate 
comparison, the maps use the same colour scales as for the EDR3 data, only shifted by 
$10~\mu$as in parallax to compensate for the different mean
biases. The standard deviations in the DR2 maps are 15.5~$\mu$as,  
26.2~$\mu$as~yr$^{-1}$, and 23.5~$\mu$as~yr$^{-1}$. Thus, in EDR3 the systematics
are reduced by the factors 0.70 ($\varpi$), 0.41 ($\mu_{\alpha*}$), and 0.46 ($\mu_\delta$), 
that is very nearly the same factors as for the random uncertainties (Sect.~\ref{sec:form}).

On much smaller scales, down to $0.1^\circ$, Fig.~\ref{fig:mapsLmcPlx} shows the 
characteristic `checkered pattern' that was very prominent in the DR2 astrometry
for the LMC and in maps of the median parallax in the Galactic bulge 
area (Sect.~4.2 in \citeads{2018A&A...616A..17A}). In EDR3 there is a similar pattern,
but with a different structure and smaller amplitude as shown in Fig.~\ref{fig:mapsLmcPlx}.
The RMS amplitude of the smoothed variations in these plots is $7.7~\mu$as for EDR3 
and $14.3~\mu$as for DR2. 

The maps in Figs.~\ref{fig:mapsQso} and \ref{fig:mapsLmcPlx} were smoothed in order to
bring out clearly the pattern of systematic errors. Although the random errors are strongly
attenuated by the smoothing, they still contribute to the standard deviations quoted above,
which are therefore somewhat higher than the actual RMS systematics on the relevant 
angular scales. In order to correct for this bias, we randomly divided the sources into 
two subsets (A and B) of roughly equal size and computed separate smoothed maps 
$s_\text{A}(\alpha,\delta)$, $s_\text{B}(\alpha,\delta)$ for the subsets. 
Because the random errors are uncorrelated between A and B, while the systematics 
are the same, an unbiased estimate of the mean square systematics is obtained as
the sample covariance between the smoothed values, 
$\text{RMS}=\langle(s_\text{A}-\langle s_\text{A}\rangle)(s_\text{B}
-\langle s_\text{B}\rangle)\rangle^{1/2}$. 
Here $\langle\,\rangle$ denotes an average over the positions (for the quasars,
only $|\,b\,|>10^\circ$ was used; for the LMC, points within a radius of $4.5^\circ$).
Averaging over 50 different random divisions, we obtain the RMS values in 
Table~\ref{tab:rmsSyst}. Compared with DR2 (values in brackets), the RMS
systematics have improved by a factor 0.7 in the quasar parallaxes and 0.44 in the proper 
motions. For the small-scale parallax systematics in the LMC the improvement is a factor 0.53.

The RMS values in Table~\ref{tab:rmsSyst} for the quasars were computed using the full sample 
down to $G=21.0$ (median $G=19.9$), without taking into account that the individual 
uncertainties increase rapidly towards the faint end. This was done in order to benefit maximally 
from the large number of faint quasars in the sample. Unfortunately there are not enough 
of the brighter quasars to determine with any certainty how the systematics depend on 
magnitude, but it appears that the they improve marginally for brighter sources. For example, 
if the sample is restricted to the 16\% quasars brighter than $G=19$ (median $G=18.4$), the 
RMS systematics are 10--15\% smaller than in Table~\ref{tab:rmsSyst}.

For the LMC, the RMS values in Table~\ref{tab:rmsSyst} were computed after subtraction 
of the mean observed parallax in the area, which means that they do not include 
systematics on angular scales $\gtrsim 4.5^\circ$. This explains why the
RMS values in the last line break the increasing trends from the previous lines. 
The magnitude dependence mentioned above could also play a role here, the
LMC sources being on average brighter than the quasars, as well as the geometrically 
favourable location near the south ecliptic pole.

\begin{table}
\caption{RMS level of systematic errors in the quasar and LMC
samples at different angular scales for EDR3 and (in brackets) DR2.  
\label{tab:rmsSyst}}
\small
\begin{tabular}{cccrrrrrr}
\hline\hline
\noalign{\smallskip}
Sample && Angular && \multicolumn{2}{c}{Parallax} && \multicolumn{2}{c}{Proper motion}\\
&& scales &&  \multicolumn{2}{c}{[$\mu$as]} && \multicolumn{2}{c}{[$\mu$as~yr$^{-1}$]} \\
\noalign{\smallskip}
\hline
\noalign{\smallskip}
QSO && $>10^\circ$ && 8.1 & (11.3) && 7.7 & (17.7)\\
QSO && $>7^\circ$ && 8.7 & (12.6) &&  8.8 & (20.1)\\
QSO && $>5^\circ$ && 9.4 & (13.6) && 9.6 & (22.6)\\
QSO && $>3^\circ$ && 10.6 & (14.8) && 10.8 & (24.7)\\
QSO && $>2^\circ$ && 11.6 & (15.5) && 11.6 & (26.3)\\
QSO && $>1^\circ$ && 13.5 & (16.9) && 12.8 & (26.0)\\
QSO && $>0.5^\circ$ && 14.3 & (23.2) && 17.2 & (23.2)\\
%QSO && $>0.1^\circ$ && 11.9 & (16.9) && 17.1 & (28.3)\\
%QSO && all && 26 & (43) && 33 & (66)\\
LMC && $0.1{-}4.5^\circ$ && 6.9 & (13.1)\\
\noalign{\smallskip}
\hline
\end{tabular}
\tablefoot{The lower limit in the second column is the standard deviation of the Gaussian 
smoothing kernel; the upper limit (for LMC)
is the diameter of the area examined. Columns~3 and 4
give unbiased estimates of the RMS variations of the systematics, computed as 
described in the text. In brackets are the corresponding estimates 
for DR2, using as far as possible the same samples. For the proper motion, 
the values refer to a single component (the mean of the RMS in 
$\mu_{\alpha*}$ and $\mu_\delta$).}
\end{table}

\begin{figure*}
\centering
  \includegraphics[width=0.85\hsize]{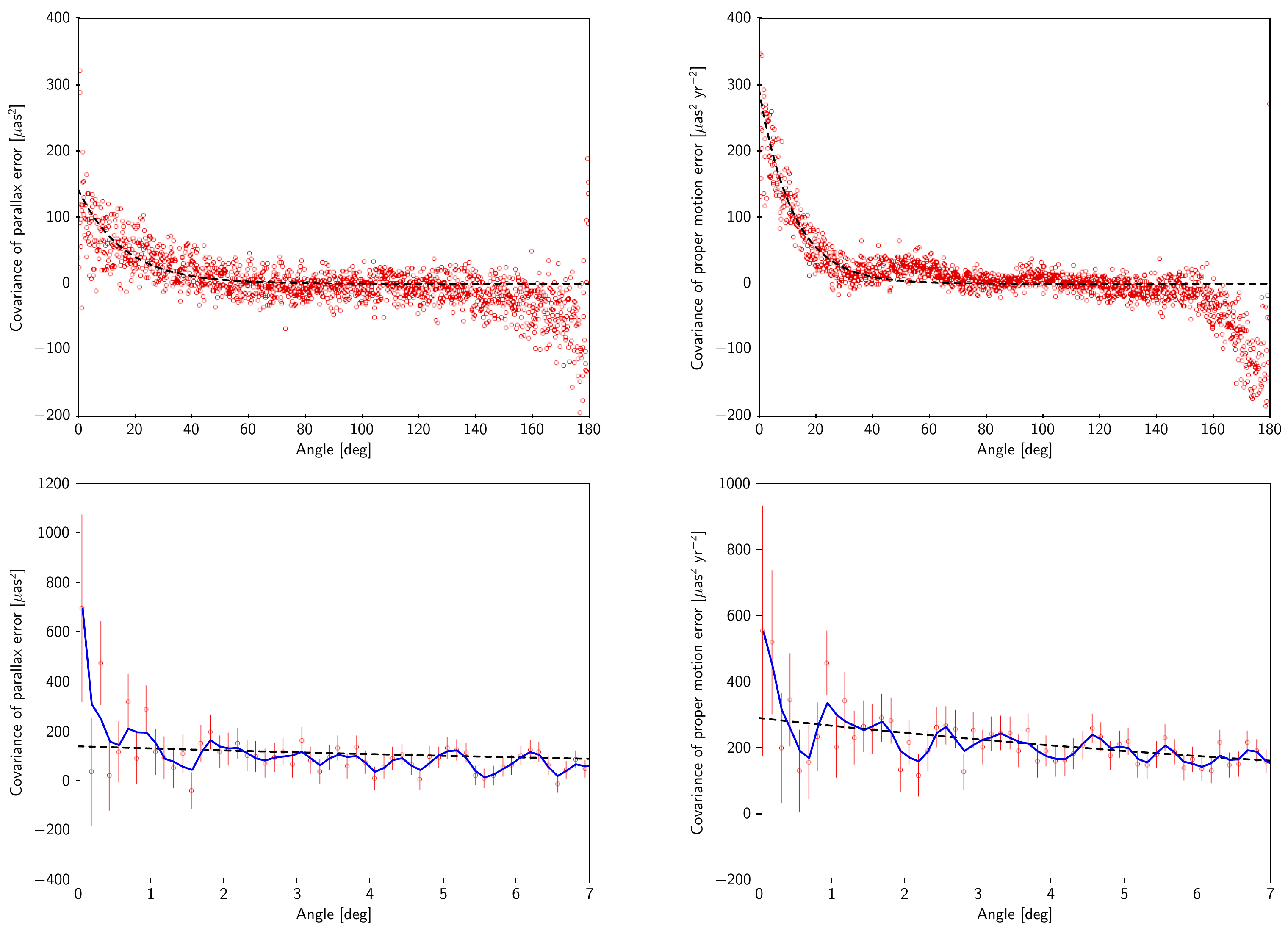}  %{Figures/covPlxPm.pdf}
    \caption{Angular covariances of the five-parameter quasar sample. 
    \textit{Left:} Covariance in parallax, $V_\varpi(\theta)$. 
    \textit{Right:} Covariance in proper motion, $V_\mu(\theta)$.
    The red circles are the individual estimates; the dashed black curves are
    fitted exponential functions. The bottom panels show the same data as in the top panels, 
    but for small separations only, with errors bars (68\% confidence intervals) and running 
    triangular mean values (blue curves).}
    \label{fig:qsoCov}
\end{figure*}

Similarly to what was done for DR2, angular covariance functions of the parallaxes and 
proper motions, $V_\varpi(\theta)$ and $V_\mu(\theta)$, have been computed for the
EDR3 quasar sample. See Sect.~5.4 in \citetads{2018A&A...616A...2L} for their definition.%
\footnote{In the DR2 paper these functions were called spatial covariance functions.
However, `angular' is a better qualifier, consistent with the established term
`angular power spectrum' and avoiding the association to (three-dimensional) spatial 
coordinates.} 
The results (Fig.~\ref{fig:qsoCov}) are qualitatively similar to the DR2 results, but
the covariances are smaller by a factor 2 to 4, consistent with other improvements. 
The black dashed curves in the upper panels are exponential fits for 
$0.5^\circ\lesssim\theta\lesssim 80^\circ$,
namely
\begin{align}\label{eq:covPlxExp}
V_\varpi(\theta) &\simeq (142~\mu\text{as}^2)\times\exp(-\theta/16^\circ) \,, \\
\label{eq:covPmExp}
V_\mu(\theta) &\simeq (292~\mu\text{as}^2\text{yr}^{-2})\times\exp(-\theta/12^\circ) \, .
\end{align}
The corresponding amplitudes for DR2 were 285~$\mu$as$^2$ and 
800~$\mu$as$^2$~yr$^{-2}$, with $e$-folding angles $14^\circ$ and $20^\circ$.
Taking the first bin ($0<\theta<0.125^\circ$) to represent the covariance of the 
systematic errors at zero separation, we have
\begin{align}\label{eq:covPlx0}
V_\varpi(0) &\simeq 700~\mu\text{as}^2 \,, \\
\label{eq:covPm0}
V_\mu(0) &\simeq 550~\mu\text{as}^2\text{yr}^{-2} \, .
\end{align}
Corresponding values for DR2 were, respectively, 1850~$\mu$as$^2$ and 
4400~$\mu$as$^2$~yr$^{-2}$. 

Both $V_\varpi(\theta)$ and $V_\mu(\theta)$
show oscillations with a period of the order of a degree, corresponding to
the checkered pattern. Consistent with the other findings, the oscillations in 
Fig.~\ref{fig:qsoCov} have significantly smaller amplitudes than their 
counterparts in DR2.

\subsection{Angular power spectrum of parallax bias}
\label{sec:aps}

A comprehensive quantification of the positional variations of systematics on all 
angular scales can be given in the form of an angular power spectrum. This section
is an attempt to estimate the power spectrum of parallax bias from EDR3 quasar data.

In astrophysics, the angular power spectrum is 
perhaps best known in the context of the cosmic microwave background
(CMB; e.g.\ \citeads{2002ARA&A..40..171H}). Any scalar field $z(\alpha,\delta)$ 
defined on the full sky
(temperature for the CMB; or in our case, the quasar parallaxes) can be 
decomposed in spherical harmonics (SH) $Y_{\ell m}(\alpha,\delta)$ as
\begin{equation}\label{eq:SH1}
z(\alpha,\delta) = \sum_{\ell=0}^\infty \sum_{m=-\ell}^\ell a_{\ell m}Y_{\ell m}(\alpha,\delta)\, ,
\end{equation}
where $\ell$ is the degree of the SH (also known as multipole), $m$ is the order (or mode),
and $a_{\ell m}$ are the (complex) coefficients
\begin{equation}\label{eq:SH2}
a_{\ell m} = \int_\Omega z(\alpha,\delta) Y_{\ell m}^*\text{d}\Omega\, .
\end{equation}
Here $^*$ is the complex conjugate and $\int_\Omega$ denotes integration over the full
sphere with solid angle element $\text{d}\Omega=\cos\delta\,\text{d}\alpha\,\text{d}\delta$.
%so that $\int_\Omega\text{d}\Omega=4\pi$. 
The equivalence of Eqs.~(\ref{eq:SH1}) and (\ref{eq:SH2}) can be verified using  
the orthonomality of the SH, 
$\int_\Omega Y_{\ell m}Y_{\ell' m'}^*\text{d}\Omega=\delta^{\ell\ell'}\delta^{mm'}$,
where $\delta^{ij}$ is the Kronecker symbol. By means of Eq.~(\ref{eq:SH1}) it is seen that 
the mean square value of $z$ on the sky is
\begin{equation}\label{eq:SH3}
\begin{split}
\overline{z^2}&=\frac{1}{4\pi}\int_\Omega |z|^2\text{d}\Omega =
\frac{1}{4\pi}\int_\Omega \left(\sum_{\ell,\, m} a_{\ell m}Y_{\ell m}\right)
\left(\sum_{\ell,\, m} a_{\ell m}^*Y_{\ell m}^*\right)\text{d}\Omega\\ 
&=\frac{1}{4\pi}\int_\Omega \sum_{\ell,\, m}|a_{\ell m}|^2 |Y_{\ell m}|^2\text{d}\Omega 
= \frac{1}{4\pi}\sum_{\ell=0}^\infty \sum_{m=-\ell}^\ell |a_{\ell m}|^2\, .
\end{split}
\end{equation}
The observed angular power spectrum, defined as 
\begin{equation}\label{eq:SH4}
C_\ell = \frac{1}{2\ell+1}\sum_{m=-\ell}^\ell |a_{\ell m}|^2\, 
\end{equation}
(\citeads{1973ApJ...185..413P}; \citeads{2003ApJS..148..135H}), thus measures the mean power
of the SH components of degree $\ell$, that is on angular scales 
\hbox{$\sim\! 180^\circ/\ell$}, integrated over the sphere. 
For our purpose it is convenient to plot the cumulative quantity
\begin{equation}\label{eq:SH5}
R(\ell_\text{max}) = \sqrt{\frac{1}{4\pi}\sum_{\ell=1}^{\ell_\text{max}} (2\ell+1)C_\ell}\, .
\end{equation}
The spherical harmonic of degree $\ell=0$, corresponding 
to the mean quasar parallax of about $-20~\mu$as, is not included in the sum. 
$R(\ell_\text{max})$ can therefore be interpreted as the RMS variation of the parallax 
systematics on angular scales $\gtrsim 180^\circ/\ell_\text{max}$.

\begin{figure}
\centering
  \includegraphics[width=\hsize]{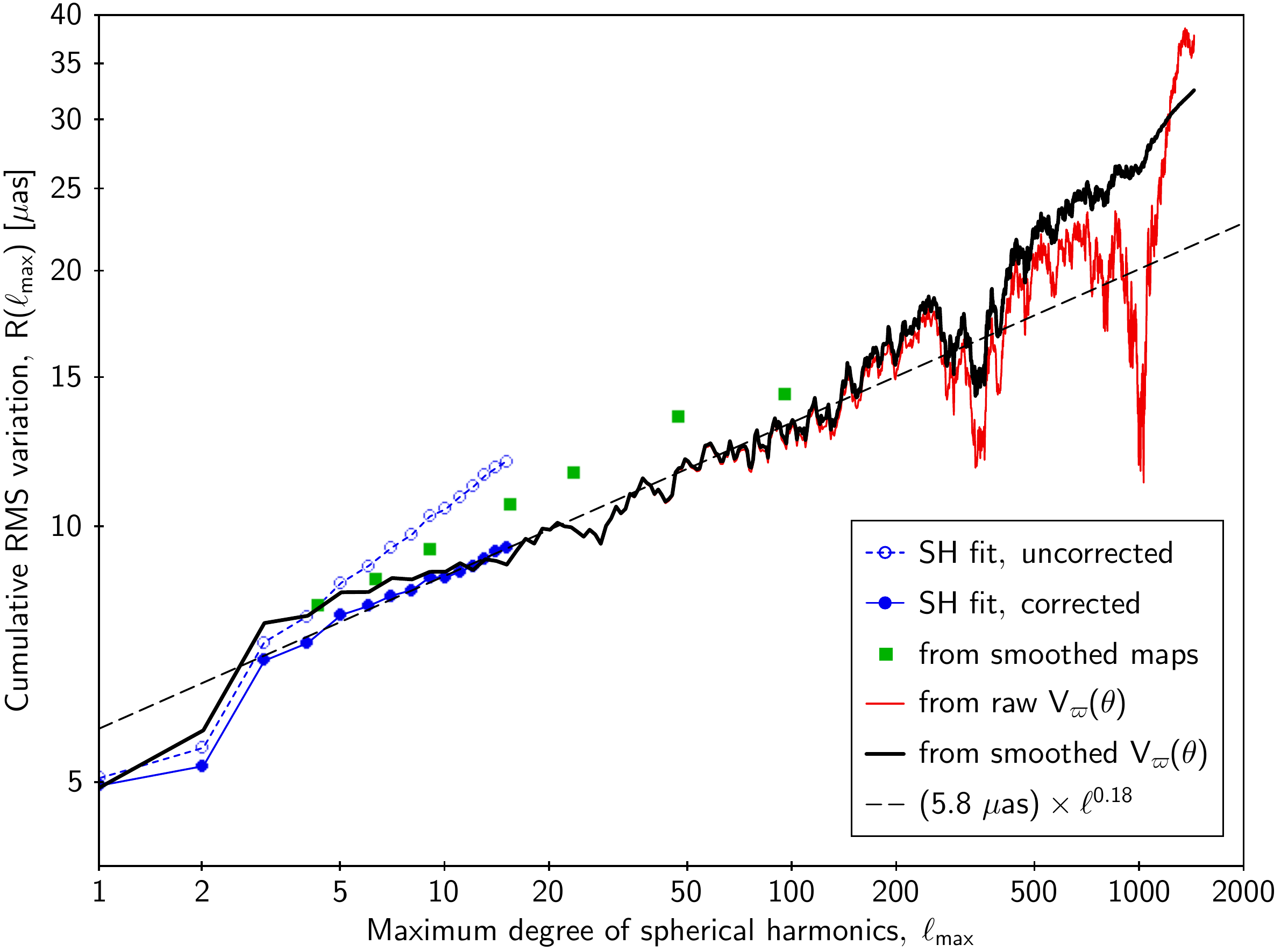}  %{Figures/apsAccNew.pdf}
    \caption{Cumulative angular power spectrum of systematics in quasar parallaxes.}
    \label{fig:caps}
\end{figure}

Figure~\ref{fig:caps} presents our estimates of the angular power spectrum of the
parallax bias in EDR3, derived from the five-parameter quasar sample using several different 
methods. Most straightforward is to determine the coefficients $a_{\ell m}$ by a weighted 
least-squares fit of Eq.~(\ref{eq:SH1}) directly to the quasar parallaxes, truncating the first 
sum at $\ell=L$ (that is, using $L^2$ unknowns). Owing to the lack of quasars at low Galactic 
latitudes ($|\sin b|<0.1$), this gives stable results only for $L\lesssim 15$ (angular scales 
$\gtrsim 12^\circ$), where the fit manages to bridge the no-data gap. 
Even so, the RMS values computed in this way (shown as open circles in
Fig.~\ref{fig:caps}) overestimate the true variations, as they include a contribution from
the random errors in the parallaxes. If the random errors are assumed to have standard
deviations equal to the formal parallax uncertainties, we can estimate their
contribution to $C_\ell$ by means of Monte Carlo simulations, and subtract from the 
power obtained in the fit. (Alternatively, the noise contributions could be estimated from
the formal variances obtained in the least-squares fits.) 
This gives the corrected RMS estimates shown as filled circles in Fig.~\ref{fig:caps}. 

The smoothed maps offer an alternative method to estimate the angular power spectrum
that is not restricted to $\ell_\text{max}\lesssim 15$, as was the case for the SH fit
owing to the no-data gap along the Galactic equator. The RMS values in Table~\ref{tab:rmsSyst}
were computed using only the smoothed points with $|\sin b|>0.1$, and are therefore 
not strongly affected by the gap.
However, the Gaussian smoothing does not correspond to a well-defined cut-off
in $\ell$, and the comparison of the smoothed RMS values with 
$R(\ell_\text{max})$ is not entirely straightforward. In Fig.~\ref{fig:caps} we have put the
RMS values from the table at the degree where the Gaussian beam transfer function,
$\exp\bigl[-(\ell+\frac{1}{2})^2\sigma^2\bigr]$ \citepads{1992PhRvD..46.4198W},
equals $\frac{1}{2}$, that is at $\ell_\text{max}\simeq 47.7^\circ/\sigma-0.5$.
The RMS values, shown as green squares, roughly continue the power-law trend of the 
corrected SH fit to higher $\ell_\text{max}$, but at RMS values that are 10--15\% higher. 
We have no explanation for this discrepancy, but conclude that the agreement is 
reasonable considering the approximations involved.  

The angular covariance function $V_\varpi(\theta)$ and the angular power spectrum $C_\ell$
contain equivalent information, and are related by the transformations \citepads{1973ApJ...185..413P}
\begin{align}
C_\ell &= 2\pi \int_{-1}^1 V_\varpi(\theta)\,P_\ell(\cos\theta)\,\text{d}(\cos\theta)\, ,\label{eq:SH6}\\
V_\varpi(\theta) &= \frac{1}{4\pi}\sum_\ell (2\ell+1)\,C_\ell\,P_\ell(\cos\theta)\, ,\label{eq:SH7}
\end{align}
where $P_\ell(x)$ are Legendre polynomials. Using the angular covariance function in
Fig.~\ref{fig:qsoCov} (left) and replacing the integral in Eq.~(\ref{eq:SH6}) by a sum over the
1440 covariance values, one can easily compute $C_\ell$ for arbitrary $\ell$. The result
is the red solid curve in Fig.~\ref{fig:caps}. Using the smoothed covariance values (blue
in Fig.~\ref{fig:qsoCov}, left) gives instead the black curve in Fig.~\ref{fig:caps}. Both curves
show some unphysical fluctuations: The cumulative RMS values cannot decrease for increasing
$\ell$. The fluctuations are caused by sampling noise and must be disregarded when 
interpreting the RMS values, although they give an impression of the statistical uncertainties
in $R(\ell_\text{max})$.

After accounting for estimation bias in the SH fit there is generally good agreement between
the different methods where they overlap. On the smallest scales, $\ell_\text{max}\gtrsim 100$, 
only the angular covariance function provides an estimate. As indicated by the dashed line 
in Fig.~\ref{fig:caps}, the overall trend can be described by a simple power law 
\begin{equation}\label{eq:SH8}
R(\ell_\text{max}) \simeq (5.8~\mu\text{as})\times \ell^{\,0.18}\, ,
\end{equation}
at least for for $\ell=3$ to ${\simeq\,}150$ (angular scales from 1.2$^\circ$ to 60$^\circ$).
For $\ell=1$ and 2 the RMS is significantly smaller than according to this relation. In particular, 
the power at $\ell=2$ (angle $\sim 90^\circ$) is remarkably small, which could be related to
the basic angle ($\Gamma=106.5^\circ$) providing a firm connection between areas
separated by angles of the order of $90^\circ$. For $\ell \gtrsim 150$ the RMS is higher 
than according to the
power law, corresponding to the `checkered pattern' in Fig.~\ref{fig:mapsLmcPlx} (left).
The value $V_\varpi(0) \simeq 700~\mu\text{as}^2$ in Eq.~(\ref{eq:covPlx0}) for
$\theta<0.125^\circ$ suggests that $R(\ell_\text{max})$ saturates at $\simeq 26~\mu$as 
for $\ell_\text{max}\gtrsim 1440$. Using a suitable (monotonic) model function
$R(\ell_\text{max})$ one can estimate the values $C_\ell$, which may be of interest 
in studies where the statistical variation of parallax bias with position is a concern.

No angular power spectrum is given for the systematics in quasar proper motions.
Large-scale systematics (for small $\ell$) are discussed in \citet{EDR3-DPACP-133} 
and \citet{EDR3-DPACP-134}.

\section{Improvements for \textit{Gaia} DR4 and beyond}
\label{sec:impr} 

Although EDR3 brings huge improvements over DR2 in terms of the overall quality of the astrometric
results, including systematics, it is obvious that the limits of \textit{Gaia}'s capability have not
been reached. At the time of writing (September 2020), \textit{Gaia} has already accumulated more 
than twice the amount of observations included in EDR3, and a solution using all these data, even 
without any improved modelling, will almost certainly bring down the random errors by a factor 0.7 for
the positions and parallaxes, and by a factor 0.35 for the proper motions. Some of the
systematics will also be reduced simply from the improved coverage in time and scan directions. 
But important advances will also come from dedicated efforts to improve and consolidate
the calibration models. In this context it is positive that many model deficiencies stand 
out clearly in the residuals (e.g.\ Appendix~\ref{sec:residT} and \ref{sec:residG}): It shows 
that the modelling errors are not degenerate with the astrometry and can be used to design 
better models. Below we list some areas where significant advances should 
be possible as a result of further model developments and analysis of the data.

\subsection{LSF and PSF modelling}
\label{sec:imprLSF}
 
The processing of CCD samples in CALIPD is intimately connected to the AGIS calibrations through 
the many instrumental effects that influence the shape and location of the image profiles. 
As detailed by \citet{EDR3-DPACP-73}, significant improvements to the LSF and PSF modelling are 
being implemented or planned for DR4 and subsequent cycles. These include improved basis 
functions for the LSF and PSF modelling, with an analytical representation of AL and AC smearing 
effects and a clear separation between parameters representing a shift of the profile and 
its shape; modelling of magnitude-dependent non-linear effects, caused for example by CTI; 
and a new bootstrapping of the attitude and geometric calibrations that will make the initial
CALIPD in a cycle more independent of the previous cycle, thus reducing the risk of 
propagating systematics from one cycle to the next. It is expected that these many developments
will further reduce systematics in the location and flux estimates per CCD coming from the IPD.
This will benefit the astrometry and photometry of all sources, but in particular the bright 
ones ($G\lesssim 13$).

\subsection{CTI modelling}
 
As shown in Appendix~\ref{sec:residG}, the AL and AC residuals exhibit strong trends with
magnitude that can be interpreted as a manifestation of CTI. CTI effects are also diagnosed, 
but not corrected, by the CTI calibration parameters displayed in Fig.~\ref{sec:residG}.
The effects may already have some impact on the astrometry, and this will become more
important with time as the accumulated radiation dose increases while random errors
in the astrometric results continue to improve. Whereas ultimately a detailed, physics-based 
modelling of CTI effects is desirable (\citeads{2011MNRAS.414.2215P}; \citeads{2013MNRAS.430.3078S}),
much progress is still possible simply by better mapping of the empirical effect as a function 
of the main variables of the observation. On this macroscopic level, CTI is comparatively easy 
to separate from other effects, thanks to the periodic charge injection, and a wealth of additional 
calibration data are available to support the modelling \citepads{2016A&A...595A...6C}. 

\subsection{Time variations}

With the exception of the spin-related (quasi-periodic) distortion discussed in Sect.~\ref{sec:glob},
all instrument calibrations are assumed to be constant or at most linearly varying over a time interval
that could be as long as 63~days. For the AL large-scale geometry the time interval is at most 3~days.
The residual normal points in Fig.~\ref{fig:rnpAllTimeNoGc1} show that this resolution is adequate 
most of the time, but not always and especially not after thermal upsets such as the (partial) eclipses
of the Sun by the Moon. Clearly a quadratic model (or perhaps an exponential model with fixed 
time constant) 
would be a big improvement at these times. The spin-related periodicity of the residuals, visible 
along most of the time axis in the diagram, may be removed by the more extended VBAC/FOC 
modelling hinted at in Sect.~\ref{sec:glob}.

\subsection{AC rate dependency}
\label{sec:acRate}

The AC rate has a big impact on the PSF by smearing the images in the AC direction during a 
CCD exposure. The width of the PSF in the AC direction is increased by the smearing, and the 
number of photoelectrons per pixel is reduced, which affects saturation and CTI. The AC rate is 
therefore relevant for the calibration of both bright and faint sources. To first order, the smearing 
is proportional to $|\,\dot{\zeta}\,|^2$, which is the only dependence considered in the AGIS calibration 
model for EDR3 (effect~7 in Table~\ref{tab:cal}). Owing to the non-linearity of saturation and CTI 
effects, it is likely that the AL centroid biases they produce also have a component that is linear in
$\dot{\zeta}$. Such terms were not included in the AL calibration model for EDR3, as they might 
be difficult to disentangle from the parallax in view of the correlation between $\dot{\zeta}$ and 
the AL parallax factor described in Appendix~\ref{sec:plxAc}. This correlation is positive throughout 
the data segments used for EDR3, but negative in data segments DS6 and DS7 where reversed 
precession was used. The reversed precession during one year (2019.536--2020.576; 
cf.\ Fig.~\ref{fig:timeline}) was introduced precisely to address this and similar issues related 
to the non-symmetry of the nominal scanning law. Together with the much improved 
PSF modelling mentioned above, this should allow the main effects of AC rate variations to 
be resolved already in the data analysis for DR4, which includes data segment DS6 obtained 
in reversed precession mode.

\subsection{Use of colour information}

A rather unsatisfactory aspect of EDR3 is the division of sources with full astrometric information 
into two distinct subsets, namely the five- and six-parameter solutions. This was a consequence
of the unavailability of good colour information for some sources, which necessitated
their special treatment in IPD and AGIS. Although much more and better colour information
from the BP and RP spectra is available for DR4 through PhotPipe~3 (cf.\ Fig.~\ref{fig:timeline}), 
it is unavoidable that this information is missing or of poor quality for some sources. A more uniform 
treatment of all sources in IPD and AGIS can be achieved by consistently using the available
colour information, weighted according to its uncertainty. For AGIS this means that all 
sources obtain six-parameter solutions, but with available BP and RP data used as prior for the 
pseudocolour. 

\section{Conclusions}
\label{sec:concl}

Compared with \textit{Gaia} DR2, the number of sources in EDR3 that have
a parallax and proper motion is only 10\% higher. However, the average improvement 
on the standard uncertainties is roughly a factor 0.8 for the positions and parallaxes, 
and 0.5 for the proper motions. These factors reflect the higher number of observations
per source, by more than 50\% on average, and the longer time span of the data, which 
make the astrometric results 
considerably more robust and help to reduce systematic errors. The astrometric solution 
for EDR3 is also the
first one in the cyclic processing of DPAC to benefit from a full reprocessing of
the LSF and PSF calibrations and the image parameter determination. 
The next full-scale astrometric solution, for \textit{Gaia} DR4,
will be based on twice as many observations as EDR3. Considerable efforts are required
and planned to ensure a matching development of models and analysis methods.

\begin{acknowledgements}

We thank the anonymous referee for constructive comments on the manuscript,
and C.~Babusiaux for pointing out a significant error in the original version.
This work has made use of data from the European Space Agency (ESA) mission
{\it Gaia} (\url{https://www.cosmos.esa.int/gaia}), processed by the {\it Gaia}
Data Processing and Analysis Consortium (DPAC,
\url{https://www.cosmos.esa.int/web/gaia/dpac/consortium}). Funding for the DPAC
has been provided by national institutions, in particular the institutions
participating in the {\it Gaia} Multilateral Agreement.
This work was financially supported by 

the European Space Agency (ESA) in the framework of the \textit{Gaia} project;

the German Aerospace Agency (Deutsches Zentrum für Luft- und Raumfahrt e.V., DLR) 
through grants 50QG0501, 50QG0601, 50QG0901, 50QG1401 and 50QG1402; 

the Spanish Ministry of Economy (MINECO/FEDER, UE) through grants ESP2016-80079-C2-1-R, 
RTI2018-095076-B-C21 and the Institute of Cosmos Sciences University of Barcelona 
(ICCUB, Unidad de Excelencia `Mar{\'i}a de Maeztu') through grants MDM-2014-0369 and CEX2019-000918-M;

the Swedish National Space Agency (SNSA/Rymdstyrelsen);

and
the United Kingdom Particle Physics
and Astronomy Research Council (PPARC), the United Kingdom Science
and Technology Facilities Council (STFC), and the United Kingdom Space
Agency (UKSA) through the following grants to the University of Bristol,
the University of Cambridge, the University of Edinburgh, the University
of Leicester, the Mullard Space Sciences Laboratory of University College
London, and the United Kingdom Rutherford Appleton Laboratory (RAL):
PP/D006511/1, PP/D006546/1, PP/D006570/1, ST/I000852/1, ST/J005045/1,
ST/K00056X/1, ST/K000209/1, ST/K000756/1, ST/L006561/1, ST/N000595/1,
ST/N000641/1, ST/N000978/1, ST/N001117/1, ST/S000089/1, ST/S000976/1,
ST/S001123/1, ST/S001948/1, ST/S002103/1, and ST/V000969/1.

The authors gratefully acknowledge the use of computer resources from MareNostrum, and the 
technical expertise and assistance provided by the Red Espa{\~n}ola de Supercomputaci{\'o}n 
at the Barcelona Supercomputing Center, Centro Nacional de Supercomputaci{\'o}n.

We thank the Centre for Information Services and High Performance Computing (ZIH) at the Technische Universität (TU) Dresden for generous allocations of computer time.

Diagrams were produced using the astronomy-oriented data handling and visualisation software 
TOPCAT \citepads{2005ASPC..347...29T}.
%
%This research has made use of the SIMBAD database, operated at CDS, Strasbourg, France.
%
%We thank A.G.A.~Brown and C.~Jordi for valuable feedback during the preparation of this paper, and 
%the referee, V.V.~Makarov, for constructive comments on the original version of the manuscript.

\end{acknowledgements}

\bibliographystyle{aa} % style aa.bst
\bibliography{refs} % your references refs.bib

\appendix

\section{Properties of the astrometric solution}
\label{sec:propSol}

This Appendix illustrates properties of the primary astrometric solution in AGIS~3.2 that 
cannot be derived from the published \textit{Gaia} EDR3 results but require access
to (unpublished) data internal to AGIS, such as calibration data and the residuals of 
individual CCD observations. Obviously, only a very limited selection from the available 
material can be shown.

\subsection{Dispersion of residuals}
\label{sec:resid}

\begin{figure}
\centering
  \includegraphics[width=8cm]{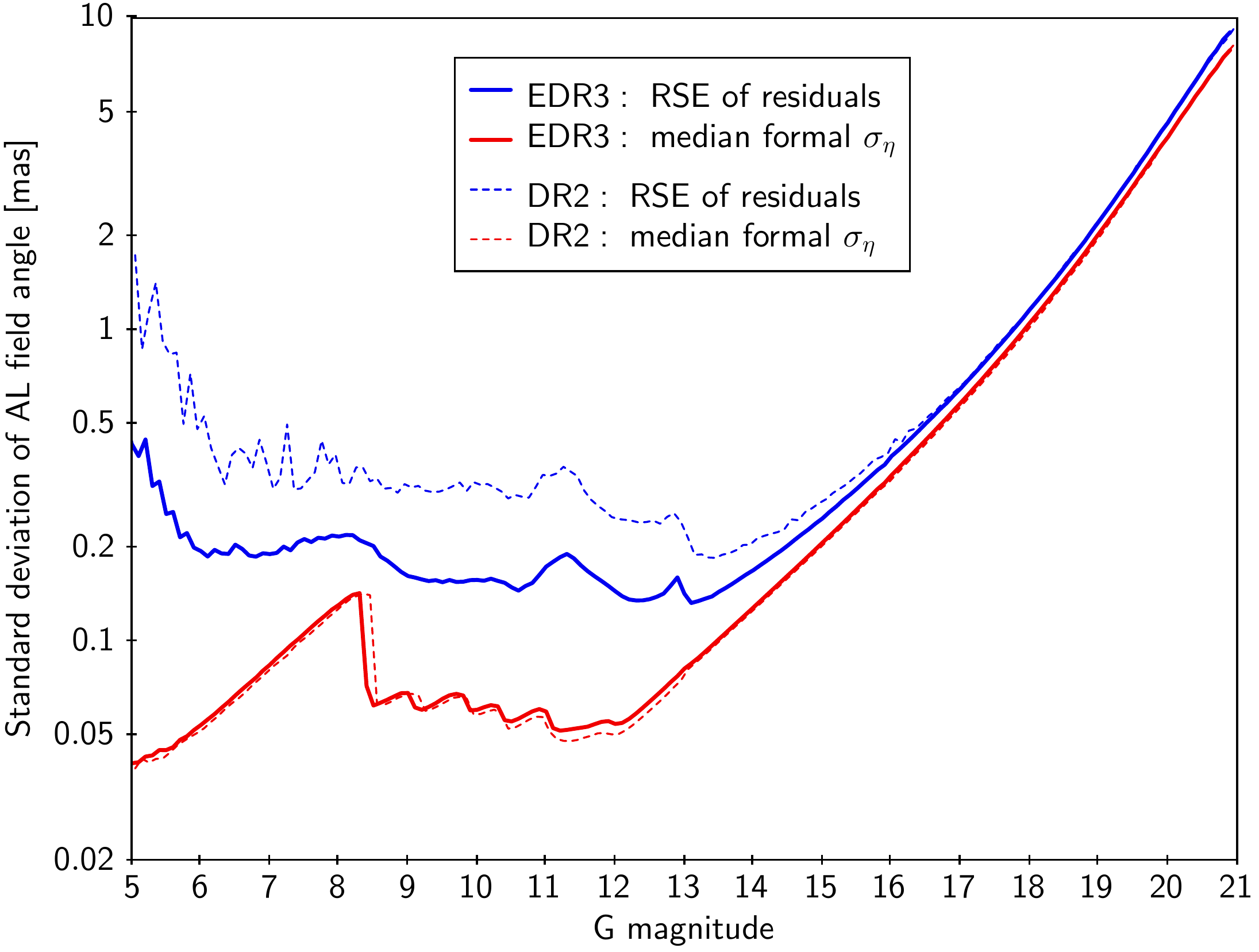}  %{Figures/alSdVsG.pdf}
  \caption{Precision of along-scan astrometric measurements as a function of magnitude. 
  Solid curves are for EDR3, dashed for DR2. The red (lower) curves show the median formal 
  precisions from the image parameter determination; the blue (upper) curves are robust 
  estimates$^{\ref{fn:RSE}}$ of the actual standard deviations of the post-fit residuals.}
  \label{fig:alSdVsG}
\end{figure}

In Fig.~\ref{fig:alSdVsG} we compare the photon-statistical uncertainties of the individual 
AL angular measurements with the scatter (RSE%
\footnote{The robust scatter estimate (RSE) is defined as $\bigl[2\sqrt{2}\,\text{erf}^{-1}(4/5)\bigr]^{-1}\approx0.390152$ times the difference between 
the 90th and 10th percentiles of the distribution of the variable. For a Gaussian distribution it 
equals the standard deviation. The RSE is used as a standardised, robust measure of dispersion 
in CU3.\label{fn:RSE}}%
) of post-fit residuals in the astrometric solution. 
For convenience, the corresponding curves for the DR2 astrometry 
(Fig.~10 in \citeads{2018A&A...616A...2L}) are shown by the dashed curves. 
While the formal precision of the individual observations is practically unchanged from DR2, 
the actual residuals have been reduced roughly by a factor two for $G\lesssim 13$, thanks 
to the improved calibration models in IDU and AGIS. This will surely continue to improve in 
future releases. For the fainter magnitudes, the improvement is successively smaller; 
for $G\gtrsim 17$ it is negligible because the residuals are completely dominated by 
photon-statistical errors.

Plotting the along-scan residuals of the individual observations from the primary solution 
(Sect.~\ref{sec:prim}) versus quantities such as time and magnitude is a powerful way to check the 
modelling of attitude and calibration in AGIS. For meaningful results, it is usually necessary to 
divide the data according to different categories such as FoV, CCD, window class, and gate. 
As the modelling errors are typically much smaller than the random errors, it is also necessary 
to reduce the random scatter, for example by plotting mean or median values. In the following 
sections we give examples of such plots versus time, magnitude, and AC rate, illustrating some 
known inadequacies of the calibration models used for EDR3. 
%More plots can be found in the on-line documentation of \textit{Gaia} EDR3 (refXX).

\subsection{Mean residual versus time}
\label{sec:residT} 
 
\begin{figure*}
\center
  \includegraphics[width=17cm]{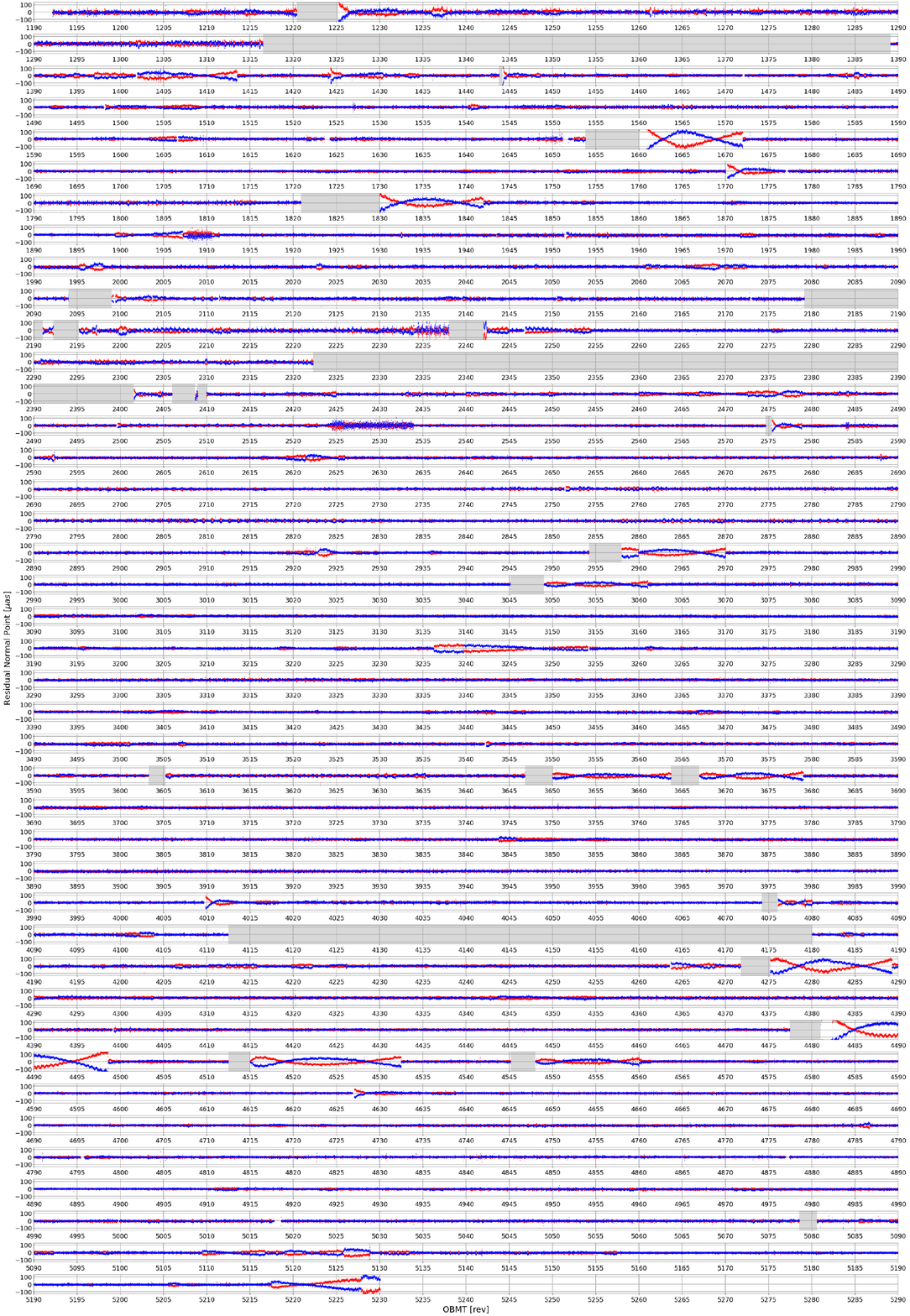}
    \caption{Residual normal points vs.\ time (OBMT). Blue and red points are, respectively, for the 
  PFoV and FFoV. Each of the 41 rows covers a time interval of 100 revolutions 
  or 25 days. The grey areas correspond to the gaps in Table~\ref{tab:gaps}.}
  \label{fig:rnpAllTimeNoGc1}
\end{figure*}

Figure~\ref{fig:rnpAllTimeNoGc1} shows residual normal points, separately for the two FoVs, 
for the entire time interval covered by the solution. The normal points are weighted averages of 
the AL residuals in the AGIS~3.2 primary solution, calculated in time bins of 87~s using the same 
weights as in the solution (Eq.~62 in \citeads{2012A&A...538A..78L}). All residuals were used, except
those in window class WC0b, which have a distinctly different (and worse) behaviour than the
other window classes. The mean number of residuals per (non-empty) bin is $\sim\,$2800, yielding a
statistical uncertainty of about 4~$\mu$as per normal point.
The figure shows at a glance not only the major gaps in the data (cf.\ Table~\ref{tab:gaps}), but also 
specific intervals where the modelling was clearly inadequate. By zooming in on the plot, a wealth of 
interesting details can be seen. Most conspicuous are the large (up to $\pm 100~\mu$as) systematic 
differences between the preceding and following FoVs seen for example after the phased array
antenna anomalies (e.g.\ for OBMT 1661--1672~rev) and eclipses by the Moon 
(e.g.\ for OBMT 2958--2970~rev),
where the AL large-scale calibration model (effect~1 in Table~\ref{tab:cal}), assuming linear variations
over an interval of 3~days (12~rev), cannot represent the non-linear behaviour of the instrument while
it is striving towards thermal equilibrium. (Not much of this effect is seen after the decontaminations, 
which are much more severe thermal upsets, because data were discarded in a much longer interval 
after these events.) The distinctly higher noise around OBMT~1908--1911 and 2525--2534~rev 
coincides
with intervals where the corrective attitude (Sect.~\ref{sec:attitude}) was missing because of a 
processing error. The overall dispersion of the normal points, as measured by the RSE,
is 14.9~$\mu$as in the PFoV and 16.6~$\mu$as in the FFoV. The
slightly better performance in the PFoV is a common feature in much of the
\textit{Gaia} data (see, for example, several plots in \citeads{EDR3-DPACP-73}). 
At most times a small residual of the 6~h and 3~h basic angle variations
can be seen.

\begin{figure*}
\center
  \includegraphics[width=17cm]{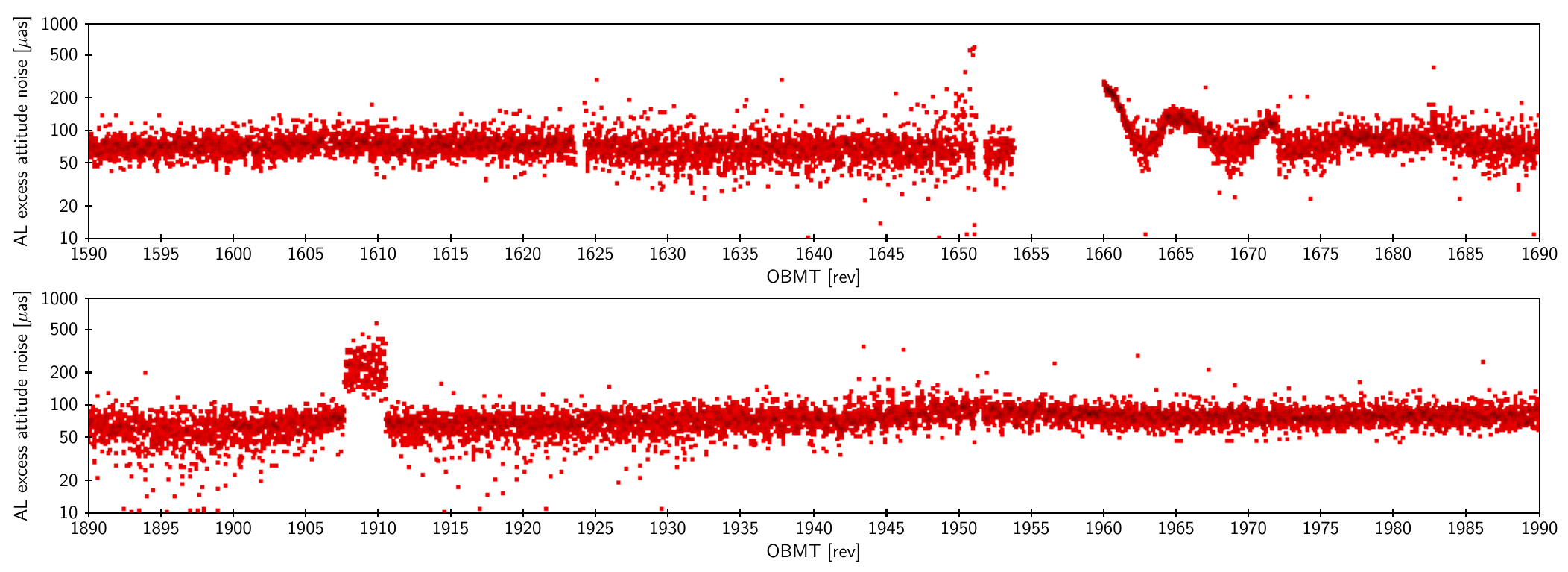}
  \caption{Excess attitude noise in two 25-day intervals.}
  \label{fig:ean}
\end{figure*}

The increased residuals at certain times, shown in Fig.~\ref{fig:rnpAllTimeNoGc1},
are reflected in the AL excess attitude noise, which is the mechanism in AGIS for applying
a time-dependent adjustment of the statistical weight of observations.
(As explained in Sect.~3.6 of \citeads{2012A&A...538A..78L}, the excess noise is 
the additional RMS noise that must be postulated in the AL error budget 
in order to account for the post-fit residuals. It consists of two parts: the 
excess source noise, which is linked to a particular
source, and the excess attitude noise, which is linked to a particular time. While the 
excess attitude noise is meant to represent attitude modelling errors, it can just as well 
represent calibration errors that affect the observations of all sources at a given time.)
This is illustrated in Fig.~\ref{fig:ean}, where the AL excess attitude noise is shown 
versus time for two 25-day intervals. In the top panel, which is the same time interval 
as row five in Fig.~\ref{fig:rnpAllTimeNoGc1}, the excess attitude noise is seen to 
exactly mirror the amplitude of the residual normal points at OBMT~1660--1672~rev.
In the bottom panel of Fig.~\ref{fig:ean}, which corresponds to row eight in 
Fig.~\ref{fig:rnpAllTimeNoGc1}, the absence of a corrective attitude at 
OBMT~1908--1911~rev triples the excess attitude noise compared with neighbouring 
times. A 6~h (or 3~h) periodicity is very often apparent in the excess attitude noise, 
as in OBMT~1940--1950~rev.  
The overall median AL excess attitude noise in EDR3 is $76~\mu$as, which represents
the average total instrument modelling error for WC1 observations.

\begin{figure*}
\center
  \includegraphics[width=17cm]{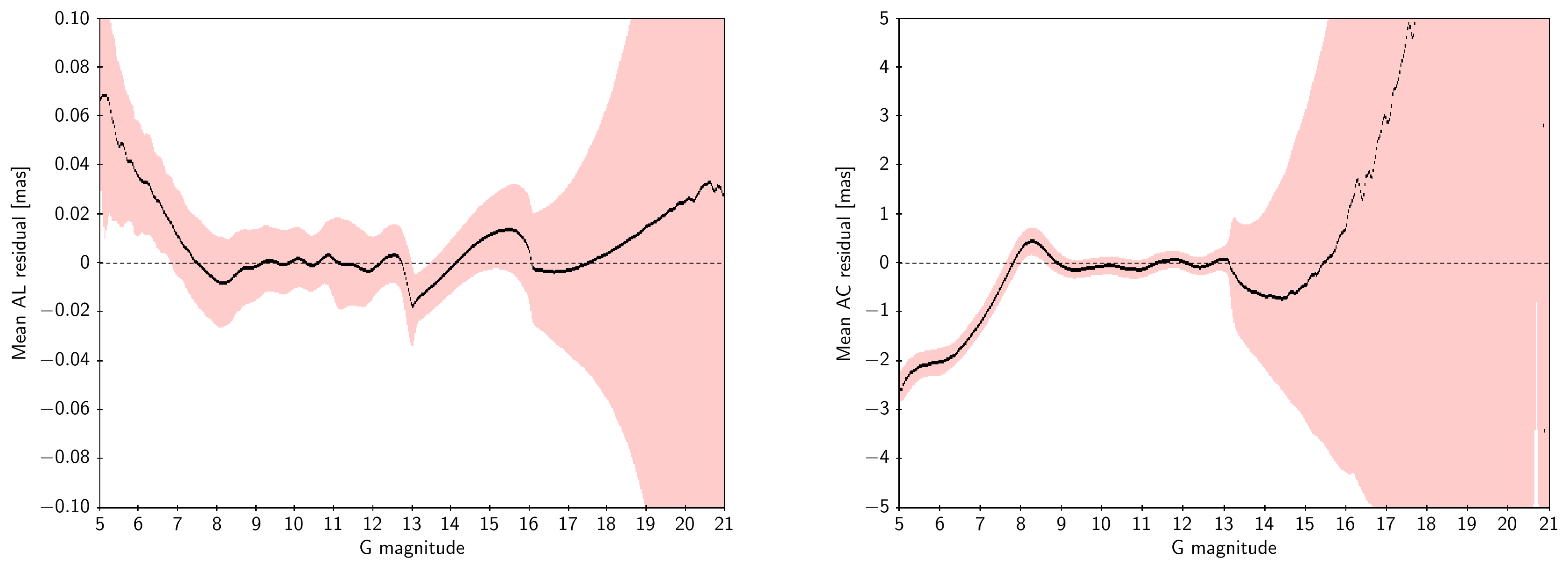}
  \caption{Mean residual AL (left) and AC (right) vs.\ magnitude for the primary sources.
  A mean residual (AL and AC) is computed for each source. The black curve is the median of
  these values, and the shaded areas indicates their 16th and 84th percentiles.}
  \label{fig:meanAlAcResVsG}
\end{figure*}

\begin{figure*}
\center
  \includegraphics[width=17cm]{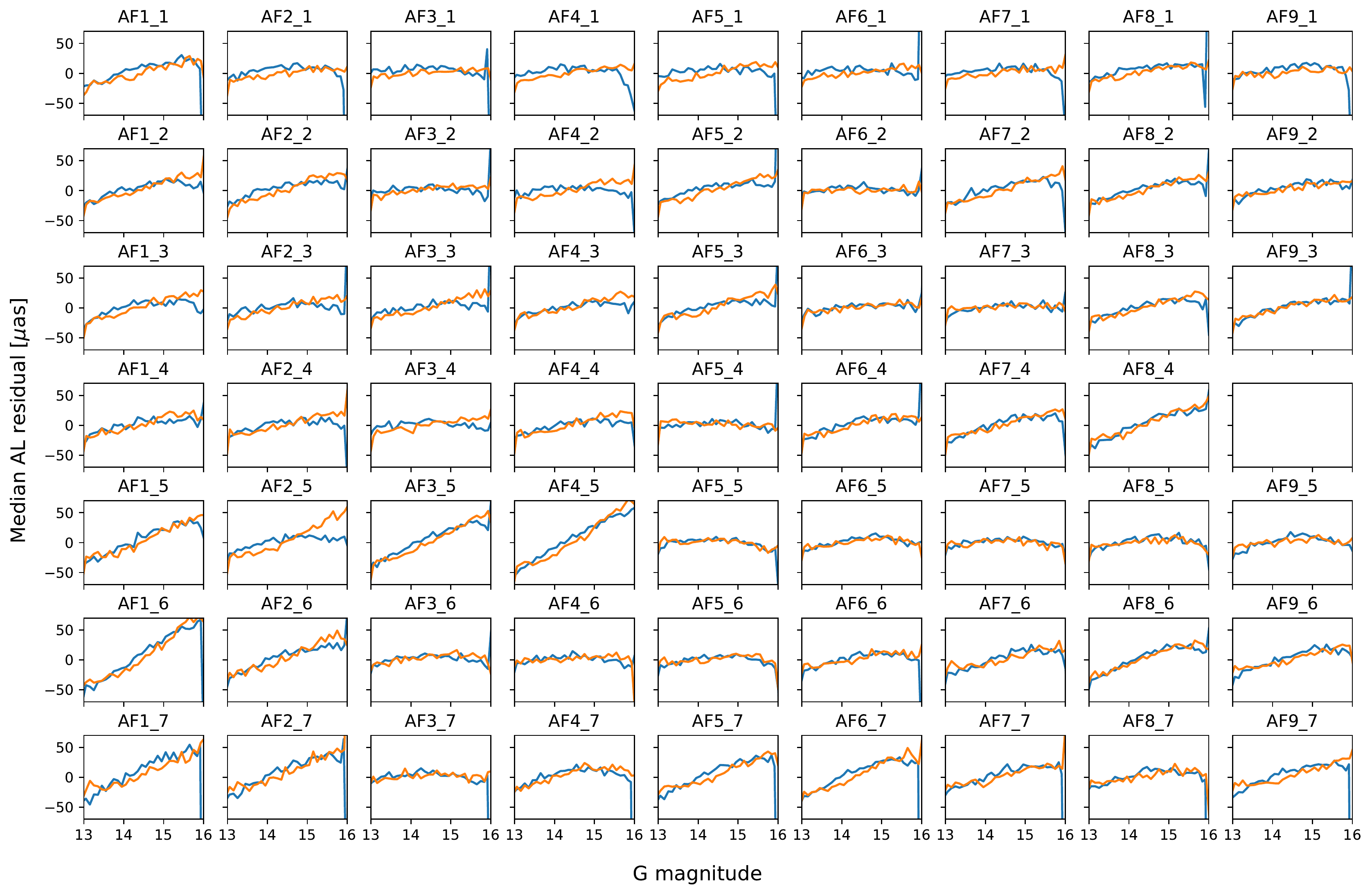} % copied from report, p.33
  \caption{Median AL residual for WC1 vs.\ $G$ (in the range 13--16)
  for each of the 62 CCDs in the AF. Blue and orange curves are for the 
  PFoV and FFoV. See Fig.~\ref{fig:fpa} for the labelling of the CCDs. 
  The layout is mirror-reversed compared to Fig.~\ref{fig:fpa}.}
  \label{fig:gMagDep}
\end{figure*}

\subsection{Mean residual versus magnitude}
\label{sec:residG}

The left panel of Fig.~\ref{fig:meanAlAcResVsG} shows the mean AL residual versus $G$ magnitude 
for the 14.3~million primary sources in AGIS~3.2 (Sect.~\ref{sec:prim}). The plot actually shows 
quantiles of the mean residual per source, so the dispersion indicated by the shaded area is 
nearly 20 times smaller than the dispersion of individual residuals shown in Fig.~\ref{fig:alSdVsG} 
(on average there are 363 AL observations per source).

The right panel of Fig.~\ref{fig:meanAlAcResVsG} is the corresponding plot for the AC residuals.
AC observations require two-dimensional windows (WC0), which are normally used only for sources
brighter than $G\simeq 13$. Only occasionally do the fainter sources by chance get 
two-dimensional windows and AC observations, which explains the sudden increase in the 
dispersion at that magnitude: The mean number of AC observations per source is 
357 for $G<13$ and 0.5 for $G>13$. The AC calibration is reasonably good in the magnitude 
interval 9--13, which includes most of the AC observations needed for the attitude determination. 

The mean AL residual is non-zero on a level of a few tens of $\mu$as, with clear and strong 
trends versus magnitude. Discontinuities are seen at the WC0/1/2 boundaries at $G=13$ and 16.
It is likely that CTI is a major factor in producing these systematics. This effect is expected to
produce a delay of the charge packages transported along the CCDs, creating a positive 
bias in the observed AL field angle $\eta$ that generally increases with magnitude. 
This is consistent with the main trends seen
within WC1 ($G=13$--16) and WC2 ($G>16$). For $G<13$ the situation is more complex
because of the gates and (for $G\lesssim 8$) the partial saturation of images. An interpretation 
in terms of CTI is supported by the similarity of the effect in the two FoVs, in spite of
the considerable variation among the different CCDs (Fig.~\ref{fig:gMagDep}). This suggests that 
the effect is not primarily driven by the shape of the LSF or PSF, which is usually quite different in the two 
FoVs, but by intrinsic properties of the CCDs.

\subsection{Astrometric calibration}
\label{sec:calData}

Of the various calibration effects summarised in Table~\ref{tab:cal}, only selected results on 
the AL chromaticity and CTI effect are shown here and briefly commented on.

\begin{figure*}
\centering
  \includegraphics[width=\hsize]{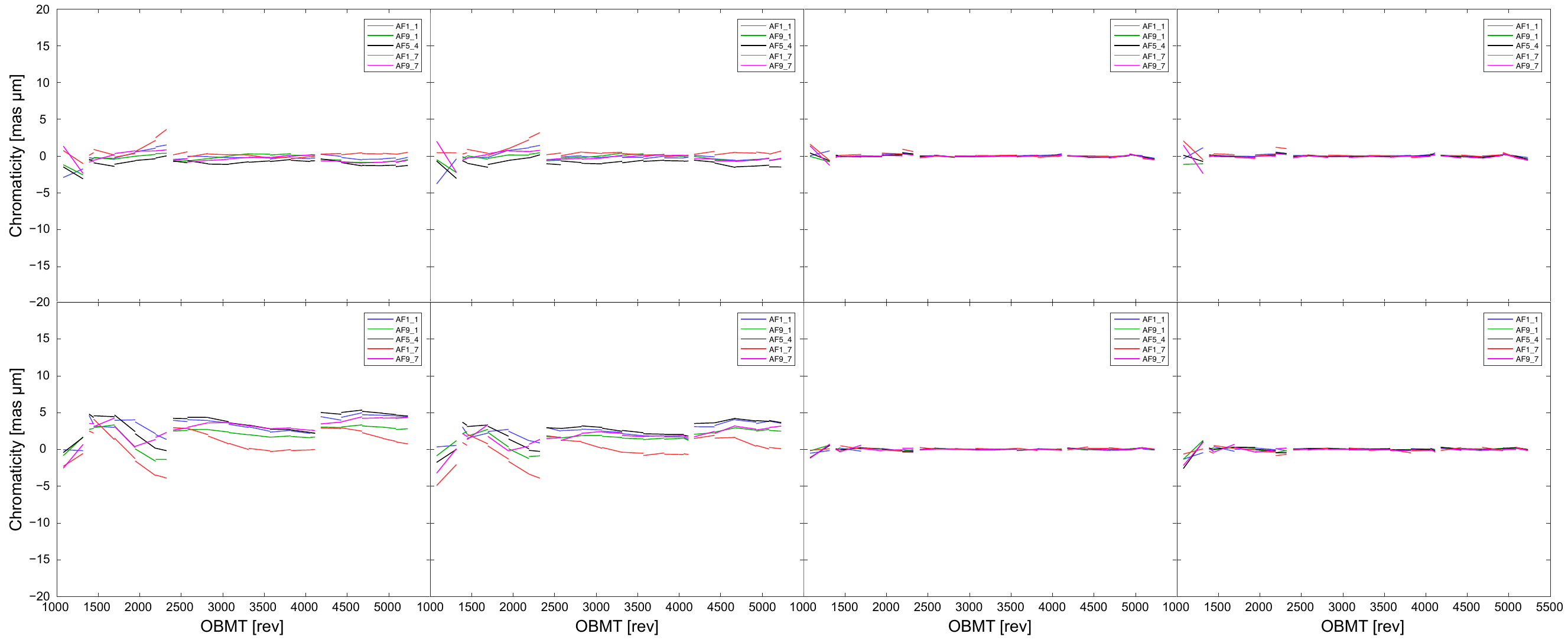}  %{Figures/effAlChrom.pdf}
    \caption{Chromaticity calibration for image parameters based on $\nu_\text{eff}$ 
    from the photometric processing (PhotPipe). This is the calibration used for sources with
    a five-parameter solution (calculated in step~\ref{step:prim} of Sect.~\ref{sec:steps}).
    \textit{Top:} Preceding FoV. \textit{Bottom:} Following FoV. From left to right: WC0a, WC0b, WC1, WC2.
    Each diagram shows the development of the chromaticity term for the five
    CCDs indicated in the legends. The chromaticity correction in IDU was very 
    successful for WC1 and WC2 (i.e.\ $G\gtrsim 13$~mag), but only partially so for 
    brighter sources.}
    \label{fig:chrom}
\end{figure*}

\begin{figure*}
\centering
  \includegraphics[width=\hsize]{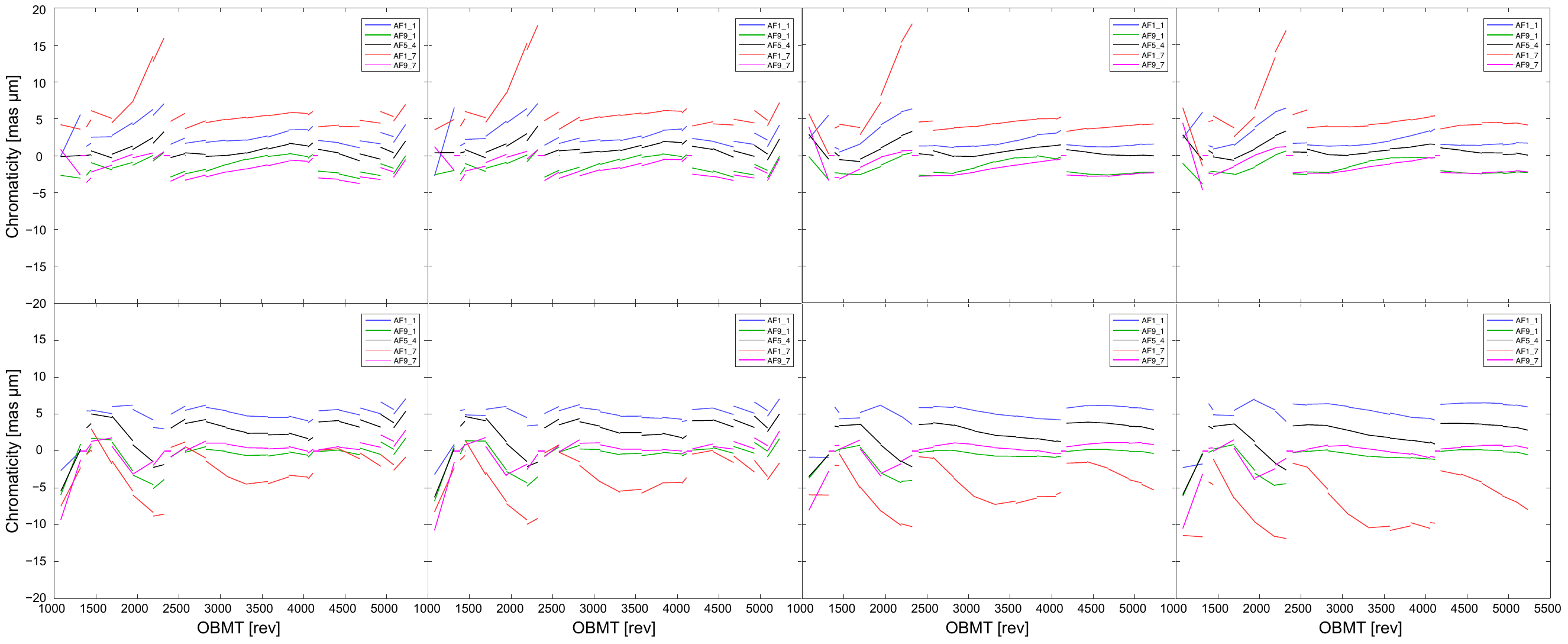}  %{Figures/effAlChromDef.pdf}
    \caption{Chromaticity calibration for image parameters based on default 
    $\nu_\text{eff}=1.43$~$\mu$m$^{-1}$. This is part of the calibration C$'$ calculated in 
    step~\ref{step:Cprime} of Sect.~\ref{sec:steps}, that is for sources that obtain six-parameter 
    solutions in AGIS. 
    \textit{Top:} Preceding FoV. \textit{Bottom:} Following FoV. From left to right: WC0a, WC0b, WC1, WC2.
    Each diagram shows the development of the chromaticity term for the five
    CCDs indicated in the legends.}
    \label{fig:chromDef}
\end{figure*}

Figures~\ref{fig:chrom} and \ref{fig:chromDef} show the AL large-scale colour calibration 
(effect~3 in Table~\ref{tab:cal})
as a function of time for five of the CCDs (in the centre of the AF and in the four 
corners), with separate plots for the four window classes (left to right) and two FoVs (top and 
bottom). Figure~\ref{fig:chrom} shows the AGIS calibration for observations where the IPD used colour 
information ($\nu_\text{eff}$) from PhotPipe to remove chromatic variations already before the data 
reached AGIS. Ideally, therefore, the remaining chromaticity found by AGIS should be negligible. 
As shown by the figure, this is almost the case for the one-dimensional images (WC1 and WC2), but 
not for the two-dimensional windows (WC0a and WC0b) used for the bright ($G\lesssim 13$) sources. 
Thus CALIPD was not fully successful in removing chromaticity by means of the PSF calibrations,
while the process worked very well for the LSF calibrations. This is one manifestation of several issues 
with the cycle~3 PSF modelling that will be resolved in the next cycle (Sect.~\ref{sec:imprLSF}
and \citeads{EDR3-DPACP-73}).

Figure~\ref{fig:chromDef} shows the corresponding AGIS calibration for observations where IPD used 
the LSF and PSF calibrations for the default colour $\nu_\text{eff}=1.43$~$\mu$m$^{-1}$. Here the
chromaticity is much stronger than in Fig.~\ref{fig:chrom} and largely similar for all four window 
classes. This figure thus illustrates intrinsic properties of the PSF while differences in the data 
processing, for example between the one- and two-dimensional windows, play a minor role. 
The calibrations are substantially different between the preceding and following fields, because 
they have different optical paths through most of the instrument and consequently different 
wavefront aberrations for a given CCD. The efficacy of the CALIPD 3.2 LSF calibration in removing 
chromaticity is striking when comparing the right-hand sides of Figs.~\ref{fig:chrom} and \ref{fig:chromDef}.

\begin{figure*}
\centering
  \includegraphics[width=\hsize]{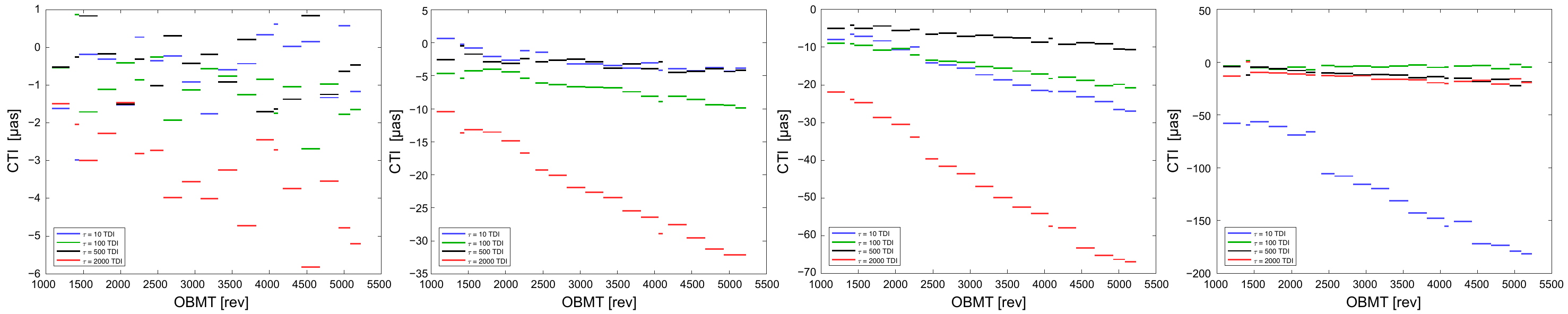}  %{Figures/effAlCti.pdf}
    \caption{CTI calibration averaged over CCDs and FoVs. From left to right: WC0a ($G\lesssim 11$), 
    WC0b ($11\lesssim G \lesssim 13$), WC1 ($13\lesssim G\lesssim 16$), and WC2 ($16\lesssim G$).
    Each diagram shows the development of the coefficients of $\exp(-\Delta t/\tau)$, where $\Delta t$
    is the time since last charge injection, for the time constants $\tau$ indicated in the legends.}
    \label{fig:cti}
\end{figure*}

Figure~\ref{fig:cti} shows the development of the AL large-scale CTI (effect~6 in Table~\ref{tab:cal}).
CTI effects are caused by the complex interaction between the build-up of charge images in the CCDs
during the TDI and the radiation-induced defects (charge traps) in the silicon lattice 
\citepads{2016A&A...595A...6C}. While the lattice defects are of course the same in the two FoVs, 
PSF shapes are different, which causes subtle differences between the FoVs in the observed CTI effects. 
These differences are generally much smaller than the calibration uncertainties, and in order to reduce 
the latter we have chosen to display in Fig.~\ref{fig:cti} only the effect averaged over the 
CCDs and the two FoVs. 
In the left-most plot (WC0a), observations are usually gated with the integration time 
reduced to less than a quarter of the maximum value, and the charge images reach close to the 
full-well capacity, or even saturate, at the end of the integration. All of these factors combine to
make the average CTI effects very small in WC0a ($\lesssim 5~\mu$as). Only for the slowest traps
($\tau = 2000$~TDI periods) does the effect become clearly stronger with time. For WC0b and WC1
(the two middle panels in Fig.~\ref{fig:cti}) the effect is clearly present at all time scales and 
increasing with time. The strongest effect is also seen here for $\tau=2000$~TDI periods.
For WC2 ($G\gtrsim 16$, in the right-most panel) the effect is mainly seen for $\tau=10$~TDI periods.
The jumps at OBMT 2400~rev in several of the data series are real and caused by the M-class solar 
flare on 21 June 2015 (see Fig.~14 in \citeads{2016A&A...595A...6C}).

\section{Parallax factor and AC rate}
\label{sec:plxAc}

\begin{figure}
\centering
  \includegraphics[width=0.8\hsize]{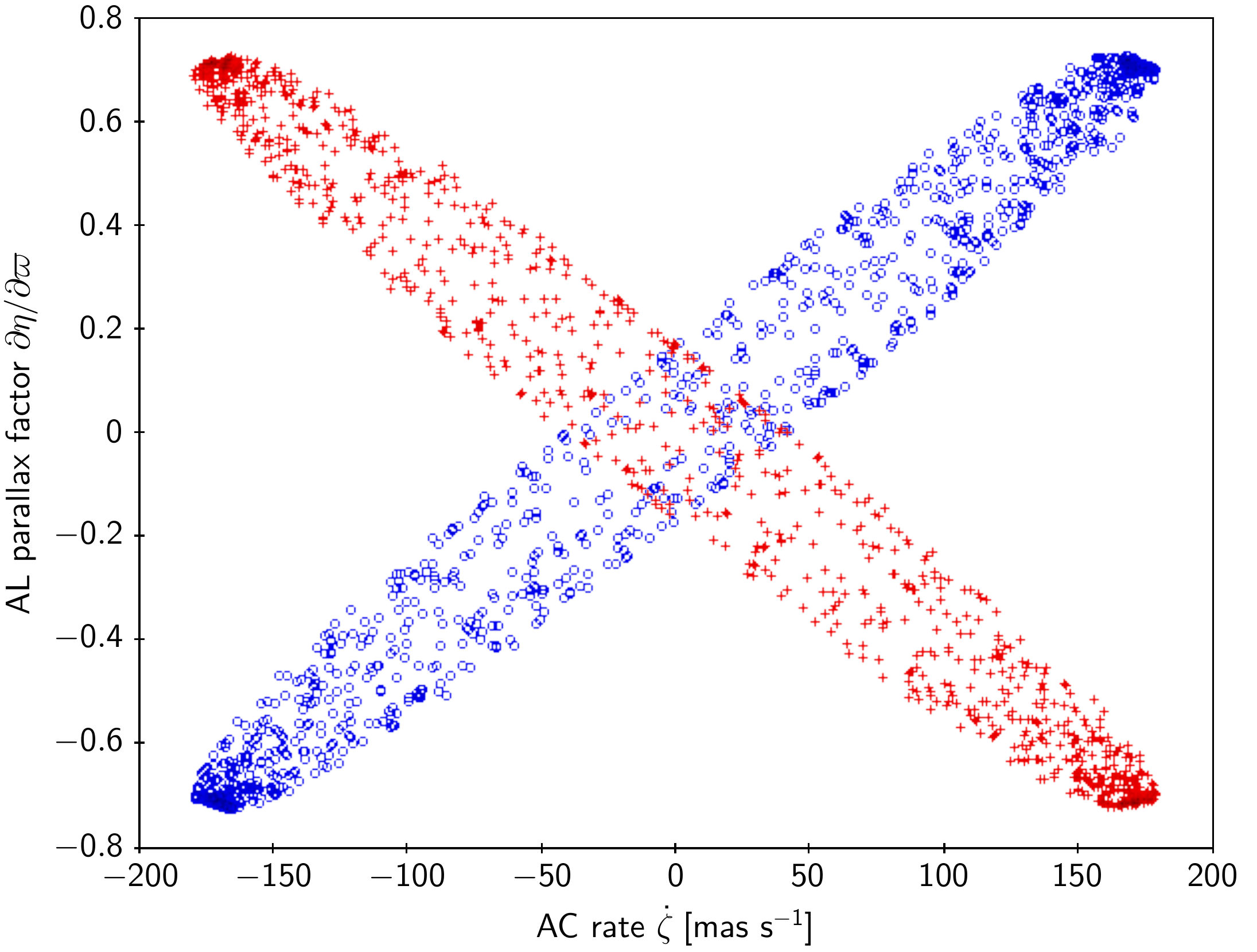}
    \caption{Correlation between AC rate and AL parallax factor.  Blue circles show a random 
    selection of 1000 FoV transits from data segments DS0--DS5, when the scanning law was 
    in its normal (forward precession) mode. Red crosses show 1000 random transits from 
    data segments DS6 and DS7, when the reversed precession mode was used. 
    EDR3 is exclusively based on observations taken in the forward precession mode.}
    \label{fig:acRatePlxFact}
\end{figure}

The scanning law of \textit{Gaia}, described in Sect.~5.2 of \citetads{2016A&A...595A...1G}, 
specifies the intended (commanded) pointing of the \textit{Gaia} telescopes as a function of time. In its 
nominal mode (the nominal scanning law, NSL), it causes a strong positive correlation between 
the AL parallax factor $\partial\eta/\partial\varpi$ and the AC scan rate $\dot{\zeta}$, where
$\eta$, $\zeta$ are the field angles of the source in either of \textit{Gaia}'s FoVs
(Fig.~\ref{fig:fpa}) and the dot signifies the time derivative. This
correlation is illustrated by the blue circles in Fig.~\ref{fig:acRatePlxFact} having a 
correlation coefficient of $+0.985$. As shown by the red crosses in the figure, the correlation 
can be reversed by changing the sense in which the spin axis revolves around the 
direction to the Sun. This mode, known as reversed precession, was used during data segments 
DS6 and DS7 (16~July 2019 to 29~July 2020; see Fig.~\ref{fig:timeline}).

\begin{figure}
\centering
  \includegraphics[width=0.65\hsize]{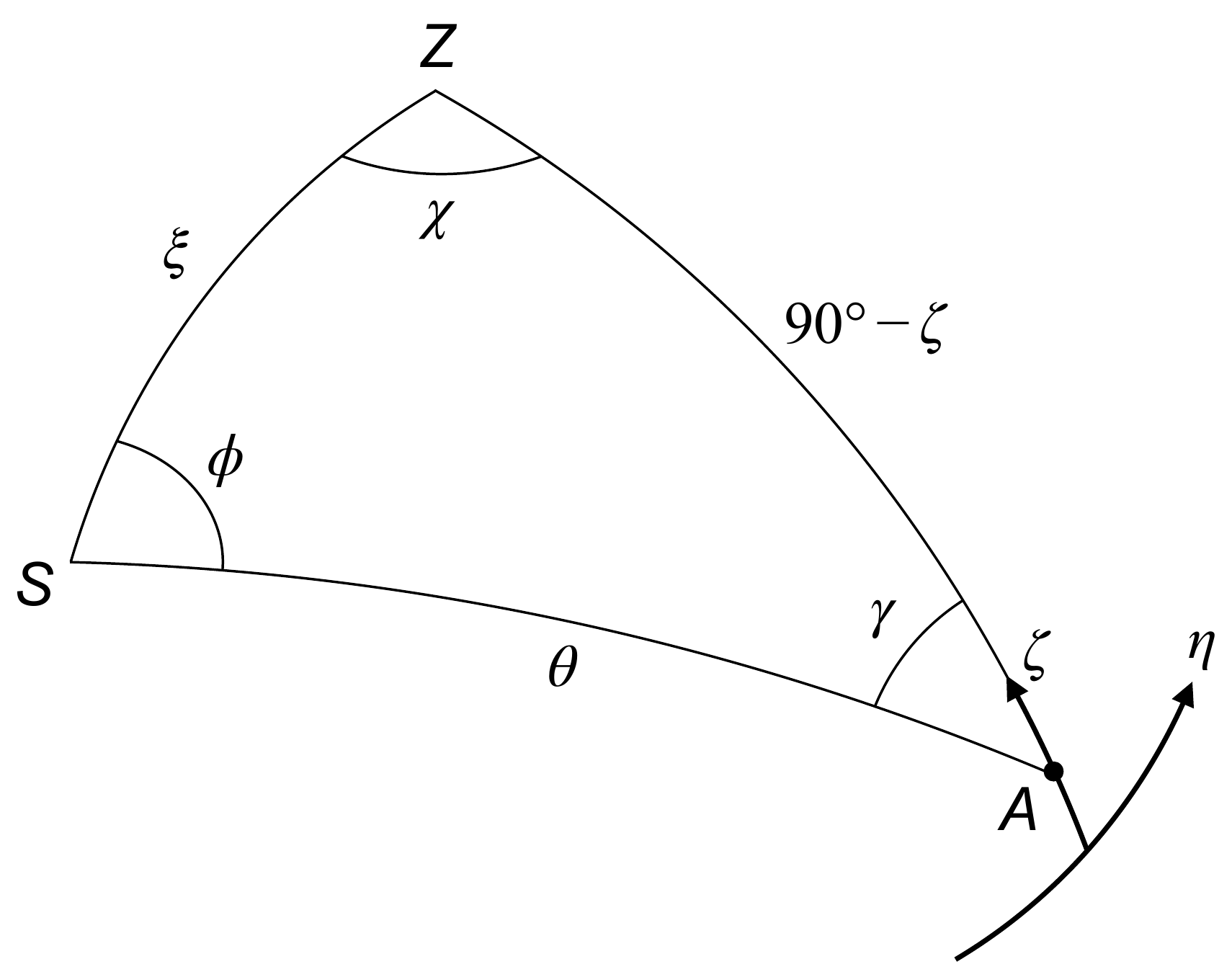}
    \caption{Spherical triangle for the parallax factor and AC rate. \textit{A} is the position 
of the star, \textit{Z} the nominal spin axis of \textit{Gaia} (perpendicular to the two 
viewing directions), and \textit{S} the position of the Sun. The directions of the AL and
AC field angles $\eta$, $\zeta$ are indicated.}
    \label{fig:C}
\end{figure}

The correlation between the AL parallax factor and the AC rate of a stellar image is a simple 
consequence of the scanning law and can be understood by considering the spherical 
triangle \textit{AZS} in Fig.~\ref{fig:C}. The diagram depicts the geometry at an instant 
when the star at \textit{A} is in the centre of the FoV in the AL direction, that is $\eta=0$. 
For the star to be inside the FoV at this time, the AC field angle must be small, 
$|\,\zeta\,|\lesssim 0.4^\circ$. According to the scanning law, the angle between 
the Sun (\textit{S}) and the spin axis (\textit{Z}) is fixed at $\xi=45^\circ$, while \textit{Z} 
revolves around \textit{S} at a rate of 5.8 revolutions per year (precession period about 
63~days). In the normal (forward) precession mode of the scanning law, used during 
most of the mission, \textit{Z} revolves in the positive sense around \textit{S}, so
$\dot{\phi}>0$; in reversed precession mode it revolves in the opposite sense, so
$\dot{\phi}<0$. It should be noted that the spin of \textit{Gaia} 
(with 6~h period) is always positive about \textit{Z} ($\dot{\chi}>0$), independent of 
the precession mode.

Parallax $\varpi$ causes a displacement of the star image by $p=\varpi d\sin\theta$
in the direction towards the Sun, that is along the great circle \textit{AS}. Here, $d$ is 
the Sun--\textit{Gaia} distance in au and $\theta$ the angle from \textit{A} to \textit{S}.
(Here, \textit{S} should be understood as the solar system barycentre, and $d$ as the 
distance from the solar system barycentre to \textit{Gaia}, that is $d_\text{b}$ in 
Sect.~\ref{sec:C10}. However, for the present discussion -- unlike the one in 
Sect.~\ref{sec:C10} -- the distinction between barycentric and heliocentric quantities 
is not important.)
The AL component of $p$ is $\Delta\eta\cos\zeta=-p\sin\gamma$, with 
$\gamma$ the angle at \textit{A} in the spherical triangle. The AL parallax factor is therefore
\begin{equation}\label{eq:plxFact}
\frac{\partial\eta}{\partial\varpi}=-d\sin\theta\sin\gamma\sec\zeta
=-d\sin\xi\sin\chi\sec\zeta\, ,
\end{equation}
where, in the last equality, we have used $\sin\theta\sin\gamma=\sin\xi\sin\chi$ 
from the law of sines.%
\footnote{Neglecting the size of the FoV, we have $\chi=\Omega+f\Gamma/2$, where
$\Omega$ is the heliotropic spin phase, $f=\pm 1$ the FoV index, and $\Gamma=106.5^\circ$
the basic angle. The differential parallax factor (preceding minus following) discussed in
Sect.~\ref{sec:C10} is then $-2d\sin\xi\sin(\Gamma/2)\cos\Omega$ 
\citepads{2017A&A...603A..45B}.\label{fn:Omega}}

The AC field angle $\zeta$ is obtained from the law of cosines, 
\begin{equation}\label{eq:zeta}
\sin\zeta = \cos\xi\cos\theta+\sin\xi\sin\theta\cos\phi \, ,
\end{equation}
which upon differentiation gives $\dot{\zeta}$ in terms of $\dot{\xi}$, $\dot{\theta}$, 
and $\dot{\phi}$. According to the scanning law we have $\dot{\xi}=0$, while expressions 
for $\dot{\theta}$ and $\dot{\phi}$ are complicated by the need to take into account the 
motion of the Sun along the ecliptic in addition to the precession. However, as the motion 
of the Sun is substantially slower than the precession, we may in a first-order approximation 
regard both Sun and star as stationary on the sky during an observation, in which case 
$\dot{\theta}\simeq 0$. Then 
\begin{equation}\label{eq:acRate}
\dot{\zeta} \simeq -\dot{\phi}\sin\xi\sin\theta\sin\varpi\sec\zeta
= -\dot{\phi}\sin\xi\sin\chi \, ,
\end{equation}
where we have used $\sin\theta\sin\phi=\cos\zeta\sin\chi$ from the law of sines. 
Comparing Eqs.~(\ref{eq:plxFact}) and (\ref{eq:acRate}), while recalling that $\zeta$ is
a small angle so $\sec\zeta\simeq 1$, we see that both the AL parallax factor and the 
AC rate vary as $\sin\chi$ with nearly constant amplitudes, yielding a very strong  
correlation between the two quantities. We also see that the correlation 
has the same sign as $\dot{\phi}$, that is positive in the nominal case and negative 
for reversed precession.

In contrast to the first-order analysis above, Fig.~\ref{fig:C} does not show a perfect
correlation between the AC rate and AL parallax factor. This is caused by the motion of
the Sun, ignored in Eq.~(\ref{eq:acRate}). A more careful analysis shows that $\dot{\zeta}$ 
is not completely in phase with $\partial\eta/\partial\varpi$, but is phase shifted 
by an amount that varies periodically with the precession period of 63~days and an 
amplitude of $\pm 13^\circ$. The elliptical envelopes of the data points in Fig.~\ref{fig:C} 
are produced by this phase shift. Additional, much smaller modulations are due to
variations in the Sun--\textit{Gaia} distance ($d$) and the neglected difference
between the nominal Sun, which regulates the scanning law, and the solar system 
barycentre, which determines parallax.

Because the astrometric solution for EDR3 could not benefit from the decorrelation
achieved with the reversed precession beginning in July 2019, it is possible that the
EDR3 parallaxes are biased for the sources where the AC smearing has a significant 
impact on the image parameter determination. From the analysis of residuals we know 
that this is the case for WC0b observations using the gates with `long' exposure times, 
which is the reason why the calibration effect depending on the square of the AC rate 
(effect~7 in Table~\ref{tab:cal}) was restricted to those observations. A subsequent 
study of the parallax biases in EDR3 \citep{EDR3-DPACP-132} indeed shows a sharp
discontinuity of the bias at $G\simeq 13$ (the faint limit of the affected observations,
according to Fig.~\ref{fig:obsFreqVsWcGate}), which could be caused by a location
bias proportional to the AC rate in the individual observations, coupled with the
positive correlation between the AC rate and parallax factor. A secure disentangling 
of the AC rate dependency from parallax will only be possible in cycle~4 with the
inclusion of observations obtained in the reversed precession mode.

\section{Adding photometric information in a six-parameter solution}
\label{sec:col6p}

In this Appendix we discuss the possibility, mentioned in Sect.~\ref{sec:nuEff},
to compute improved estimates of the astrometric parameters for a source
with a six-parameter solution, when a better colour estimate is available than the 
astrometrically determined pseudocolour $\hat{\nu}_\text{eff}$. The colour could 
be $G_\text{BP}-G_\text{RP}$, if available, or a colour index from a
different instrument. A prerequisite for the method is that the photometric
colour index, and its uncertainty, can be transformed into an estimate $\nu_\text{p}$
of the effective wavenumber, and a corresponding uncertainty $\sigma(\nu_\text{p})$.

For a given source with six-parameter solution, let $\hat{\alpha}$, $\hat{\delta}$, 
$\hat{\varpi}$, $\hat{\mu}_{\alpha*}$, $\hat{\mu}_\delta$, and $\hat{\nu}_\text{eff}$ 
be the parameters as published in EDR3 and $\phat{\vec{K}}$ the
$6\times 6$ covariance matrix computed as in Eq.~(\ref{eq:cov}).
Given also the photometric estimate $\nu_\text{p}\pm\sigma(\nu_\text{p})$, we
seek a vector of updates,
\begin{equation}\label{eq:x}
{\vec{x}} = \begin{bmatrix} 
({\alpha}-\hat{\alpha})\cos\hat{\delta} \\ 
{\delta}-\hat{\delta} \\ 
{\varpi}-\hat{\varpi} \\ 
{\mu}_{\alpha*}-\hat{\mu}_{\alpha*} \\ 
{\mu}_\delta-\hat{\mu}_\delta \\
{\nu}_\text{eff}-\hat{\nu}_\text{eff} \end{bmatrix} \, ,
\end{equation}
that optimally combine the six-parameter solution with the photometric data. We use the
tilde to indicate the updated solution, thus $\smash{\widetilde{\vec{x}}}$ for the optimal update
and $\smash{\widetilde{\vec{K}}}$ for its covariance matrix. We use here the notation
$\sigma(\hat{\varpi})$, $\sigma(\widetilde{\varpi})$, etc.\ for the uncertainties.

On the assumption of a multivariate normal distribution of the errors, the problem can be 
solved in a Bayesian framework, taking the original and updated parameters as prior and 
posterior estimates, and the colour information as the data. The same result can be obtained 
by considering the normal equations for the corresponding least-squares problems, which
is the approach taken here.

The original six-parameter solution may be represented by the update vector $\hat{\vec{x}}=\vec{0}$
with covariance $\phat{\vec{K}}$. The corresponding system of normal equations is
\begin{equation}\label{eq:Nhat}
\phat{\vec{K}}^{-1}\vec{x} = \vec{0} \, .
\end{equation}
It is assumed that observation equations are normalised to unit variance, so that the covariance 
of the least-squares estimate is obtained as the inverse of the normal matrix. 

The unit variance observation equation representing the photometric estimate 
$\nu_\text{p}\pm\sigma(\nu_\text{p})$ is
\begin{equation}\label{eq:u}
\vec{u}'\vec{x}\,\sigma(\nu_\text{p})^{-1} 
= (\nu_\text{eff}-\hat{\nu}_\text{eff})\,\sigma(\nu_\text{p})^{-1} \, ,
\end{equation}
where $\vec{u}$ is the column vector $(0,0,0,0,0,1)'$ (the prime indicates transpose). 
The system of normal equations obtained by adding this observation to the original system reads
\begin{equation}\label{eq:Nhatp}
\Bigl(\phat{\vec{K}}^{-1}\!+\vec{u}\vec{u}'\,\sigma(\nu_\text{p})^{-2}\Bigr) \,\vec{x} 
= \vec{u}\,(\nu_\text{eff}-\hat{\nu}_\text{eff})\,\sigma(\nu_\text{p})^{-2} \, .
\end{equation}
By means of the Sherman--Morrison formula (e.g.\ \citeads{press2007numerical}), the square matrix in
the left member can be inverted to give the covariance matrix of the updated solution,
\begin{equation}\label{eq:Ctilde}
\begin{split}
\widetilde{\vec{K}} &= \Bigl(\phat{\vec{K}}^{-1}+\vec{u}\vec{u}'\sigma(\nu_\text{p})^{-2}\Bigr)^{-1}\\
&= \phat{\vec{K}} - \phat{\vec{K}}\vec{u}\left(\vec{u}'\phat{\vec{K}}\vec{u}
+\sigma(\nu_\text{p})^2\right)^{-1}\!\vec{u}'\phat{\vec{K}}\, ,
\end{split}
\end{equation}
and hence the updated solution
\begin{equation}\label{eq:xtilde}
\widetilde{\vec{x}} 
%= \widetilde{\vec{K}}\vec{u}\,(\nu_\text{eff}-\hat{\nu}_\text{eff})\,\sigma(\nu_\text{p})^{-2}
= \phat{\vec{K}}\vec{u}\left(\vec{u}'\phat{\vec{K}}\vec{u}+\sigma(\nu_\text{p})^2\right)^{-1}
(\nu_\text{eff}-\hat{\nu}_\text{eff})\, .
\end{equation}
The last two equations are readily written in component form thanks to the simple structure of 
$\vec{u}$; thus,
%Using subscripts $i,~j=0\dots 5$ for the components of $\widetilde{\vec{x}}$ and 
%$\widetilde{\vec{K}}$ we find
\begin{equation}\label{eq:xtildeComp}
\widetilde{x}_i = \phat{K}_{i5}\frac{\nu_\text{p}-\hat{\nu}_\text{eff}}{\phat{K}_{55}+\sigma(\nu_\text{p})^2}\, ,
\qquad i=0\dots 5  
\end{equation}
and
\begin{equation}\label{eq:CtildeComp}
\widetilde{K}_{ij} = \phat{K}_{ij}-\frac{\phat{K}_{i5}\,\phat{K}_{j5}}{\phat{K}_{55}+\sigma(\nu_\text{p})^2}\, ,
\qquad i,\,j=0\dots 5\, .
\end{equation}
For example, the updated parallax ($i=2$) is
\begin{equation}\label{eq:varpiTilde}
\widetilde{\varpi} = \hat{\varpi}+\rho(\hat{\varpi},\hat{\nu}_\text{eff})\,\sigma(\hat{\varpi})\,\sigma(\hat{\nu}_\text{eff})
\frac{\nu_\text{p}-\hat{\nu}_\text{eff}}{\sigma(\hat{\nu}_\text{eff})^2+\sigma(\nu_\text{p})^2}\, ,
\end{equation}
with uncertainty $\sqrt{\widetilde{K}_{22}}$, that is
\begin{equation}\label{eq:varVarpiTilde}
\sigma(\widetilde{\varpi}) = \sigma(\hat{\varpi})
\sqrt{1 - \frac{\rho(\hat{\varpi},\hat{\nu}_\text{eff})^2}{1+\sigma(\nu_\text{p})^2/\sigma(\hat{\nu}_\text{eff})^2}}\, . 
\end{equation}
Corresponding expressions hold for the other parameters. For the effective wavenumber ($i=5$),
they can be written
\begin{equation}\label{eq:nuTilde}
\widetilde{\nu}_\text{eff} = \frac{\sigma(\hat{\nu}_\text{eff})^{-2}\hat{\nu}_\text{eff}+
\sigma(\nu_\text{p})^{-2}\nu_\text{p}}{\sigma(\hat{\nu}_\text{eff})^{-2}+\sigma(\nu_\text{p})^{-2}} \, ,
\end{equation}
with uncertainty
\begin{equation}\label{eq:varNuTilde}
\sigma(\widetilde{\nu}_\text{eff}) = \left(\sigma(\hat{\nu}_\text{eff})^{-2}+\sigma(\nu_\text{p})^{-2}\right)^{-1/2}\, . 
\end{equation}

A few interesting observations can be made concerning the last four equations.
We note that the parallax and its uncertainty are unchanged if 
$\sigma(\nu_\text{p})\gg\sigma(\hat{\nu}_\text{eff})$,
or if there is no correlation between the parallax and pseudocolour, $\rho(\hat{\varpi},\hat{\nu}_\text{eff})=0$. If $\nu_\text{p}=\hat{\nu}_\text{eff}$, the parallax value is also unchanged, 
but its uncertainty will decrease if the correlation is non-zero.
We note, furthermore, that the parallax uncertainty is at most reduced by the factor 
$[1-\rho(\hat{\varpi},\hat{\nu}_\text{eff})^2]^{1/2}$, which is 
reached in the limit when $\sigma(\nu_\text{p})\ll\sigma(\hat{\nu}_\text{eff})$.
This means that the potential gain in precision by the procedure 
may be significant (say, more than 5\%),
only if the correlation between the pseudocolour and the astrometric parameter of interest is 
$\gtrsim 0.3$ in absolute value. For the parallax, this is the case for only about 6\% of the sources with 
six-parameter solutions. The median $|\,\rho(\cdot,\hat{\nu}_\text{eff})\,|$ is about 0.1 for all five
astrometric parameters, giving a median improvement in their uncertainties of at most 0.5\%. 
Finally, we note that $\widetilde{\nu}_\text{eff}$ is the mean of $\hat{\nu}_\text{eff}$ 
and $\nu_\text{p}$ weighted by their
inverse variances, with uncertainty corresponding to the sum of weights.

Equation~(\ref{eq:varpiTilde}) shows that the updated parallaxes $\widetilde{\varpi}$ are formally 
more precise than the original values $\hat{\varpi}$ (for non-zero correlations); however, we want 
to determine whether they actually are better. The sample of the quasars in 
\gacs{agn\_cross\_id} with six-parameter solutions offers an opportunity to test this, 
although the sample is not representative for most six-parameter solutions in EDR3, 
as the quasars are usually not in crowded areas. Of the 398\,231 
sources in \gacs{agn\_cross\_id} with six-parameter solutions, 396\,445 have colour indices 
$G_\text{BP}-G_\text{RP}$ in the main table. To transform the colours to $\nu_\text{p}$ we use
Eq.~(\ref{eq:nuEff1}), from which we also have an expression for the uncertainty:
\begin{equation}\label{eq:bpRpErr}
\sigma(\nu_\text{p})=\frac{1.61}{\pi}\,\frac{0.531\,\sigma(G_\text{BP}-G_\text{RP})}%
{1+\bigl[0.531(G_\text{BP}-G_\text{RP})\bigr]^2}\, .
\end{equation}
In this sample the effective wavenumber derived from the photometric colour is usually much more 
precise than the pseudocolour: The median $\sigma(\nu_\text{p})$ is 0.03~$\mu$m$^{-1}$ against
a median $\sigma(\hat{\nu}_\text{eff})$ of 0.18~$\mu$m$^{-1}$. Thus most of the sources should
benefit from the procedure, which is confirmed by the statistics in Table~\ref{tab:6pC}. The median
formal uncertainty is reduced by 1.2\% for the full sample and by 7.7\% for the subsample with 
correlations exceeding $\pm 0.3$. That these improvements are actual and not only formal is shown 
by the dispersions (RSE) of the parallaxes, which are reduced by, respectively, 1.7\% and 8.1\%. 
The dispersions of the normalised parallaxes (last line in the table) are practically unchanged
by the update.

The quasar sample thus demonstrates that the procedure is capable of bringing a real and possibly
significant improvement to the astrometry of a six-parameter solution under specific circumstances.
Necessary conditions are that the correlation coefficients with pseudocolour are significant, and that
a reliable colour is available. Although these conditions hold for a number of the quasars analysed above,
they may not apply to more than a small fraction of the sources with six-parameter solutions. 
It should also be remembered that 
$G_\text{BP}-G_\text{RP}$, if available in EDR3, may be problematic for these sources. After all, 
they received six-parameter solutions because they did not have reliable BP and RP photometry in 
DR2, and the reason for that, such as crowding, may still be present in EDR3. The necessary colour 
information could of course come from a different instrument with better angular resolution than 
the BP and RP photometers of \textit{Gaia}.

\begin{table}
\caption{Statistics of original and updated parallaxes for quasars with six-parameter solutions.
\label{tab:6pC}}
\small
\begin{tabular}{lcrrrr}
\hline\hline
\noalign{\smallskip}
&&\multicolumn{2}{c}{Full sample} & \multicolumn{2}{c}{Subsample} \\
 && \multicolumn{2}{c}{(396\,445 sources)} & \multicolumn{2}{c}{(28\,806 sources)} \\
Quantity  & Unit &  $\hat{\varpi}$~~ & $\widetilde{\varpi}$~~ & $\hat{\varpi}$~~ & $\widetilde{\varpi}$~~ \\ 
\noalign{\smallskip}
\hline
\noalign{\smallskip}
$\text{med}(x)$ &  [$\mu$as] & $-28.6$ & $-28.3$ & $-41.5$ & $-40.8$\\
$\text{med}\left(\sigma(x)\right)$ &  [$\mu$as]  &  836 & 826 & 1297 & 1197 \\
$\text{RSE}(x)$ &  [$\mu$as]  & 973 & 956 & 1426 & 1311 \\
$\text{RSE}\left(\frac{x-\text{med}(x)}{\sigma(x)}\right)$ &  -- & 1.073 & 1.074 & 1.062 & 1.067 \\
\noalign{\smallskip}
\hline
\end{tabular}
\tablefoot{Columns~3 and 4 give statistics on the original parallaxes ($\hat{\varpi}$) and updated
values ($\widetilde{\varpi}$) for the full sample of quasars having six-parameter solutions and
colours in EDR3. 
Columns~5 and 6 give statistics on the subsample with $|\,\rho(\hat{\varpi},\hat{\nu}_\text{eff})\,|>0.3$. 
The first two lines of data give the median parallax and median parallax uncertainty, the last two lines 
the robust scatter estimate (RSE; see footnote~\ref{fn:RSE}) of the parallaxes and of the median-centred 
parallaxes normalised by their uncertainties.}
\end{table}

\end{document}